\documentclass[a4paper,12pt]{article}
\usepackage[top=2cm,bottom=2cm,left=2.5cm,right=2.5cm]{geometry}
\usepackage[utf8]{inputenc}
\usepackage{amsmath,amssymb}
\usepackage[colorlinks=true]{hyperref}
\usepackage{graphicx}
\usepackage{color}
\usepackage{float}
\usepackage{sidecap}

\usepackage{epsfig,cancel,amsmath,amsthm,amssymb}
\usepackage{color}

\usepackage{tikz}
\usetikzlibrary{decorations.fractals}
\usepackage{wrapfig}
\usepackage{color}
\usepackage{graphicx,framed}
\definecolor{shadecolor}{rgb}{0.95, 0.95, 0.86}

\newcommand{\bt}{\beta}

\newcommand{\G}{\Gamma}

\newcommand{\g}{\gamma}
\newcommand{\la}{\lambda}

\renewcommand{\part}{\partial}

\newcommand{\br}{{\mathbb R}}

\newtheorem{theorem}{Theorem}[section]
\newtheorem{example}{Example}[section]
\newtheorem{exercise}{Exercise}[section]

\newtheorem{lemma}{Lemma}[section]
\newtheorem{remark}{Remark}[section]

\newtheorem{proposition}{Proposition}[section]
\newtheorem{corollary}{Corollary}[section]
\newtheorem{definition}{Definition}[section]
\newtheorem{conjecture}{Conjecture}[section]

\def\le{\left}
\def\ri{\right}

\def\bt{\begin{theorem}}
\def\et{\end{theorem}}
\def\bc{\begin{corollary}}
\def\ec{\end{corollary}}
\def\bx{\begin{example}\small}
\def\ex{\end{example}}
\def\bxr{\begin{exercise}\small}
\def\exr{\end{exercise}}
\def\bl{\begin{lemma}}
\def\el{\end{lemma}}
\def\bd{\begin{definition}}
\def\ed{\end{definition}}
\def\bp{\begin{proposition}}
\def\ep{\end{proposition}}

\def\br{\begin{remark}}
\def\er{\end{remark}}

\def\be{\begin{equation}}
\def\ee{\end{equation}}
\def\bea{\begin{eqnarray}}
\def\eea{\end{eqnarray}}
\def\beas{\begin{eqnarray*}}
\def\eeas{\end{eqnarray*}}

\def\R{{\mathbb R}}

\def\g{\gamma}

\def\m{\mu}
\def\l{\lambda}

\def\1{{\bf 1}}

\def\s{ {\sigma}}

\def\x{\xi}
\def\z{\zeta}

\def\hf{\frac{1}{2}}

\newcommand{\Rscr}{\mathcal R}
\numberwithin{equation}{section}

\newcommand{\bs}{\boldsymbol}

\title{Dispersive hydrodynamics of soliton condensates for the Korteweg-de Vries equation}

\author{T.~Congy, G.A.~El, G.~Roberti, and A.~Tovbis}

 \date{}

\begin{document}

\maketitle


\begin{abstract}
  We consider large-scale dynamics of non-equilibrium dense soliton gas for the Korteweg-de Vries (KdV) equation in the special ``condensate'' limit. We prove that in this limit  the integro-differential kinetic equation for the spectral density of states  reduces to the $N$-phase  KdV-Whitham modulation equations derived by Flaschka, Forest and McLaughlin (1980) and Lax and Levermore (1983). We consider   Riemann problems  for  soliton condensates and construct 
  explicit solutions of the kinetic equation describing generalized rarefaction and dispersive shock waves.  We then present  numerical results for ``diluted'' soliton condensates exhibiting  rich incoherent  behaviours associated with integrable turbulence. 
\end{abstract}

\section{Introduction}
Solitons represent the fundamental localised solutions of integrable nonlinear dispersive equations such as the Korteweg-de Vries (KdV), nonlinear Schr\"odinger (NLS), sine-Gordon, Benjamin-Ono and other equations. Along with the remarkable localisation properties solitons exhibit particle-like  elastic  pairwise  collisions accompanied by definite phase/position shifts.  A comprehensive description of solitons and their interactions is achieved within the inverse scattering transform (IST) method framework,  where each soliton is characterised by a certain spectral parameter related to the soliton's amplitude, and the phase related to its position (for the sake of definiteness we refer here to the properties of KdV solitons).
Generally, integrable equations support $N$-soliton solutions which can be viewed as  nonlinear superpositions of $N$ solitons.  Within the IST framework $N$-soliton solution is characterised by a finite set of spectral and phase parameters  completely determined by the initial conditions for the integrable PDE. 

The particle-like properties of solitons suggest  some  natural questions pertaining to the realm of statistical mechanics, e.g. one can consider a {\it soliton gas} as an infinite ensemble of interacting solitons characterised by random  spectral (amplitude) and phase distributions.  The key question  is to understand the emergent macroscopic dynamics (i.e. hydrodynamics or kinetics) of soliton gas given the  properties of  the elementary, ``microscopic''  two-soliton interactions. It is clear that, due to the presence of an infinite number of conserved quantities and the lack of thermalisation in integrable systems the properties of soliton gases will be very different compared to the properties of  classical gases whose particle interactions are non-elastic.
Invoking the wave aspect of the soliton's dual identity, soliton gas can be viewed as a prominent example of {\it integrable turbulence} \cite{zakharov_turbulence_2009}.  The pertinent questions arising in this connection are related to 
the determination of the parameters of the random nonlinear wave field in the soliton gas such as probability density function,  autocorrelation function,  power spectrum etc.

The IST-based phenomenological construction of a {\it rarefied}, or diluted, gas of KdV solitons was  proposed in 1971 by V. Zakharov \cite{zakharov_kinetic_1971} who has formulated an approximate spectral kinetic equation for such a gas based on the  properties of soliton collisions: the conservation of the soliton spectrum (isospectrality) and the accumulation of phase shifts in pairwise collisions that results in the modification of an effective average soliton's velocity in the gas. 
Zakharov's spectral kinetic equation was generalised in \cite{el_thermodynamic_2003} to the case of a dense gas using the finite gap theory and the thermodynamic, infinite-genus, limit of the KdV-Whitham modulation equations  \cite{flaschka_multiphase_1980}.  The results of \cite{el_thermodynamic_2003} were used in \cite{el_kinetic_2005} for the formulation of a phenomenological construction of kinetic equations for dense soliton gases for integrable systems describing both unidirectional and bidirectional soliton propagation and including the focusing, defocusing and resonant NLS equations, as well as  the Kaup-Boussinesq system for  shallow-water waves \cite{congy_soliton_2021}. The detailed spectral theory of soliton and breather gases for the focusing NLS equation has been developed in \cite{el_spectral_2020}.  

The spectral kinetic equation for a dense soliton gas represents a
nonlinear integro-differential  equation describing the evolution of  the {\it density of states} (DOS)---the density function $u(\eta; x,t)$ 
in the phase space $(\eta, x) \in \Gamma^+ \times \mathbb{R}$, 
where  $\eta \in \Gamma^+ \subset \mathbb{R}^+$ is 
the spectral parameter in the Lax pair  associated with the nonlinear integrable PDE,
  \begin{equation} \label{kineq}
u_t + (us)_x=0, \quad
s(\eta, x,t)=s_0(\eta)+ \int_{\Gamma^+} G(\eta, \mu)u(\mu, x,t)[s(\eta, x,t)-s(\mu, x,t)]d\mu\,   .
\end{equation}
Here $s_0(\eta)$ is the velocity of a ``free'' soliton, and  the integral term in the second equation represents the effective modification of the soliton velocity in the gas due to   pairwise soliton collisions that are accompanied by  the phase-shifts described by the kernel $G(\eta, \mu)$. Both $s_0(\eta)$ and $G(\eta, \mu)$ are system specific. In particular, for KdV $s_0=4\eta^2$ and $G(\eta, \mu)= \tfrac{1}{\eta}\ln \le|\tfrac{\m+\eta}{\m-\eta}\ri|$. The spectral support $\Gamma^+$ of the DOS is determined by initial conditions. 
 We note that, while 
 $\Gamma^+ \subset \mathbb{R}^+$ for the KdV equation, one can have
 $\Gamma^+ \subset \mathbb{C}^+$ for other equatons, e.g. the focusing
 NLS equation, see \cite{el_spectral_2020} . Equation \eqref{kineq}
 describes the DOS evolution in a {\it dense} soliton gas and
 represents a broad generalisation of Zakharov's  kinetic equation for
 rarefied gas  \cite{zakharov_kinetic_1971}.  The existence, uniqueness and properties of solutions to the ``equation of state'' (the integral equation in \eqref{kineq} for fixed $x,t$)  for the focusing NLS and KdV equations were studied in \cite{kuijlaars_minimal_2021}.

The original spectral theory of the KdV soliton gas \cite{el_thermodynamic_2003} has been developed under the assumption that the spectral support $\Gamma^+$ of the DOS is a fixed, simply-connected interval of $\mathbb{R}^+$;  without loss of generality one can assume $\Gamma^+=[0,1]$. In \cite{el_spectral_2020}, \cite{kuijlaars_minimal_2021}  this restriction has been removed by allowing the spectral support $\Gamma^+$ to be a union of $N+1$ disjoint intervals $\g_j=[\lambda_{2j-1}, \lambda_{2j}]$, termed here {\it s-bands}: $\Gamma^+= \cup_{j=0}^{N} \gamma_j$, ($\gamma_i \cap \gamma_j= \emptyset$, $i \ne j$). 
 In this paper we introduce a further generalization of the existing theory by  allowing the endpoints $\lambda_i$ of the s-bands be functions of $x,t$.  We show that this generalization has profound implications for soliton gas dynamics, in particular, the kinetic equation implies certain nonlinear evolution of the endpoints $\lambda_j(x,t)$ of the s-bands. We determine this evolution for
 a special type of soliton gases, termed in \cite{el_spectral_2020} {\it soliton condensates}. Soliton condensate represents the ``densest possible'' gas whose DOS is uniquely defined by a given spectral support $\Gamma^+$. The number $N$ of disjoint s-bands in $\Gamma^+$ determines the {\it genus} $g=N-1$ of the soliton condensate.
 We show that the evolution of $\lambda_j$'s in a soliton condensate is governed by the $g$-phase averaged KdV-Whitham modulation equations \cite{flaschka_multiphase_1980}, also derived in the context of the semi-classical, zero-dispersion limit of the KdV equation 
 \cite{lax_small_1983}. 
 
 We then consider the soliton condensate dynamics arising in the Riemann problem initiated by a rapid jump in the DOS. Our results  suggest that in the condensate limit the KdV dynamics of soliton gas is almost everywhere equivalent to the (deterministic) generalised rarefaction waves (RWs) and  generalized dispersive shock waves (DSWs) of the KdV equation. We prove this statement for the genus zero case and present a strong numerical evidence for genus one.
Our results also suggest direct connection of the ``deterministic KdV soliton gases''  constructed in the recent paper \cite{girotti_rigorous_2021} with modulated soliton condensates.

Our work puts classical results of integrable dispersive hydrodynamics (Flaschka-Forest-McLaughlin \cite{flaschka_multiphase_1980}, Lax-Levermore \cite{lax_small_1983}, Gurevich-Pitaevskii \cite{gurevich_nonstationary_1974}) in a  broader  context of the soliton gas theory. Namely, we show that the KdV-Whitham modulation equations   describe  the emergent  hydrodynamic motion  of a special soliton gas---a condensate---resulting from the accumulated effect of ``microscopic'' two-soliton interactions. This new interpretation of the Whitham equations is particularly pertinent in the context of  generalized hydrodynamics, the emergent hydrodynamics of quantum and classical many-body systems \cite{doyon_lecture_2020}. The direct connection between the kinetic theory of KdV soliton gas and generalized hydrodynamics was established recently  in \cite{bonnemain_generalized_2022} (see also \cite{bettelheim_} where the Whitham equations for the defocusing NLS equation were shown to arise in the semi-classical limit of the generalised hydrodynamics of the quantum Lieb-Liniger model).

Our work also paves the way to a major extension of the existing dispersive hydrodynamic theory by including the random aspect of soliton gases. To this end we consider ``diluted'' soliton condensates whose DOS has the same spectral distribution as in genuine condensates but allows for a wider spacing between solitons giving rise to rich incoherent dynamics associated with ``integrable turbulence'' \cite{zakharov_turbulence_2009}. In particular, we show numerically that evolution of initial discontinuities in diluted soliton condensates  results in the development of incoherent oscillating rarefaction and dispersive shock waves.

An important aspect of our work is the numerical modelling of soliton condensates using $n$-soliton KdV solutions with large $n$  configured according to the condensate density of states. The challenges of the numerical implementation of standard $n$-soliton formulae for sufficiently large $n$ due to rapid accumulation of roundoff errors are known very well.  Here we use the efficient algorithm  developed in \cite{huang1992darboux},
which relies on the Darboux transformation.  We improve this algorithm following the recent methodology 
 developed in \cite{gelash2018strongly} for the focusing NLS equation with the implementation of  high
precision arithmetic routine.  Our numerical simulations show excellent agreement with analytical predictions for  the solutions of soliton condensate Riemann problems and  provide a strong support to the basic conjecture about the connection of KdV soliton condensates with finite-gap potentials. 

It should be noted that soliton condensates have been recently  studied for the focusing NLS equation, where they represent incoherent wave fields  exhibiting distinct statistical properties. In particular, it was shown in \cite{gelash_prl2019} that the so-called bound state soliton condensate dynamics underies  the long-term behavior of  spontaneous modulational instability, the fundamental physical phenomenon that gives rise to the statistically stationary integrable turbulence \cite{agafontsev_2015, suret_2021}.    

The paper is organised as follows. In
Section~\ref{sec:specrtal_theory_summary} we present a brief outline
of the spectral theory of soliton gas for the KdV equation and
introduce the notion of soliton condensate for the simplest genus zero
case. In Section~\ref{sec:gen_sol_cond}, following \cite{kuijlaars_minimal_2021}, 
we generalize the spectral
definition of soliton condensate to an arbitrary genus case and prove
the main Theorem~\ref{theo-Whit} connecting spectral dynamics of
non-uniform soliton condensates with multiphase Whitham modulation
theory \cite{flaschka_multiphase_1980} describing slow deformations of the spectrum of 
periodic and quasiperiodic KdV
solutions. Section~\ref{sec:gen_sol_cond} is concerned with properties
of KdV solutions corresponding to the condensate spectral DOS,
i.e. the soliton condensate realizations. We formulate
Conjecture~\ref{prop4.1} that any realization of an equilibrium soliton
condensate almost surely coincides with a finite-gap potential defined
on the condensate's hyperelliptic spectral curve. This proposition is
proved for genus zero condensates and a strong numerical evidence is
provided for genus one and two. In Section~\ref{sec:Riemann} we construct
solutions to Riemann problems for the soliton gas kinetic equation
subject to discontinuous condensate initial data. These solutions
describe evolution of generalized rarefaction and dispersive shock
waves. In Section~\ref{sec:numeric} we present numerical simulations of  the Riemann problem for the KdV soliton condensates and compare them with analytical  solutions from Section~\ref{sec:Riemann}. 
 Finally, in Section~\ref{sec:dilute} we consider basic properties of
 ``diluted'' condensates having a scaled condensate DOS and exhibiting rich
incoherent behaviors. In particular, we present numerical solutions to Riemann problems for such diluted condensates.  Appendix A contains details of the
numerical implementation of dense soliton gases. In Appendix B we present results of the numerical realization of the genus 2 soliton condensate and its comparison with two-phase solution of the KdV equation.

\section{Spectral theory of KdV soliton gas:  summary of results}
\label{sec:specrtal_theory_summary}

Here we present an outline of the spectral theory of KdV soliton gas developed in \cite{el_thermodynamic_2003, el_soliton_2021}.
We consider the  KdV equation in the form
\begin{equation}
  \label{kdv}
  \varphi_t +6\varphi \varphi_x+\varphi_{xxx} = 0.
\end{equation}
The inverse scattering theory  associates  soliton  of the KdV equation~\eqref{kdv} with a point $z=z_1=-\eta_1^2$,  $\eta_1>0$  of the discrete spectrum  of the  Lax operator
\begin{equation}\label{schr}
  \mathcal{L}= -\partial_{xx}^2  - \varphi(x,t),
\end{equation}
with sufficiently rapidly decaying potential $\varphi(x,t)$: $\varphi(x,t) \to 0$ as $|x| \to \infty$.
The corresponding KdV soliton solution  is given by
\begin{equation}\label{1sol}
  \varphi_{\rm s}(x,t;\eta_1) = 2 \eta_1^2 \hbox{sech}^2 [\eta_1(x-4\eta_1^2 t - x_1^0)],
\end{equation}
where the soliton amplitude $a_1= 2\eta_1^2$,  the  speed $s_1= 4\eta_1^2$, and $x_1^0$ is the initial position or  `phase'.
Along with the simplest single-soliton solution~\eqref{1sol} the KdV equation supports $N$-soliton solutions $\varphi_n(x,t)$ characterized by $n$ discrete spectral parameters $0< \eta_1 < \eta_2 < \dots <\eta_n$ and the set of initial positions  $\{x_i^0 | i=1, \dots, n\}$ associated with the phases of the so-called norming constants \cite{novikov_theory_1984}. It is also known that $n$-soliton solutions can be realized as special limits of more general  $n$-gap solutions, whose Lax spectrum $\mathcal{S}_n$ consists of $N$ finite and one semi-infinite bands separated by $n$ gaps \cite{novikov_theory_1984},
\begin{equation}\label{spectralset}
  z \in \mathcal{S}_n=[\zeta_1, \zeta_2] \cup [\zeta_3, \zeta_4] \cup \dots \cup [\zeta_{2n+1}, \infty) .
\end{equation}
The $n$-gap solution of the KdV equation~\eqref{kdv} represents a multiphase quasiperiodic function
\begin{equation} \label{N-gap}
  \begin{split}
    &\varphi(x,t)=F_n(\theta_1, \theta_2, \dots, \theta_n), \quad \theta_j=k_jx-\omega_j t + \theta_j^0, \\
    &F_n(\dots,  \theta_j+ 2 \pi, \dots) = F_n(\dots,  \theta_j, \dots),
  \end{split}
\end{equation}
where $k_j$ and $\omega_j$ are the wavenumber and frequency associated with the $j$-th phase $\theta_j$, and $\theta_j^0$ are the initial phases. Details on the explicit representation of the solution \eqref{N-gap} in terms of Riemann theta-functions can be found in classical papers and monographs on finite-gap theory, see  \cite{matveev_30_2008} and references therein.

The $n$-phase ($n$-gap) KdV solution~\eqref{N-gap} is parametrized by $2n+1$ spectral parameters---the endpoints $\{\z_j \}_{j=1}^{ 2n+1}$ of the spectral bands.
The nonlinear dispersion relations  (NDRs) for finite gap potentials can be represented in the general form, see  \cite{flaschka_multiphase_1980} for the concrete expressions,
\begin{equation} \label{fg_NDR}
  k_j=K_j(\z_1, \dots, \z_{2n+1}), \quad \omega_j=\Omega_j(\z_1, \dots, \z_{2n+1}), \quad j=1, \dots, n,
\end{equation}
---and connect the wavenumber-frequency set $\{k_j, \omega_j\}_{j=1}^n$ of~\eqref{N-gap} with the spectral set $\mathcal{S}_n$~\eqref{spectralset}.
These are complemented by the relation $\langle{\varphi}\rangle =\Phi (\z_1, \dots, \z_{2n+1})$, where
$\langle \varphi \rangle = \int F_n d\theta_1 \dots d\theta_n$ is the mean obtained by averaging of $F_n$ over the phase $n$-torus $\mathbb{T}^n=[0, 2\pi) \times \dots \times [0, 2 \pi)$, assuming respective non-commensurability of $k_j$'s and  $\omega_j$'s and, consequently, ergodicity of the KdV flow on the torus.

The $n$-soliton limit of an $n$-gap solution is achieved by collapsing all the finite bands $[\z_{2j-1}, \z_{2j}]$ into double points corresponding to the soliton discrete spectral values,
\begin{equation} \label{fgap-sol}
  \z_{2j} -\z_{2j-1} \to 0, \ \ \z_{2j},  \z_{2j-1} \to - \eta_j^2,  \ \ j=1, \dots ,n.
\end{equation}
It was proposed in \cite{el_thermodynamic_2003} that the special infinite-soliton limit of the spectral $n$-gap KdV solutions, termed the thermodynamic limit, provides spectral description the KdV soliton gas. The thermodynamic limit  is achieved by assuming a special band-gap distribution (scaling) of the spectral set $\mathcal{S}_n$ for $n \to  \infty$ on a fixed interval $[\z_{1}, \z_{2n+1}]$ (e.g. $[-1, 0]$). Specifically, we set  the spectral bands to be exponentially narrow compared to the gaps
so that $\mathcal{S}_n$ is asymptotically characterized by two continuous nonnegative functions on some fixed interval  $\G^+ \subset \mathbb{R}^+$: the density $\phi(\eta)$ of the lattice points $\eta_j \in \G^+$ defining the band centers via $-\eta_j^2=(\z_{2j}+\z_{2j-1})/2$, and the logarithmic bandwidth distribution $\tau(\eta)$ defined for $n \to \infty$ by
\begin{equation} \label{therm_scale}
  \eta_j-\eta_{j+1}  \sim  \frac{1}{n \phi(\eta_j)}, \quad  \tau(\eta_j) \sim -\frac{1}{n} \ln (\z_{2j}-\z_{2j-1}).
\end{equation}
The scaling~\eqref{therm_scale} was originally introduced by Venakides \cite{venakides_continuum_1989} in the  context of the continuum limit of theta functions.

Complementing the spectral distributions~\eqref{therm_scale}  with the uniform distribution of  the initial phase vector  ${\bs \theta}^{0}$  on the torus
$ \mathbb{T}^n$ we say that the resulting random finite gap solution  $\varphi(x,t)$ approximates {\it soliton gas} as $n \to \infty$. An important consequence of this definition of soliton gas  is ergodicity, implying that spatial averages of the KdV field in a soliton gas are equivalent to the  ensemble averages, i.e. the averages  over $\mathbb{T}^n$  in the thermodynamic limit  $n \to \infty$.
We shall use the notation $\langle F[\varphi] \rangle$ for ensemble averages and $\overline{F[\varphi]}$ for spatial averages.

From now on we shall refer to $\eta$ as the spectral parameter and $\G^+$--the spectral support. The density of states (DOS)  $u(\eta)$ of a spatially homogeneous (equilibrium) soliton gas is phenomenologically introduced in such a way that $u(\eta_0)d\eta dx$ gives the number of solitons with the spectral parameter $ \eta \in [\eta_0; \eta_0 + d \eta]$  contained  in the portion of soliton gas over a macroscopic (i.e. containing sufficiently many solitons) spatial interval  $x \in [x_0, x_0+dx] \subset \mathbb{R}$ for any $x_0$ (the individual solitons can be counted by cutting out the relevant portion of the gas and letting them separate with time). The corresponding spectral flux density $v(\eta)$ represents the temporal counterpart of the DOS i.e. $v(\eta_0) d \eta$ is the number of solitons with the spectral parameter $ \eta \in [\eta_0; \eta_0 + d \eta]$ crossing any given  point $x=x_0$ per unit interval of time. These definitions are physically suggestive in the context of rarefied soliton gas where solitons are identifiable as individual localized wave structures.  The general mathematical definitions of $u(\eta)$ and $v(\eta)$ applicable to dense soliton gases   are introduced  by applying the thermodynamic limit  to the finite-gap NDRs~\eqref{fg_NDR}, leading to the integral equations \cite{el_thermodynamic_2003, el_soliton_2021}:
\begin{eqnarray}
  &&\int_{\Gamma^+} \ln \le|\frac{\m+\eta}{\m-\eta}\ri|
  u(\m)d\m+u(\eta)\s(\eta)= \eta,  \label{ndr_kdv1} \\
  &&\int_{\Gamma^+} \ln \le|\frac{\m+\eta}{\m-\eta}\ri|
  v(\m)d\m+v(\eta)\s(\eta)= 4  \eta^3  \label{ndr_kdv2},
\end{eqnarray}
for all $\eta\in \Gamma^+ $.
Here the {\it spectral scaling function}
$\sigma: \Gamma^+ \to [0,\infty)$ is a continuous non-negative function that encodes the Lax spectrum of soliton gas  via $\sigma(\eta)=\phi(\eta)/\tau(\eta)$.  Equations~\eqref{ndr_kdv1}, \eqref{ndr_kdv2}  represent the soliton gas NDRs.

Eliminating $\sigma(\eta)$ from the NDRs~\eqref{ndr_kdv1}, \eqref{ndr_kdv2} yields the {\it equation of state} for KdV soliton gas:
\begin{equation} \label{s_kin}
  s(\eta) = 4\eta^2 + \frac{1}{\eta} \int_{\Gamma^+} \log \left|
  \frac{\eta+\mu }{\eta-\mu}
  \right| u(\mu) [s(\eta) - s(\mu)] d\mu,
\end{equation}
where $s(\eta)= v(\eta)/u(\eta)$ can be interpreted as the velocity of a {\it tracer soliton} in the gas. It was shown in \cite{el_thermodynamic_2003} that
for  a weakly non-uniform (non-equilibrium) soliton gas, for which
$u(\eta)\equiv u(\eta;x,t)$, $s(\eta) \equiv s(\eta; x, t)$, the DOS satisfies the continuity equation
\begin{equation} \label{cont_1}
  u_t + (us)_x=0,
\end{equation}
so that $s(\eta; x, t)$ acquires the natural meaning of the transport velocity in the soliton gas. Equations~\eqref{cont_1}, \eqref{s_kin} form the spectral kinetic equation for soliton gas. One should note that the typical scales of spatio-temporal variations in the kinetic equation~\eqref{cont_1} are much larger than in the KdV equation~\eqref{kdv}, i.e. the kinetic equation describes macroscopic evolution, or hydrodynamics, of soliton gases.

Let the spectral support $\Gamma^+$ be fixed. Then, differentiating equation~\eqref{ndr_kdv1} with respect to $t$,  equation~\eqref{ndr_kdv2}  with respect to $x$, and using the continuity equation~\eqref{cont_1} we obtain the evolution equation for the spectral scaling function
\begin{equation} \label{sigma}
  \sigma_t + s \sigma_x = 0,
\end{equation}
which shows that $\sigma(\eta;x,t)$ plays the role of the Riemann invariant for the spectral kinetic equation.

Finally, the ensemble averages of the conserved densities  of the KdV wave field in the soliton gas (the Kruskal integrals) are evaluated in terms of the DOS as  $\langle {\cal P}_n[\varphi] \rangle = C_n \int _{\G^+} \eta^{2n-1} u(\eta)d \eta$,
where  ${\cal P}_n[\varphi]$ are conserved quantities of the KdV equation and $C_n$ constants \cite{el_thermodynamic_2003, el_soliton_2021} (see also \cite{TovbisWang} for rigorous derivation in the NLS context).
In particular, for  the two  first moments we have, on dropping the $x,t$-dependence \cite{el_thermodynamic_2003, el_soliton_2021},
\begin{equation} \label{eq:moments}
  \langle \varphi \rangle = 4\int_{\Gamma^+} \eta \, u(\eta) d\eta,  \quad
  \langle \varphi^2 \rangle = \frac{16}{3} \int_{\Gamma^+} \eta^3 \,
  u(\eta) d\eta .
\end{equation}
We note that in the original works on KdV soliton gas it was  assumed (explicitly or implicitly) that the spectral support $\Gamma^+$ of the KdV soliton gas is a fixed, simply connected interval  (without loss of generality one can assume that in this case $\Gamma^+=[0,1]$).  In what follows we will be considering a more general configuration where  $\Gamma^+$ represents a union of $N+1$ disjoint intervals.

A special kind of soliton gas, termed {\it soliton condensate}, is realized spectrally by letting $\sigma \to 0$ in the NDRs~\eqref{ndr_kdv1}, \eqref{ndr_kdv2}. This limit was first considered in \cite{el_spectral_2020} for the soliton gas in the focusing NLS equation and then in \cite{kuijlaars_minimal_2021} for KdV. Loosely speaking, soliton condensate can be viewed as the ``densest possible'' gas (for a given spectral support $\G^+$) whose  properties  are fully determined by the interaction (integral) terms in the NDRs~\eqref{ndr_kdv1}, \eqref{ndr_kdv2}.

For the KdV equation, setting $\sigma = 0$ and, considering the simplest  case $\Gamma^+=[0,1]$ in~\eqref{ndr_kdv1}, \eqref{ndr_kdv2}, we
obtain the soliton condensate NDRs \cite{el_soliton_2021}:
\begin{equation} \label{NDR_cond0}
  \int_0^1 \ln \le|\frac{\m+\eta}{\m-\eta}\ri|
  u(\m)d\m = \eta,  \quad \int_0^1 \ln \le|\frac{\m+\eta}{\m-\eta}\ri|
  v(\m)d\m = 4  \eta^3.
\end{equation}
These are solved by
\begin{equation}\label{uv0}
  u(\eta)= \frac{\eta}{\pi\sqrt{1-\eta^2}},\quad
  v(\eta)=  \frac{6\eta(2\eta^2-1)} {\pi \sqrt{1-\eta^2}},
\end{equation}
as verified by direct substitution (it is advantageous to  first differentiate equations~\eqref{NDR_cond0} with respect to $\eta$). The formula~\eqref{uv0} for $u(\eta)$ is sometimes called the Weyl distribution, following the terminology from the semiclassical theory of linear differential operators \cite{lax_small_1983}.

\br \label{r:backflow}
The meaning of the zero $\eta_0=  1/{\sqrt{2}}$ of $v(\eta)$ is that all the tracer solitons with the spectral parameter $\eta>\eta_0$
move to the right, whereas  all the tracer solitons  with $\eta<\eta_0$ move  in the opposite
direction while the tracer soliton with $\eta=\eta_0$ is stationary.   The somewhat counter-intuitive ``backflow'' phenomenon (we remind that KdV solitons considered in isolation move to the right)  has been observed    in the numerical simulations of the KdV soliton gas  \cite{pelin} and can be readily understood  from the phase shift formula of two interacting solitons, where the the larger  soliton gets a kick forward upon the interaction while the smaller soliton is pushed back.
As a matter of fact, the  KdV soliton backflow  is general and can be observed  for a broad range of sufficiently dense gases (see Fig.~\ref{fig:backflow} in Section~\ref{sec:dilute_eq} for the numerical illustration).
\er

\section{Soliton condensates and their modulations}
\label{sec:gen_sol_cond}

We now consider  the general case of the soliton gas NDRs~\eqref{ndr_kdv1}, \eqref{ndr_kdv2}  by letting the support $\G^+  \subset \mathbb{R}^+$ of $u(\eta), v(\eta)$ to be  a union of disjoint intervals $\g_k \subset \R^+$ with endpoints $\lambda_j>0$, $j=1,2,\dots,2N+1$,
where $\g_0=[0,\lambda_1]$  and $\g_k=[\lambda_{2k},\lambda_{2k+1}]$, $k=1,\dots,N$, i.e.
\begin{equation} \label{G^+N}
  \Gamma^+  = [0, \lambda_1] \cup [\lambda_2, \lambda_3] \cup \dots \cup[\lambda_{2N}, \lambda_{2N+1}].
\end{equation}
We shall call the intervals $\g_k$ the {\it s-bands}, and the soliton gas spectrally supported on $\Gamma^+$~\eqref{G^+N}--- the {\it genus $N$ soliton gas}.
Correspondingly, we refer to  the intervals $c_j=(\lambda_{2j-1}, \lambda_{2j})$ separating
the s-bands as to s-gaps.
Note that the s-bands  and s-gaps are different from the original bands and gaps in the spectrum $\mathcal{S}_n$ of finite-gap potential (cf.~\eqref{spectralset}) as they emerge {\it after} the passage to the thermodynamic limit: loosely speaking, one can view the s-bands  as a continuum limit of the ``thermodynamic band clusters'',  each representing an isolated dense subset of $\mathcal{S}_\infty$ consisting of the collapsing original bands.
The existence and uniqueness of solutions $u(\eta)$,  $v(\eta)$ for~\eqref{ndr_kdv1},
\eqref{ndr_kdv2} respectively,  as well as the fact that $u(\eta)\geq 0$
on $\G^+$ with some mild constraints, was established in \cite{kuijlaars_minimal_2021}.
Our goal here is to find explicit expressions for $u,v$ for the genus $N$ soliton condensate, that is, solutions of~\eqref{ndr_kdv1}, \eqref{ndr_kdv2} for the particular case $\s\equiv 0$ on $\G^+$.

Denote by $\G^-$ the symmetric image of $\G^+$ with respect to the origin, i.e., $\G^-=-\G^+$.
If we take the odd continuation of $u,v$ to  $\G^-$ (preserving the same notations), we observe
that equations~\eqref{ndr_kdv1}, \eqref{ndr_kdv2} become
\begin{eqnarray}
  &&-\int_\G\ln |\m-\eta|
  u(\m)d\m+u(\eta)\s(\eta)=  \eta,  \label{NDR-KdV-sym1} \\
  &&-\int_\G\ln|\m-\eta|
  v(\m)d\m+v(\eta)\s(\eta)= 4  \eta^3,  \label{NDR-KdV-sym2}
\end{eqnarray}
where $\G:=\G^+\cup\G^-$,
for all $\eta\in \G^+$.
In fact, if we symmetrically extend $\s(\eta)$ from $\G^+$ to $\G$, equations~\eqref{NDR-KdV-sym1}, \eqref{NDR-KdV-sym2} should be valid on $\G$ since every term in these equations is odd.  The expressions~\eqref{eq:moments} for the first two moments (ensemble averages) of the KdV wave field in the soliton gas become
\begin{equation}
  \label{eq:moment1}
  \langle \varphi \rangle =  2 \int_\Gamma \eta \, u(\eta) d\eta,\ \qquad
  \langle \varphi^2 \rangle =  \frac{8}{3}  \int_{\Gamma}\eta^3 \,
  u(\eta)
  d\eta.
\end{equation}
We now consider soliton condensate of genus $N$ by setting $\s\equiv 0$ in~\eqref{NDR-KdV-sym1}, \eqref{NDR-KdV-sym2}. Then,
differentiating in $\eta$ we obtain
\begin{equation}\label{Hilbert-eq}
  H[u]=\frac 1 \pi, \quad H[v]=\frac{12\eta^2}{\pi} \quad\text{on $\G$},
\end{equation}
where $H$ denotes the Finite Hilbert Transform (FHT) on $\G$,  see for example \cite{tricomi_finite_1951}, \cite{okada_finite_1991},
\begin{equation}
  H [f] (\xi)= \frac{1}{\pi} \int \limits_{\G}  \frac{ f(y)dy}{y-\x}.
\end{equation}
Equations~\eqref{Hilbert-eq} are the (transformed) NDRs for the KdV soliton condensate.

To find $u,v$ for the soliton condensate, it is sufficient to invert the FHT $H$ on $\G$.
Denote by $\Rscr_{2N}$ the hyperelliptic  Riemann surface   of the genus $2N$, defined by the branchcuts (s-bands) $\g_k$,
$k=0,\pm 1,\dots, \pm N$, where $\g_{-k}=-\g_k$. Define  two  meromorphic differentials of second kind, $dp$ and $dq$ on $\Rscr_{2N}$ by
\begin{equation}\label{KdVuv}
  dp=\frac{iP(\eta)}{2\pi R(\eta)} d \eta, \qquad dq= \frac{2iQ(\eta)}{\pi R(\eta)}d\eta,
\end{equation}
where
\begin{equation} \label{RS_N}
  R(\eta)=\sqrt{ (\eta^2-\lambda_1^2)(\eta^2-\lambda_2^2)\dots (\eta^2-\lambda_{2N+1}^2)},
\end{equation}
and $P,Q$ are odd monic polynomials of degree $2N+1$ and $2N+3$ respectively that are chosen so
that all their s-gap integrals are zero,
i.e.
\begin{equation} \label{dpdq_norm}
  \int \limits_{\la_{2j-1}}^{\la_{2j}} dp = \int \limits_{\la_{2j-1}}^{\la_{2j}} dq  =0, \quad j=1, \dots, N.
\end{equation}

Equivalently, one can say that  $dp,dq$ are real normalized differentials. Note that Equations~\eqref{KdVuv}, \eqref{dpdq_norm}   uniquely define $dp,dq$.

\bt \label{theo-2N}
Functions $u(\eta)=dp/d\eta$  and $v(\eta)=dq/d\eta$ defined by~\eqref{KdVuv} and~\eqref{dpdq_norm} satisfy the respective equations~\eqref{Hilbert-eq} and  are odd and real valued on $\G$. Thus $u, v$ are the solutions of NDRs~\eqref{ndr_kdv1}, \eqref{ndr_kdv2} for $\sigma=0$. Moreover, $u(\eta)\geq 0$ on $\G^+$. Here the value of $R(\eta)$ for $\eta\in\G$ is taken on the positive (upper) shore  of the branchcut.
\et

Theorem \ref{theo-2N} for $u$ was proven in  \cite{kuijlaars_minimal_2021}, Section 6, for the so-called bound state soliton condensate. The proof for KdV is analogous. The proof for $v$ goes along the same lines, except $v(\eta)$ attains different signs.

\br
The normalization~\eqref{dpdq_norm} requires that  both   polynomials $P,Q$ have zeros in every of the $2N$ gaps on $[-\lambda_{2N+1},\lambda_{2N+1}]$.
Note also that $P(0)=Q(0)=0$.
That takes care of all the zeros of $P$. The polynomial $Q$ has two additional symmetric real zeros $\pm \eta_0$ that must be located
on some band $\g_{k}$ and its symmetrical image $\g_{-k}$, see below. In the case $N=0$ such zeros are $\eta_0=\pm\frac{1}{\sqrt{2}}$, see~\eqref{uv0}.
Let us prove that $\eta_0$ belongs to a band.
It is easy to see that the zero  level curves  $\Im \int_0^\eta dp=0$  consist of all bands and the imaginary axis,
whereas the zero  level curves  $\Im \int_0^\eta dq=0$ consist of that of   $\Im \int_0^\eta dp=0$ with an extra two curves
crossing $\R$ at $\pm \eta_0$ and approaching $z=\infty$ with angles $\pm \frac \pi 6$  and $\pm \frac {5\pi} 6$
respectively. Note that there must be four zero level curves passing through $\pm \eta_0$ and, therefore,
they must be on the bands.
\er

Thus,  for the soliton condensate of genus $N$, we obtain, on using Theorem \ref{theo-2N} and Equation~\eqref{KdVuv},
\begin{equation}\label{uv_c}
  u(\eta) \equiv u^{(N)}(\eta; \la_1, \dots \la_{2N+1}) = \frac{iP(\eta)}{2\pi R(\eta)}, \quad v(\eta) \equiv v^{(N)}(\eta; \la_1, \dots, \la_{2N+1})= \frac{2iQ(\eta)}{\pi R(\eta)}.
\end{equation}
\begin{figure}[h]
  \centering
  \includegraphics[width=7.5cm]{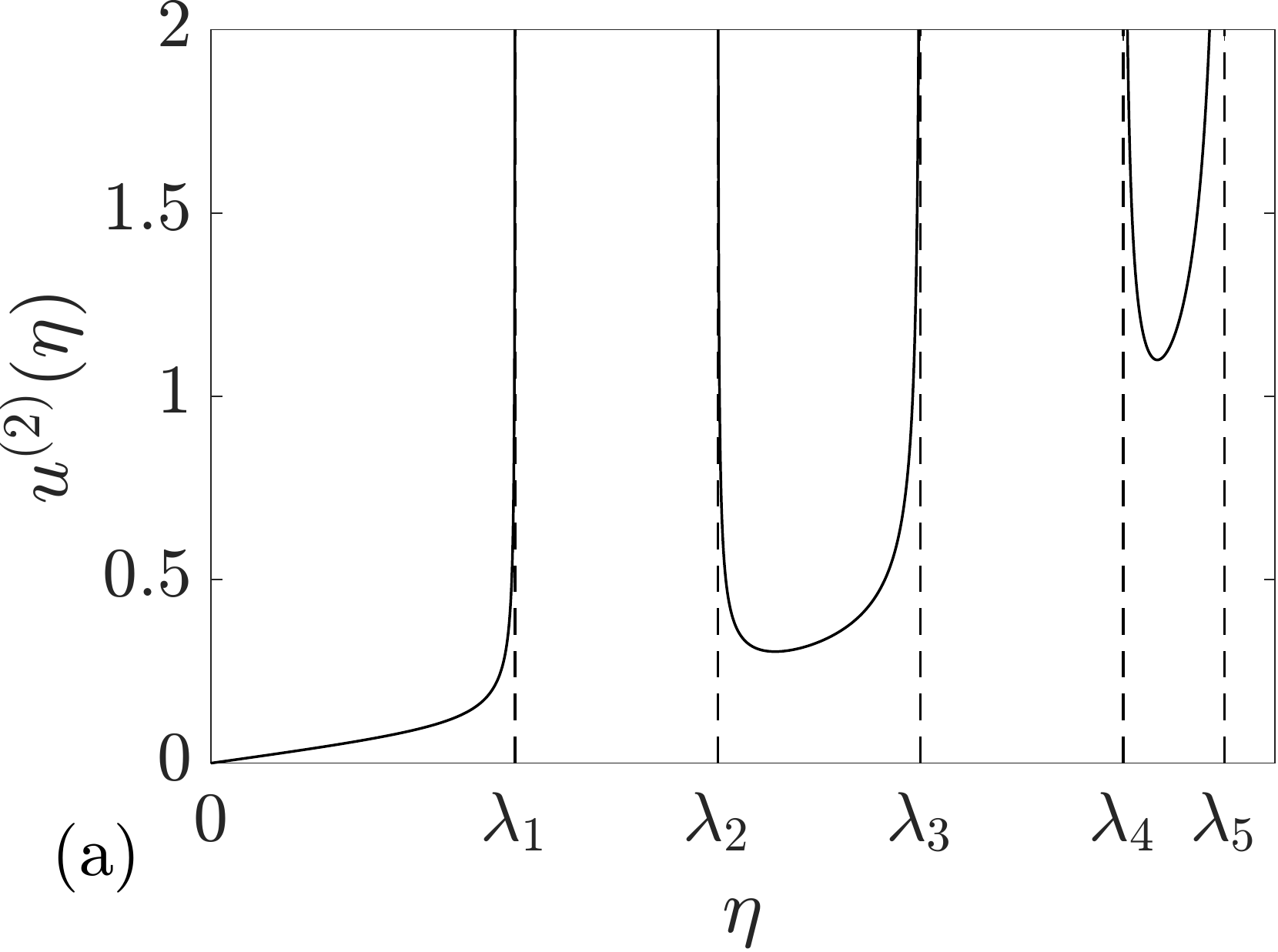} \hfill    \includegraphics[width=7.5cm]{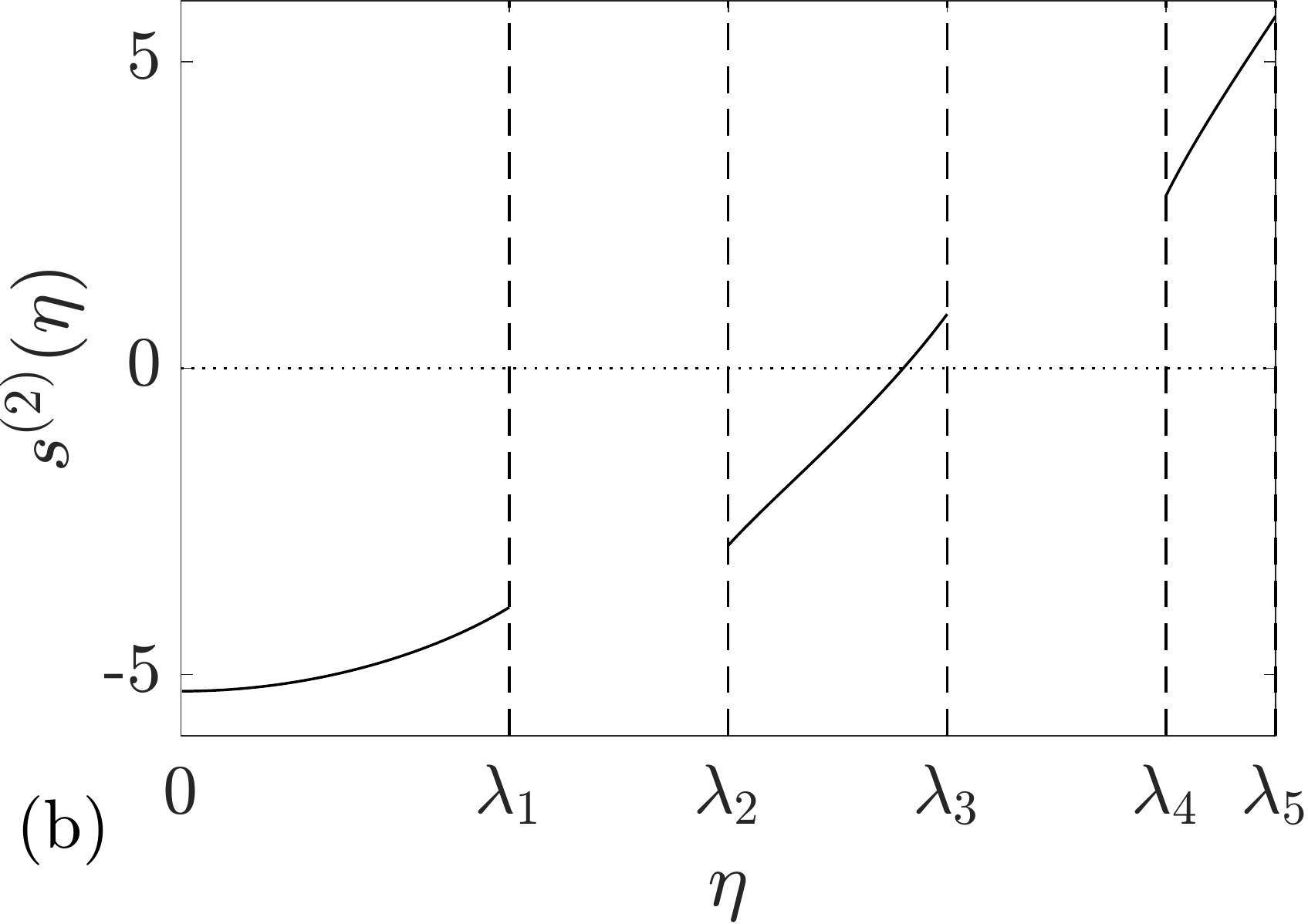}
  \caption{Spectral distributions~\eqref{uv_c} for genus 2 soliton condensate.  a) Density of states $u^{(2)}(\eta; {\bs \la})$. b) Tracer velocity $s^{(2)}(\eta; {\bs \la})$. Here
  ${\bs \la}=(\la_1, \la_2, \la_3, \la_4, \la_5)=(0.3, 0.5, 0.7, 0.9, 1)$.  }
  \label{fig:cond2}
\end{figure}
The velocity of a tracer  soliton with the spectral parameter $\eta \in \G^+$ propagating in the soliton condensate with DOS $u(\eta)$ is then found as
\begin{equation}\label{s_c}
  s(\eta) \equiv s^{(N)}(\eta; \la_1, \dots, \la_{2N+1}) = \frac{v(\eta)}{u(\eta)} = \frac{4Q(\eta)}{P(\eta)}.
\end{equation}
As an illustrative example we present in Fig.~\ref{fig:cond2} the plots of the DOS  and  tracer velocity  for the genus 2 soliton condensate.

We now consider slow modulations of  non-equilibrium (non-uniform) soliton condensates by assuming  $u \equiv u( \eta; x, t)$,
$v \equiv v( \eta; x, t)$, $\Gamma \equiv \Gamma(x,t)$.
Equations~\eqref{cont_1}, \eqref{uv_c} then yield the kinetic equation for genus $N$ soliton condensate:
\begin{equation}\label{KdV-kin}
  \le(\frac{P}{R}\ri)_t + \le(\frac{4Q}{R}\ri)_x =0,
\end{equation}
that is valid for $\eta \in \G=\cup_{k=-N}^N (\g_k)$.
The velocity~\eqref{s_c} then assumes the meaning of the tracer, or transport, velocity in a non-uniform genus $N$ soliton condensate.

\bt\label{theo-Whit}
The kinetic equation~\eqref{KdV-kin} for soliton condensate   implies the evolution of the endpoints $\l_j$, $j=1,\dots,2N+1$ according to the Whitham modulation equations
\begin{equation}\label{kdv-whitham}
  \partial_t  \lambda_j + V_j({\bs \lambda}) \partial_x \lambda_j = 0, \quad j=1, \dots, 2N+1,
\end{equation}
where
${\bs \lambda} = (\la_1, \dots, \la_{2N+1})$ and
\begin{equation} \label{Vj}
  V_j(\bs \lambda)=s^{(N)}(\lambda_j; \la_1, \dots, \la_{2N+1}) = \frac{4Q(\lambda_j)}{P(\lambda_j)}.
\end{equation}
\et

{\it Proof.}  (See \cite{dubrovin_hydrodynamics_1989}) Multiplying~\eqref{KdV-kin} by $(\eta^2 - \lambda_j^2)^{3/2}$  and passing to the limit $\eta \to \lambda_j$ we obtain equations~\eqref{kdv-whitham}, \eqref{Vj} for the evolution of the   spectral $s$-bands (i.e. the evolution of $\G(x,t)$).   These are the KdV-Whitham modulation equations \cite{flaschka_multiphase_1980}, \cite{lax_small_1983} (see also Remark~\ref{remark_3.2} below).

\qed

\bc
The endpoints of the ``special'' band $\g_k=[\la_{2k}, \la_{2k+1}]$, $k\neq 0$, containing the point $\eta_0$ of zero tracer speed, $s(\eta_0)=0$, are moving in  opposite directions, whereas all the  endpoints on the same side from $\eta_0$ are moving in the same direction. See Fig.~\ref{fig:cond2} (right) for $N=2$\ec

\br \label{remark_3.2}
Modulation equations~\eqref{KdV-kin}, \eqref{kdv-whitham} were originally derived by Flaschka, Forest and  McLaughlin \cite{flaschka_multiphase_1980}  by  averaging the KdV equation over the multiphase (finite-gap) family of solutions. These equations  along with the condensate NDRs~\eqref{Hilbert-eq}, also appear in the seminal work of  Lax and Levermore \cite{lax_small_1983}  in the context of the  semiclassical (zero-dispersion) limit of multi-soliton KdV ensembles (see Section 5 and, in particular, Equation (5.23) in \cite{lax_small_1983}).  A succinct  exposition of the spectral Whitham  theory for the KdV equation can be found in Dubrovin and Novikov \cite{dubrovin_hydrodynamics_1989}). \er

\smallskip
\br \label{cor_cond}The Whitham modulation equations~\eqref{kdv-whitham}, \eqref{Vj} are locally integrable for any $N$ via Tsarev's generalized hodograph transform \cite{tsarev_geometry_1991, dubrovin_hydrodynamics_1989}.
Moreover, by allowing the genus $N$ to take different values in different regions of $x,t$-plane, $N=N(x,t)$, global solutions of the KdV-Whitham system can be constructed for a broad class of initial data (see Section~\ref{sec:mod_dyn} for further details). Invoking the definitive property $\sigma \equiv 0$ of a soliton condensate, the existence of the solution to an initial value problem for the Whitham system for all $t>0$ implies that this property will remain invariant under the $t$-evolution, i.e. soliton condensate will remain a condensate during the evolution, however its genus can change.
\er

\smallskip
The finite-genus Whitham modulation system~\eqref{kdv-whitham}, \eqref{Vj} can be viewed as an exact hydrodynamic reduction of the full kinetic equation~\eqref{cont_1}, \eqref{s_kin} under the ansatz~\eqref{uv_c}, \eqref{s_c}.  Recalling the origin of the soliton gas kinetic equation  as a singular, thermodynamic  limit of the Whitham equations \cite{el_thermodynamic_2003} the recovery of the finite-genus Whitham dynamics in the condensate limit might not look surprising.   On the other hand, viewed from the general soliton gas perspective
the condensate reduction notably shows that the highly nontrivial nonlinear modulation (hydro)dynamics  emerges as a collective effect of the elementary two-soliton scattering events. This understanding is in line with ideas of generalised hydrodynamics, a powerful theoretical framework for the description of non-equilibrium macroscopic dynamics  in many-body quantum and classical integrable systems \cite{doyon}. The connection of the KdV soliton gas theory with generalised hydrodynamics has been recently established  in \cite{bonnemain_generalised_2022}.
Relevant to the above, it was shown in \cite{bettelheim} that the semiclassical limit of the generalised hydrodynamics  for the Lieb-Liniger model of Bose gases yields the Whitham modulation system  for the defocusing NLS equation.

A different type of hydrodynamic reductions of the soliton gas kinetic equation defined by the multi-component delta-function ansatz  $u(\eta, x,t)= \sum_{i=1}^{m} w_i(x,t)\delta(\eta -\eta_j)$ for the DOS  has been  studied  in \cite{el_kinetic_2011} for $\eta_j = const$ and  in \cite{pavlov_generalized_2012, ferapontov} for $\eta_j =\eta_j(x,t)$. One of the definitive properties of the multicomponent hydrodynamic reductions of this type is their linear degeneracy which, in particular, implies the absence of the wavebreaking and the  occurrence of contact discontinuities in the solutions of  Riemann problems \cite{carbone_macroscopic_2016}.  In contrast, the condensate (Whitham) system~\eqref{kdv-whitham}, \eqref{Vj} obtained under the condition $\sigma \equiv 0$ is  known to be {\it genuinely nonlinear}, $\partial V_j /\partial \la_j \ne 0$, $j=1, \dots 2N+1$ \cite{levermore_hyperbolic_1988} implying the inevitability of wavebreaking  for general initial data, which is in stark contrast with linear degeneracy of the  multicomponent ``cold-gas'' hydrodynamic reductions.  Reconciling  the  genuine nonlinearity property of soliton condensates with linearly degenerate non-condensate multicomponent cold-gas dynamics is an interesting problem which will be considered in future publications.

\medskip
Thus, we have shown that the spectral dynamics of soliton condensates are equivalent to those of finite gap potentials, which naturally suggests a close connection (or even  equivalence) of these two objects at the level of realizations, i.e. the corresponding solutions $\varphi(x,t)$ of the KdV equation. This connection will be explored in the next section using a combination of analytical results and numerical simulations for genus 0 and genus 1 soliton condensates.

\section{Genus 0 and genus 1 soliton condensates}
Having developed the spectral description of KdV soliton condensates, we now look closer at the two simplest representatives: genus 0 and genus 1 condensates. In particular, we shall be interested in the characterization of the {\it realizations}
of soliton condensates, i.e. the KdV solutions, denoted $\varphi_{\rm c}^{(N)}(x,t)$, corresponding to the condensate spectral DOS $u^{(N)}(\eta)$ for $N=0,1$.  We do not attempt here to  construct
the soliton gas realizations explicitly via the thermodynamic limit of finite gap potentials (see Section~\ref{sec:specrtal_theory_summary}), instead, we  infer some of their key  properties from the expressions~\eqref{eq:moment1} for the ensemble averages as integrals over the spectral DOS. We then conjecture the exact form of soliton condensate realizations and support our conjecture by detailed numerical simulations.

\subsection{Equilibrium properties}
\label{sec:equilibrium}

\subsubsection{Genus 0}
For  $N=0$  equations~\eqref{uv_c} for the DOS and the spectral flux density yield (cf.~\eqref{uv0})
\begin{equation}\label{uv}
  u(\eta)=u^{(0)}(\eta; \la_1) \equiv \frac{\eta}{\pi\sqrt{\la_1^2-\eta^2}},\quad
  v(\eta)= v^{(0)}(\eta;\la_1) \equiv  \frac{6\eta(2\eta^2-\la_1^2)} {\pi \sqrt{\la_1^2-\eta^2}},
\end{equation}
so that the tracer velocity (cf.~\eqref{s_c})
\begin{equation} \label{s_gen0}
  s(\eta)=s^{(0)}(\eta;\la_1) = 6(2\eta^2-\la_1^2).
\end{equation}

Next, substituting~\eqref{uv} in~\eqref{eq:moment1}
(where $\Gamma = [-\la_1,\la_1]$ or equivalently, $\Gamma^+ = [0,\la_1]$), we obtain for the ensemble averages:
\begin{equation}
  \langle \varphi \rangle = \la_1^2,\quad \langle \varphi^2 \rangle = \la_1^4,
\end{equation}
where $\varphi \equiv \varphi_{\rm c}^{(0)}(x,t)$.
Thus the variance $\Delta=\sqrt{\langle \varphi^2 \rangle -   \langle \varphi \rangle^2}=\sqrt{\langle( \varphi - \langle \varphi \rangle)^2 \rangle }= 0$,  which implies (see, e.g. \cite{rohatgi2015introduction}) that  genus 0 soliton condensate is
  {\it almost surely} described by a constant solution of the KdV equation , i.e.
\begin{equation}
  \varphi =   \langle \varphi \rangle = \la_1^2
\end{equation}
(note that constant solution is classified as a genus 0 KdV potential).

This result can be intuitively understood by identifying soliton condensate with the ``densest possible'' soliton gas for a given spectral support $\Gamma$.  The densest  ``packing''  for genus 0 is achieved  by distributing soliton parameters according to the spectral DOS $u(\eta)$~\eqref{uv} which results in the individual solitons ``merging'' into a uniform KdV field of amplitude $\la_1^2$.  The numerical implementation of soliton condensate realizations, using $n$-soliton KdV solution with
$n$ large, shows that the condensate DOS~\eqref{uv} is only achievable within this framework if all $n$ solitons in the solution have the same phase of the respective norming constants. Invoking the interpretation of the phase of the norming constant as the soliton position in space \cite{novikov_theory_1984, drazin_solitons_1989} one can say that in the condensate all solitons are placed at the same point, say $x=0$ (cf. Appendix~\ref{sec:numerical} for a mathematical justification).
Details of the numerical implementation of KdV soliton gas using $n$-soliton solutions can be found in Appendix \ref{sec:numerical}. Fig.~\ref{fig:cond0}  displays the realization $\varphi_{\rm c}^{(0)}(x)$ of genus 0 soliton condensate with $\la_1=1$ modeled by $n$-soliton solutions $\varphi_n(x)$ with $n=100$ and $n=200$, along with the absolute errors $\varphi_n(x)-1$; in the following we refer to these $n$-soliton solutions as ``numerical realizations'' of the soliton gas. One can see that the error at the center of the numerical domain, where the gas is nearly uniform, is very small: Fig.~\ref{fig:error_genus0} displays the variation of this error with $n$ and shows that it decreases with $1/n^2$.
\begin{figure}[h]
  \centering
  \includegraphics[width=7cm]{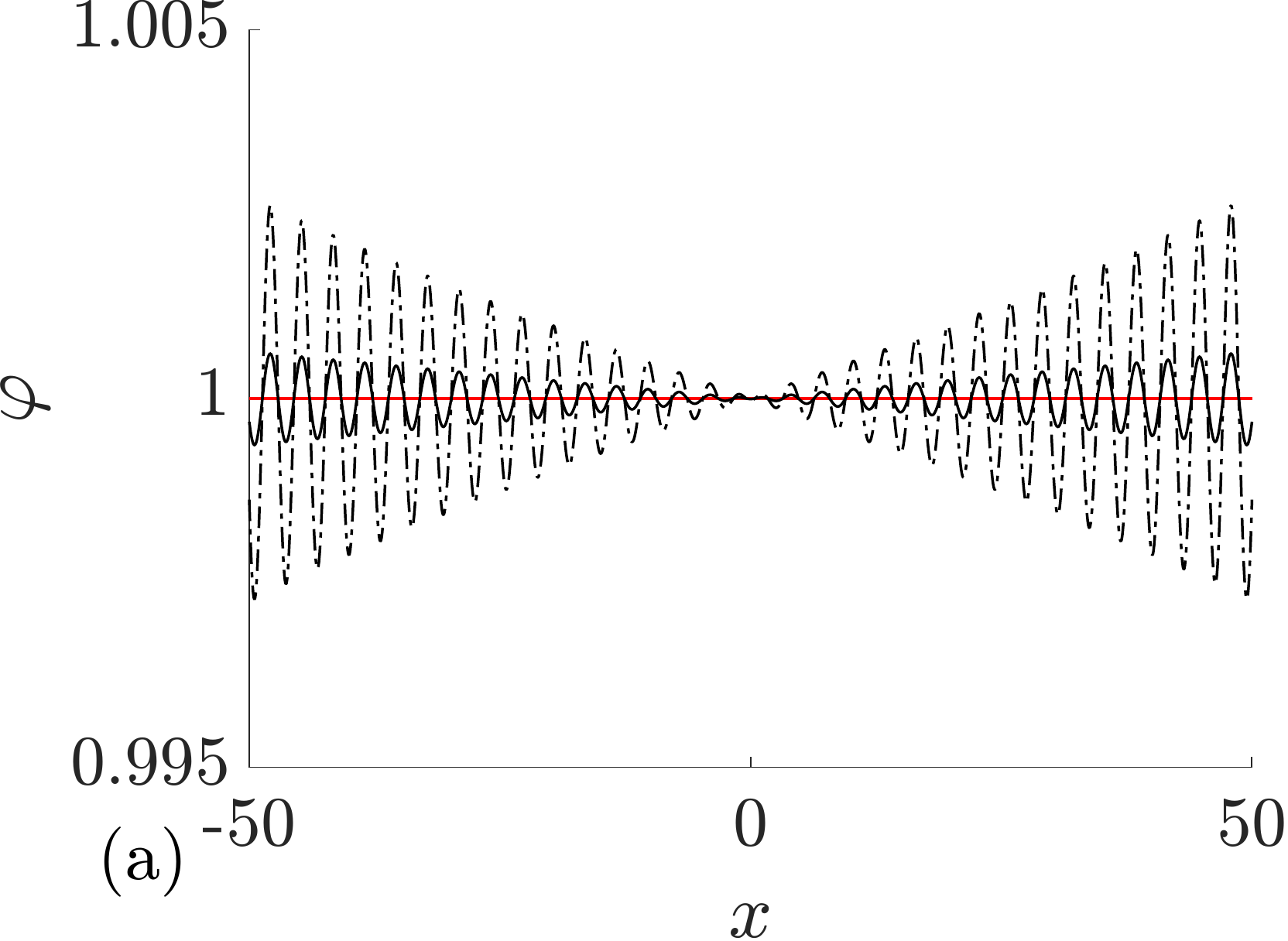} \hspace{1cm}
  \includegraphics[width=7cm]{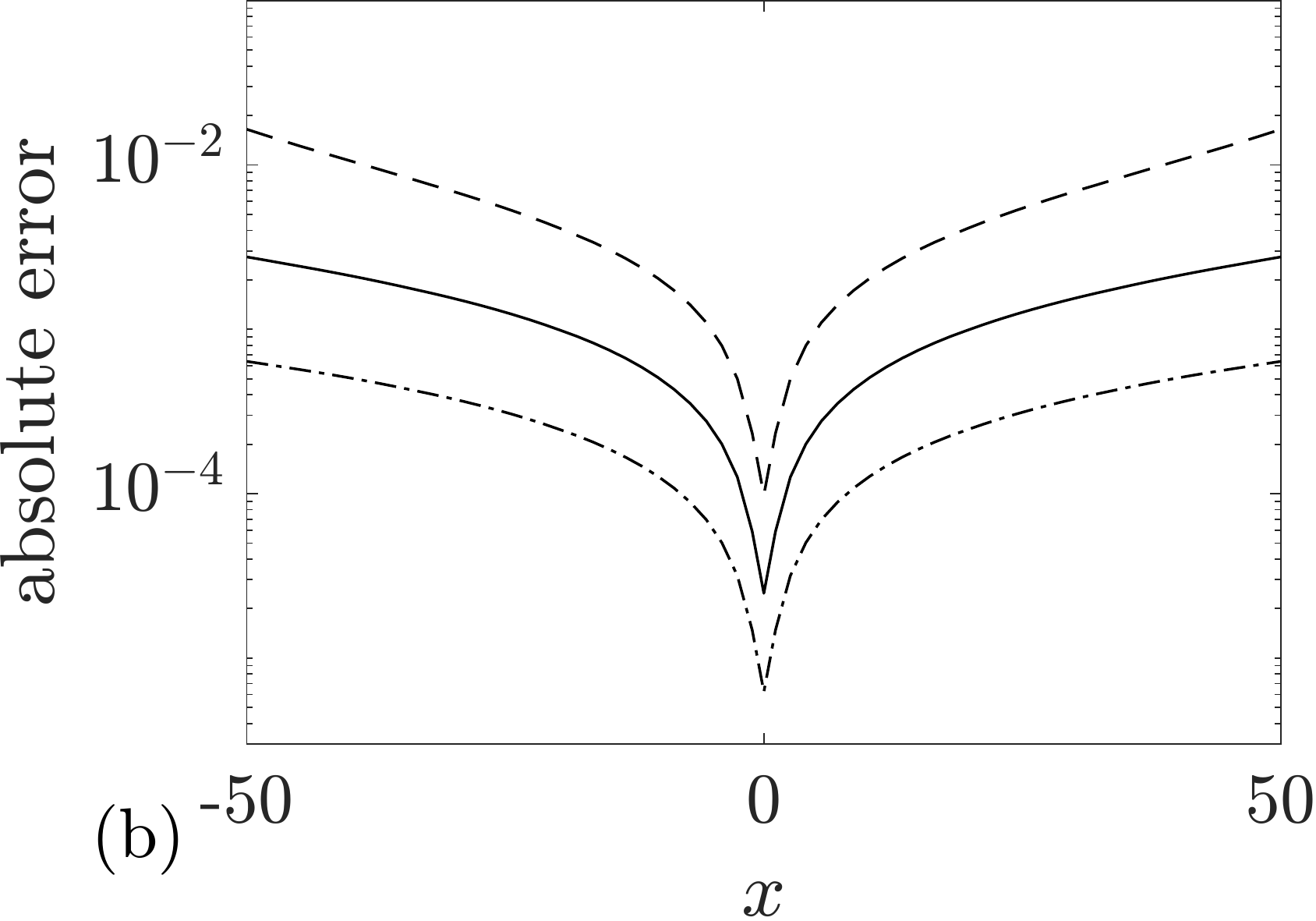}
  \caption{a) Comparison between numerical realizations of genus 0 condensate generated with $100$ solitons (dashed line), $200$ solitons (black solid line), and the constant KdV
    solution~$\varphi=1$ (red solid line).
    b) Corresponding absolute
    errors $|\varphi_n(x)-1|$ obtained with $50$ solitons (dashed line),
    $100$ solitons (solid line) and $200$ (dash-dotted line); the absolute error is evaluated at the extrema of the oscillations.
  }
  \label{fig:cond0}
\end{figure}
\begin{figure}[h]
  \centering
  \includegraphics[width=7cm]{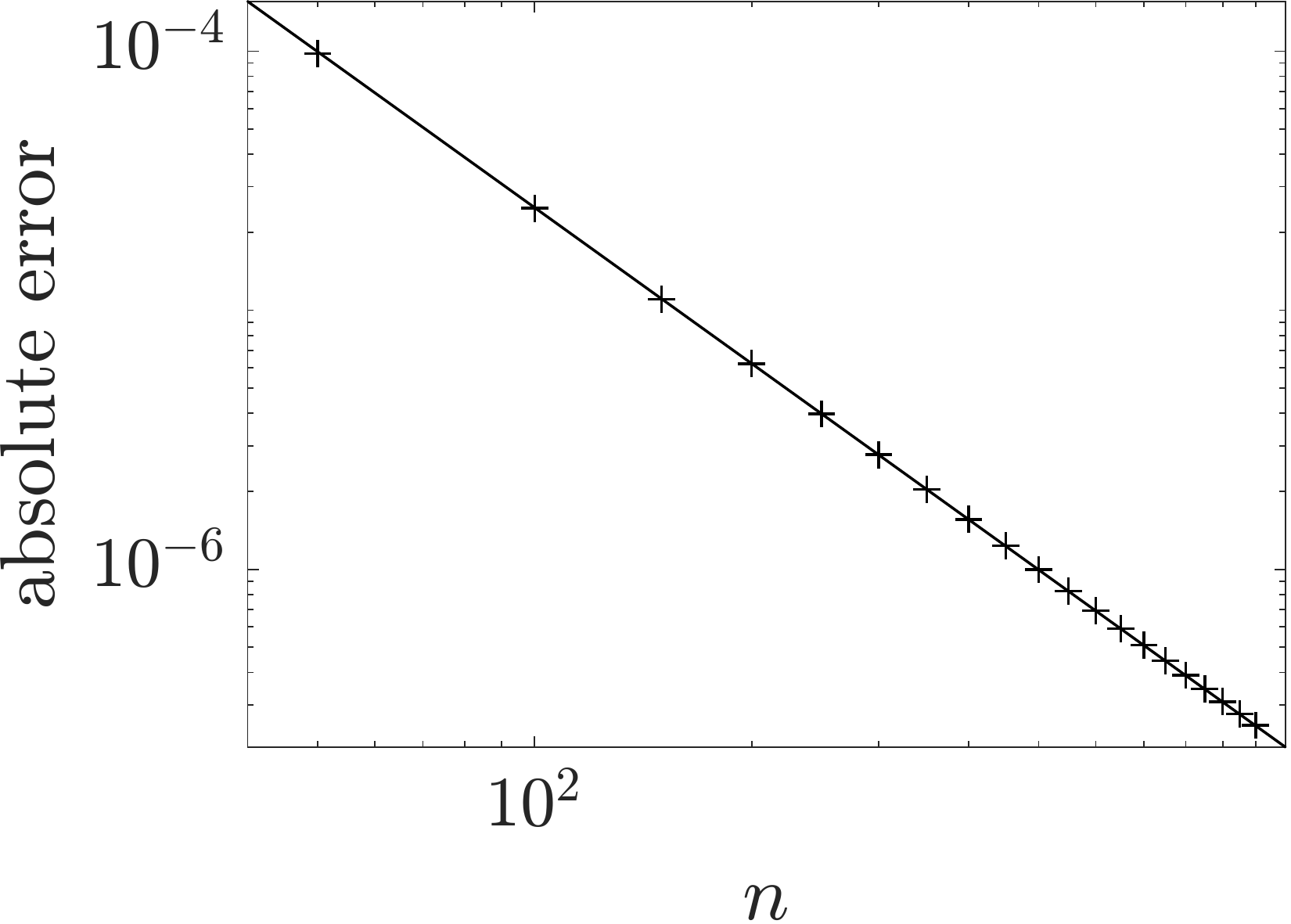}
  \caption{Variation of the absolute error $|\varphi_n(x)-1|$ at the center of the numerical domain $x=0$ (cf. Fig.~\ref{fig:cond0}). The markers correspond to the error obtained numerically and the solid line the corresponding fit $\alpha/n^2$ where $\alpha \approx 0.25$.
  }
  \label{fig:error_genus0}
\end{figure}
The numerical approximation used here is similar to the approximation of the soliton condensate of the focusing NLS equation via a $n$-soliton solution presented in~\cite{gelash2021}. 
In the latter case the uniform wavefield limit as a central part of the so-called ``box potential'', is also reached when the complex phases of the norming constants are  chosen deterministically. The absolute error---the difference between the $n$-soliton solution and the expected constant value of the wavefield---measured at the center of the numerical realization---follows a different scaling law and is proportional to $n^{-1/2}$.

\subsubsection{Genus 1}

We now consider the case of genus 1 soliton condensate. For $N=1$
\begin{equation}\label{RSgen1}
  R(\eta)=\sqrt{(\eta^2-\lambda_1^2)(\eta^2-\lambda_2^2)(\eta^2-\lambda_3^2)}
\end{equation}
is purely imaginary on $\G=[-\lambda_3,-\lambda_2]\cup[-\lambda_1,\lambda_1]\cup[\lambda_2,\lambda_3]$. According to Theorem \ref{theo-2N}
\begin{align}
  \label{eq:N1}
   & u(\eta)=u^{(1)}(\eta;\lambda_1,\lambda_2,\lambda_3) \equiv
  \frac{i\eta(\eta^2-w^2)}{\pi R(\eta)},                                                                          \\
   & v(\eta)=v^{(1)}(\eta;\lambda_1,\lambda_2,\lambda_3) \equiv\frac{12i\eta(\eta^4-h^2\eta^2-r^2)}{\pi R(\eta)},
\end{align}
where $h^2=\frac{\lambda_1^2+\lambda_2^2+\lambda_3^2}{2}$ follows from the fact that $\left.-i{\rm Res}\frac{Q(\z)}{(\z-\eta)R(\z)}\right|_{\z=\infty}=-6\eta^2$. The normalization conditions~\eqref{dpdq_norm}  imply that
\begin{equation}\label{mr} w^2=\frac{\int_{\lambda_1}^{\lambda_2}\frac{y^3dy}{R(y)}}{\int_{\lambda_1}^{\lambda_2}\frac{ydy}{R(y)}},\quad
  r^2= \frac{\int_{\lambda_1}^{\lambda_2}\frac{y^5-\frac{\lambda_1^2+\lambda_2^2+\lambda_3^2}{2}y^3}{R(y)}dy}{\int_{\lambda_1}^{\lambda_2}\frac{ydy}{R(y)}}.
\end{equation}
Using 3.131.3 and 3.132.2 from \cite{Gradshteyn_1988}, we calculate
\begin{equation}\label{m-ans}
  w^2=\lambda_3^2-(\lambda_3^2-\lambda_1^2)\mu(m), \quad {\rm where}\quad
  \mu(m)=\frac{{\rm E} \left(m \right)}{{\rm
        K} \left(m \right)} \quad {\rm
    and}\quad
  m = \frac{\lambda_2^2-\lambda_1^2}{\lambda_3^2-\lambda_1^2}.
\end{equation}
Calculation of $r^2$ is a bit more involved as it is based on the observation
\begin{equation}\label{r-form}
  \begin{split}
    \int_{\lambda_1}^{\lambda_2} \frac{y^5}{R(y)}dy&=\hf \int_{\lambda_1^2}^{\lambda_2^2}\frac{z^2dz}{R(z^\hf)}, \\
    &=
  \frac{\lambda_1^2+\lambda_2^2+\lambda_3^2}3\int_{\lambda_1^2}^{\lambda_2^2}\frac{zdz}{R(z^\hf)}-\frac{\lambda_1^2\lambda_2^2+\lambda_1^2\lambda_3^2+\lambda_2^2\lambda_3^2}6\int_{\lambda_1^2}^{\lambda_2^2}\frac{dz}{R(z^\hf)}.
  \end{split}
\end{equation}
Using~\eqref{mr}, \eqref{r-form}, we obtain after some algebra
\begin{equation}\label{r-ans}
  r^2=\frac 16\left[\lambda_3^2(\lambda_3^2-\lambda_2^2-\lambda_1^2)-2\lambda_2^2\lambda_1^2 - (\lambda_3^2+\lambda_1^2+\lambda_2^2)(\lambda_3^2-\lambda_1^2)\mu(m)\right].
\end{equation}
Thus, the velocity of a tracer soliton with spectral parameter $\eta \in \G^+$ in the genus 1 soliton condensate, characterized by DOS~\eqref{eq:N1}, is given by
\begin{equation}\label{s_N1}
  \begin{split}
    s(\eta) & \equiv s^{(1)}(\eta;\lambda_1,\lambda_2,\lambda_3) = 12\frac{\eta^4-
      \frac{\lambda_2^2+\lambda_3^2+\lambda_1^2}{2}\eta^2-r^2}{\eta^2-w^2}\\
    &=
    12\frac{\eta^4- \frac{\lambda_2^2+\lambda_3^2+\lambda_1^2}{2}\eta^2-\frac {\lambda_3^2(\lambda_3^2-\lambda_2^2-\lambda_1^2)-2\lambda_2^2\lambda_1^2 - (\lambda_3^2+\lambda_1^2+\lambda_2^2)(\lambda_3^2-\lambda_1^2)\mu(m)}6}
    {\eta^2-\lambda_3^2+(\lambda_3^2-\lambda_1^2)\mu(m)}.
  \end{split}
\end{equation}
We note that a similar expression for the tracer velocity in a dense soliton gas was obtained in \cite{girotti2} in the context of  the modified KdV (mKdV) equation.

For $N=1$ the integrals~\eqref{eq:moment1}  for the mean and mean square of the soliton condensate  wave field $\varphi \equiv \varphi_{\rm c}^{(1)}(x,t)$  can
be explicitly evaluated using (253.11) and (256.11) from
\cite{byrd_handbook_1954} and 19.7.10 from \cite{DLMF}:
\begin{align}
   & \langle \varphi \rangle = \lambda_1^2+\lambda_2^2-\lambda_3^2+2(\lambda_3^2-\lambda_1^2)
  \mu(m), \label{meanphi}                                                                         \\
   & \langle \varphi^2 \rangle = \frac{2(\lambda_1^2+\lambda_2^2+\lambda_3^2)}{3} \langle \varphi
  \rangle +\frac{\lambda_1^4+\lambda_2^4+\lambda_3^4-2\lambda_1^2\lambda_2^2-2\lambda_2^2\lambda_3^2-2\lambda_1^2\lambda_3^2}{3}, \label{meansquarephi}
\end{align}
with $\mu(m)$ and $m$ given by~\eqref{m-ans}.   It is not difficult to verify that, unlike in the case  of genus 0 condensates,
the variance
$\Delta=\sqrt{\langle \varphi^2 \rangle -   \langle \varphi \rangle^2}$ does not vanish identically implying that all realizations of the genus 1 soliton condensate are almost surely non-constant.

A key observation is, that formulae~\eqref{meanphi}, \eqref{meansquarephi}  coincide with the period averages  $\overline{\varphi}$ and $\overline{\varphi^2}$ of the genus 1  KdV
solution  associated with the spectral Riemann surface  $\mathcal{R}_2$ of~\eqref{RSgen1} (see e.g. \cite{kamchatnov_nonlinear_2000, el_dispersive_2016}):
\begin{equation}
  \label{eq:cnoidal}
  \begin{split}
    &\varphi (x,t) \equiv F_1(\theta; \la_1, \la_2, \la_3)=
    \lambda_1^2+\lambda_2^2-\lambda_3^2 +2(\lambda_3^2-\lambda_1^2){\rm
        dn}^2 \left(\frac{\sqrt{\lambda_3^2-\lambda_1^2}}{k} \theta;m \right),\\
    &\theta = k(x-Ut) +\theta^0 ,\quad U=2(\lambda_1^2+\lambda_2^2+\lambda_3^2), \quad
    k= \frac{\pi \sqrt{\la_3^2 - \la_1^2}}{{\rm K}(m)},
  \end{split}
\end{equation}
where $\theta^0 \in [0, 2\pi)$ is an arbitrary initial phase.
\begin{figure}[h]
  \centering
  \includegraphics[width=7cm]{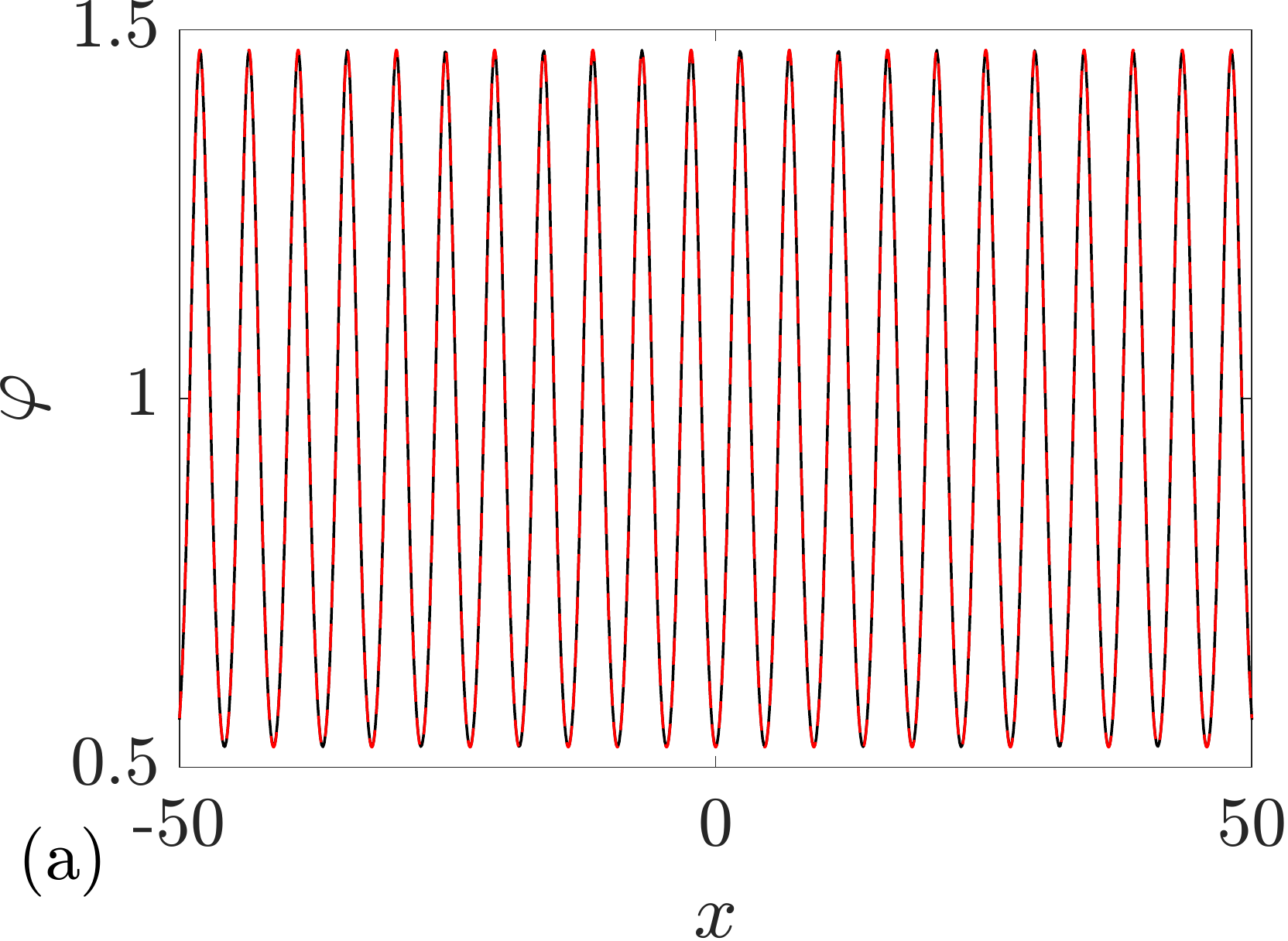}\hspace{1cm}
  \includegraphics[width=7cm]{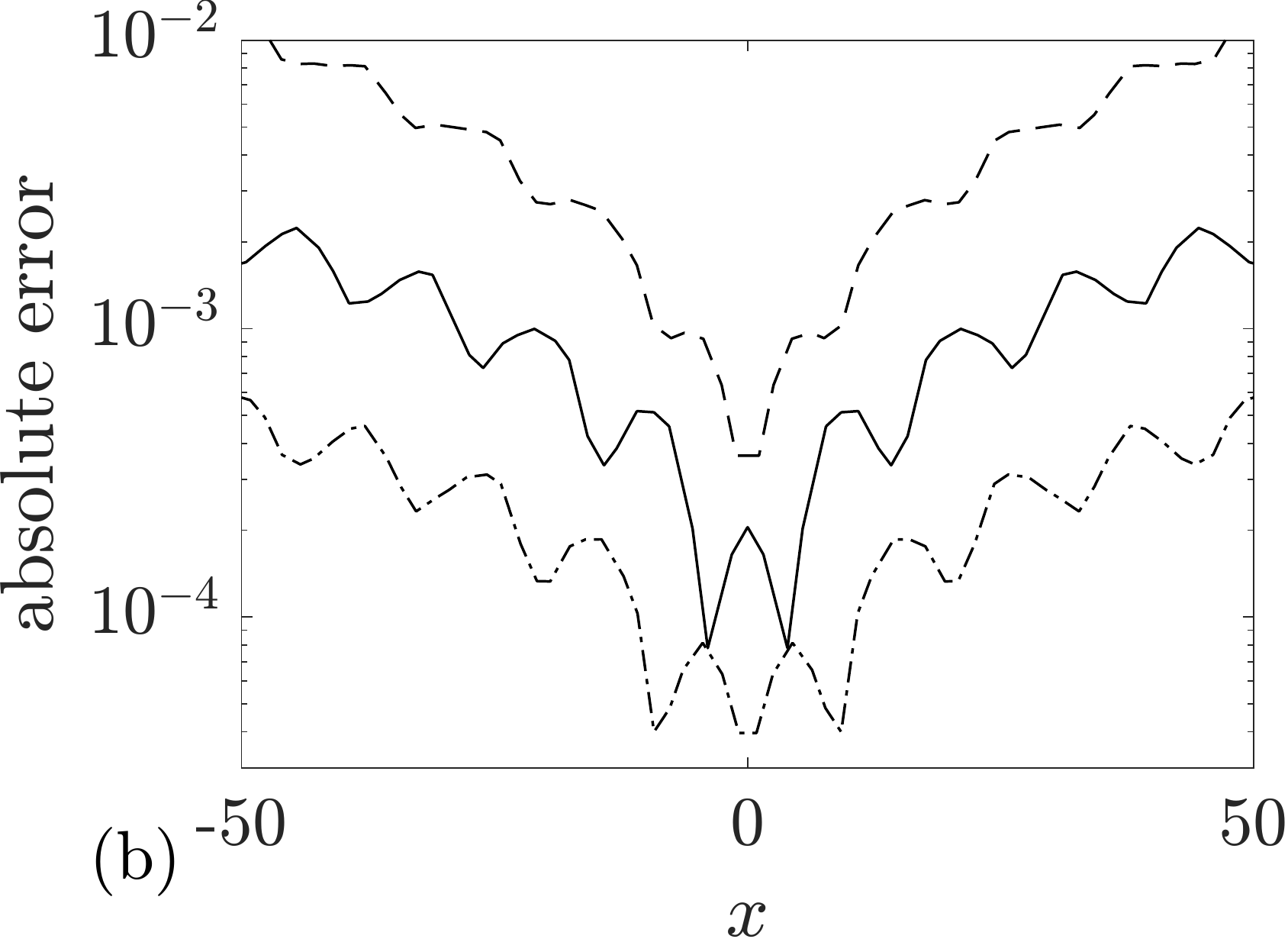}
  \caption{Comparison between the numerical realization of genus 1 condensate generated with $200$ solitons (black solid line), $200$ solitons (black solid line), and the exact cnoidal wave
    solution $F_1(kx)$~\eqref{eq:cnoidal} (red dashed line) for
    $\lambda_1=0.5,\lambda_2=0.85,\lambda_3=1$ ($m=0.63$); the two plots are visually indistinguishable from one another.  
    b) Corresponding absolute
    errors $\varphi_n(x)-F_1(kx)$ obtained with $50$ solitons (dashed line),
    $100$ solitons (solid line) and $200$ (dash-dotted line); the absolute error is evaluated at the extrema of the oscillations.
  }
  \label{fig:cnoidal3}
\end{figure}
The  equivalence between the ensemble averages in genus 1 KdV soliton condensates  and the period averages in single-phase KdV solutions, along with the established in Section~\ref{sec:gen_sol_cond} equivalence between the respective modulation dynamics,  strongly suggest that  realizations of the genus 1 soliton condensates are described by  the periodic solutions $F_1(\theta)$~\eqref{eq:cnoidal}  of the KdV equation. This motivates  the following

\begin{conjecture} \label{prop4.1}
  For any realization $\varphi=\varphi_{\rm c}^{(1)}(x,t)$ of the genus 1 KdV soliton condensate associated with the spectral curve $\mathcal{R}_2$ of~\eqref{RSgen1} one can find the initial phase $\theta^0 \in [0, 2 \pi)$ in the periodic solution $F_1 (\theta; \la_1, \la_2, \la_3)$~\eqref{eq:cnoidal} such that almost surely $\varphi_{\rm c}^{(1)}(x,t) = F_1 (\theta; \la_1, \la_2, \la_3)$.
\end{conjecture}

We support  Conjecture~\ref{prop4.1}  by a detailed comparison of a numerical realization of KdV soliton condensate (as $n$-soliton solution with $n$ large) spectrally configured according to the DOS~\eqref{eq:N1}, and the periodic KdV  solution~\eqref{eq:cnoidal}, defined on the same spectral curve $\mathcal{R}_2$, with the appropriately chosen initial phase $\theta^0$  (see Appendix~\ref{sec:numerical} for the details of the numerical implementation of soliton condensate). The comparison is presented in Fig.~\ref{fig:cnoidal3} and reveals a remarkable agreement, which further improves as $n$ increases.

Conjecture~\ref{prop4.1} can be naturally generalized to an arbitrary genus $N$:  for any realization of the KdV soliton condensate of genus $N$ corresponding to  the density of states $u^{(N)}(\eta; {\bs \la})$~\eqref{uv_c} and associated with the spectral Riemann surface $\mathcal{R}_{2N}$ of~\eqref{RS_N}, one can find $N$-component initial phase vector ${\bs \theta}^0 \in \mathbb{T}^N$  so that  $\varphi^{(N)}_{\rm c}(x,t)$ almost surely coincides with  $N$-phase  KdV solution $\varphi=F_N({\bs \theta}; {\bs \la})$~\eqref{N-gap}. To support this generalization we performed a comparison of a numerical realization of the genus 2 soliton condensate with the respective two-phase (two-gap) KdV solution, see Appendix~\ref{sec:gen_two_cond}.

A rigorous mathematical proof of  Conjecture~\ref{prop4.1} and its  generalization for an arbitrary genus will be the subject of future work.

In conclusion we note that Conjecture~\ref{prop4.1}  correlates with the results of \cite{girotti_rigorous_2021} where a particular ``deterministic soliton gas'' solution of the KdV equation has been constructed by considering the $n$-soliton solution  with the discrete spectrum confined within two symmetric intervals---the analogs of s-bands of our  work---and letting $n \to \infty$.   This solution was shown in  \cite{girotti_rigorous_2021}  to represent  a {\it primitive potential}  \cite{dyachenko_primitive_2016} whose long-time asymptotics is described at leading order by a modulated genus 1 KdV solution.  A similar construction was realized for the mKdV equation in \cite{girotti2}.

\subsection{Modulation dynamics}
\label{sec:mod_dyn}

The dynamics of DOS in non-equilibrium (weakly non-homogeneous) soliton condensates is determined by the evolution of the endpoints $\la_j$ of the spectral bands of
$\Gamma$ (the s-bands). As proven in Section~\ref{sec:gen_sol_cond}, this evolution is governed by the Whitham modulation equations~\eqref{kdv-whitham}. Properties of the KdV-Whitam modulation systems are well studied: in particular, system~\eqref{kdv-whitham}  is strictly hyperbolic and genuinely nonlinear for any genus $N \ge 1$  \cite{levermore_hyperbolic_1988}. This implies inevitability of  wavebreaking for a broad class of initial conditions. What is the meaning of the wavebreaking in the context of soliton condensates, and how is the solution of the kinetic equation continued beyond the wavebreaking time?

We first invoke the definitive property of a soliton condensate---the vanishing of the spectral scaling function, $\sigma(\eta) \equiv 0$ in the soliton gas NDRs~\eqref{ndr_kdv1}. According to Remark~\ref{cor_cond},  if $\sigma(\eta; x, 0) \equiv 0$ for all $x \in \mathbb{R}$, then $\sigma(\eta; x, t) \equiv 0$ for all $x \in \mathbb{R}$,  $\forall t>0$ implying  that soliton condensate necessarily remains a condensate during the evolution (at least of some class of initial data). The only qualitative modification that is permissible during the evolution is the change of the genus $N$.  The description of the evolution of a soliton condensate is then reduced to the determination of  the spectral support $\Gamma(x,t)$, parametrizing the DOS  via the band edges $\la_j(x,t)$: \  $u=u^{(N)}(\eta; \la_1, \dots \la_{2N+1})$~\eqref{uv_c}.

In view of the above,  the evolution of soliton condensates   can be naturally put in the framework  of the problem of hydrodynamic evolution of multivalued functions originally  formulated by Dubrovin and Novikov \cite{dubrovin_hydrodynamics_1989}. Let $\Lambda_N (x,t)=\{\la_1(x,t), \dots, \la_{2N+1} (x,t)\}$ be a smooth multivalued curve whose  branches $\la_j (x,t)$ satisfy the Whitham modulation equations~\eqref{kdv-whitham}.  Then, if wavebreaking occurs within one of the branches it  results in a change of  the genus $N$   so that $\Lambda_{N} \to \Lambda_{N+1}$ in some space-time region $[x^-(t), x^+(t)]$ that includes the wavebreaking point. The curves $\Lambda_N$ and $\Lambda_{N+1}$ are   glued together  at free boundaries $x^{\pm}(t)$. Details of the implementation of this procedure can be found in
\cite{dubrovin_hydrodynamics_1989,  dubrovin_functionals_1997, el_unified_2001, grava_generation_2002}.  The simplest case of the multivalued curve evolution  arises when the initial data for $\Lambda_N$ is a piecewise-constant distribution (both for $\la_j$'s and for $N$), with a discontinuity at $x=0$ --- a Riemann problem. In this special case the wavebreaking occurs at $t=0$ (subject to appropriate sign of the initial jump) and smoothness of $\Lambda_N$ is not a prerequisite.

In this paper, we restrict ourselves to Riemann problems  involving only genus 0 and genus 1 modulation solutions and show how the resulting spectral dynamics are interpreted in terms of soliton condensates. For that we will need explicit expressions  for  the Whitham characteristic velocities for $N=0$ and $N=1$. These expressions are known very well (see e.g. \cite{gurevich_nonstationary_1974, kamchatnov_nonlinear_2000, el_dispersive_2016}) but here we obtain them as  transport velocities for the respective soliton condensates, using the  expressions~\eqref{s_gen0}, and~\eqref{s_N1} respectively.

\medskip

(i) \ $N=0$.  Consider a non-equilibrium (non-uniform) soliton condensate of genus 0, characterized by a space-time dependent DOS $u(\eta; x, t)$.
To this end we set $\eta = \la_1(x,t)$   in~\eqref{s_gen0}, then the Whitham system~\eqref{kdv-whitham}, \eqref{Vj}  assumes the form of the Hopf (inviscid Burgers)
equation
\begin{equation}\label{Hopf}
  (\la_1)_t +6\la_1^2(\la_1)_x =0.
\end{equation}
Note that this is exactly the result obtained by Lax and Levermore \cite{lax_small_1983} for the pre-breaking evolution of semi-classical soliton ensembles.

\smallskip
(ii) $N=1$.  We obtain on using~\eqref{s_N1},
\begin{equation}\label{wh_g1}
  (\lambda_j)_t + V_j(\lambda_1,\lambda_2,\lambda_3) (\lambda_j)_x = 0,\quad j=1,2,3,
\end{equation}
where
\begin{equation}
  \label{eq:mod3}
  \begin{split}
    &V_1(\lambda_1,\lambda_2,\lambda_3) \equiv s^{(1)}(\lambda_1;\lambda_1,\lambda_2,\lambda_3) = 2(\lambda_1^2+\lambda_2^2+\lambda_3^2)+\frac{4(\lambda_2^2-\lambda_1^2)}{\mu(m)-1},\\
    &V_2(\lambda_1,\lambda_2,\lambda_3) \equiv s^{(1)}(\lambda_2;\lambda_1,\lambda_2,\lambda_3) = 2(\lambda_1^2+\lambda_2^2+\lambda_3^2) + \frac{4(\lambda_3^2-\lambda_2^2)(\lambda_2^2-\lambda_1^2)}{\lambda_3^2-\lambda_2^2-(\lambda_3^2-\lambda_1^2)\mu(m)},\\
    &V_3(\lambda_1,\lambda_2,\lambda_3) \equiv s^{(1)}(\lambda_3;\lambda_1,\lambda_2,\lambda_3) = 2(\lambda_1^2+\lambda_2^2+\lambda_3^2) + \frac{4(\lambda_3^2-\lambda_2^2)}{\mu(m)},
  \end{split}
\end{equation}
and $\mu(m)$ is defined in~\eqref{m-ans}.
System~\eqref{wh_g1}, \eqref{eq:mod3} coincides with the original Whitham modulation equations  derived for  $r_j =6\lambda_j^2$  in \cite{whitham_non-linear_1965} by averaging KdV conservation laws over the  single-phase, cnoidal  wave family of solutions
(see also \cite{gurevich_nonstationary_1974, dubrovin_hydrodynamics_1989, kamchatnov_nonlinear_2000, el_dispersive_2016}).

\section{Riemann problem for soliton condensates}
\label{sec:Riemann}

The classical Riemann problem consists of finding solution
to a system of hyperbolic conservation laws subject to piecewise-constant initial
conditions exhibiting discontinuity at $x = 0$. The distribution solution of such Riemann
problem generally represents a combination of constant states, simple (rarefaction)
waves and strong discontinuities (shocks or contact discontinuities) \cite{lax_hyperbolic_1973}. In dispersive hydrodynamics, classical shock waves are replaced by dispersive shock waves (DSWs) --- nonlinear expanding wavetrains with a certain, well-defined structure \cite{el_dispersive_2016}. Here we generalize the Riemann problem formulation to
the soliton gas kinetic equation by considering~\eqref{kineq} subject to discontinuous initial DOS:
\begin{equation}
  \label{eq:ustep}
  u(\eta, x,t=0) =
  \begin{cases}
    u^{(N_-)}(\eta;\lambda_1^-, \dots, \lambda_{2N_-+1}^-), & x<0, \\
    u^{(N_+)}(\eta;\lambda_1^+, \dots, \lambda_{2N_++1}^+), & x>0,
  \end{cases}
\end{equation}
where  $u^{(N)}(\eta; \la_1, \dots, \la_{2N+1})$ is the DOS~\eqref{uv_c} of genus $N$ condensate  and $\lambda_j^\pm >0$.

As discussed in Section~\ref{sec:mod_dyn}, soliton condensate necessarily retains its definitive property  $\sigma=0$ during the evolution, with the only qualitative modification  permissible being the change of the genus $N$.  The evolution of the soliton condensate is then determined by the motion of the s-band  edges $\la_j$ according to the Whitham modulation equations~\eqref{kdv-whitham} subject to discontinuous initial conditions  following from~\eqref{eq:ustep}:
\begin{equation}
  \label{eq:laNstep}
  \{N; {\bs \la}\}( x,t=0) =
  \begin{cases}
    \{N_-; (\lambda_1^-, \dots, \lambda_{2N_-+1}^-) \}, & x<0, \\
    \{N_+; (\lambda_1^+, \dots, \lambda_{2N_+ +1}^+)\}, & x>0.
  \end{cases}
\end{equation}
Thus the Riemann problem  for soliton gas kinetic equation   is effectively reduced in the condensate limit  to the  Riemann problem~\eqref{eq:laNstep} for the Whitham modulation equations~\eqref{kdv-whitham}. Depending on the  sign of the jump $\la_j^- - \la_j^+$ the regularization of the discontinuity in $\la_j$  can occur in two ways: (i)   if $(\la_j^- - \la_j^+)>0$ then the regularization occurs via the generation of a rarefaction wave for $\la_j$ without changing the genus $N$ of the condensate; (ii) if $(\la_j^- - \la_j^+)<0$ (which implies immediate wavebreaking for $\la_j$) the regularization occurs via  the generation of a higher genus condensate whose evolution is governed by the modulation equations.

Below we consider several particular cases of Riemann problems describing some prototypical features of the soliton condensate dynamics.

\subsection{$N_-=N_+=0$}

Consider the initial condition for the kinetic equation in the form of a  discontinuous genus 0  condensate DOS,
\begin{equation}
  \label{eq:ustep0}
  u(\eta, x,t=0) =
  \begin{cases}
    u^{(0)}(\eta;q_-), & x<0, \\
    u^{(0)}(\eta;q_+), & x>0,
  \end{cases}
\end{equation}
where $q_\pm = \lambda_1^\pm$, and $u^{(0)}>0$ is defined
in~\eqref{uv}. The DOS  distribution~\eqref{eq:ustep0} implies the step initial conditions for the Whitham modulation  system~\eqref{kdv-whitham}:
\begin{equation}\label{whitham-riemann0}
  N(x,t=0)=0, \quad  \la_1(x,t=0) =
  \begin{cases}
    q_- ,  & x<0, \\
    q_+  , & x>0,
  \end{cases}
\end{equation}
with $q_- \ne q_+$. Additionally, since the
wave field in a genus 0 soliton condensate is almost surely a constant, $\varphi(x,t)=(\la_1)^2$,
we conclude that the DOS distribution~\eqref{eq:ustep0} gives rise to the Riemann step data
\begin{equation}\label{kdv_riemann}
  \varphi(x,t=0) =
  \begin{cases}
    q_-^2 , & x<0, \\
    q_+^2 , & x>0,
  \end{cases}
\end{equation}
for the KdV equation~\eqref{kdv} itself.

The Riemann problem for the KdV equation was originally studied by Gurevich and Pitaevskii (GP) \cite{gurevich_nonstationary_1974} in the context of the description of dispersive shock waves. The key idea of GP construction was to replace the  dispersive Riemann
problem~\eqref{kdv_riemann}  for the KdV equation  by an appropriate boundary value
problem for the  hyperbolic KdV-Whitham system~\eqref{wh_g1}
which is then solved in the class of $x/t$-self-similar
solutions. Here we take advantage of the GP  modulation solutions and their higher genus analogues to describe  dynamics of soliton condensates. The choice of the genus of the Whitham system and, correspondingly, the genus of the associated soliton condensate, depends on whether $q_->q_+$ or $q_+ < q_-$.

\subsubsection{Rarefaction wave ($q_-<q_+$)}

\begin{figure}[h]
  \centering
  \includegraphics[width=7cm]{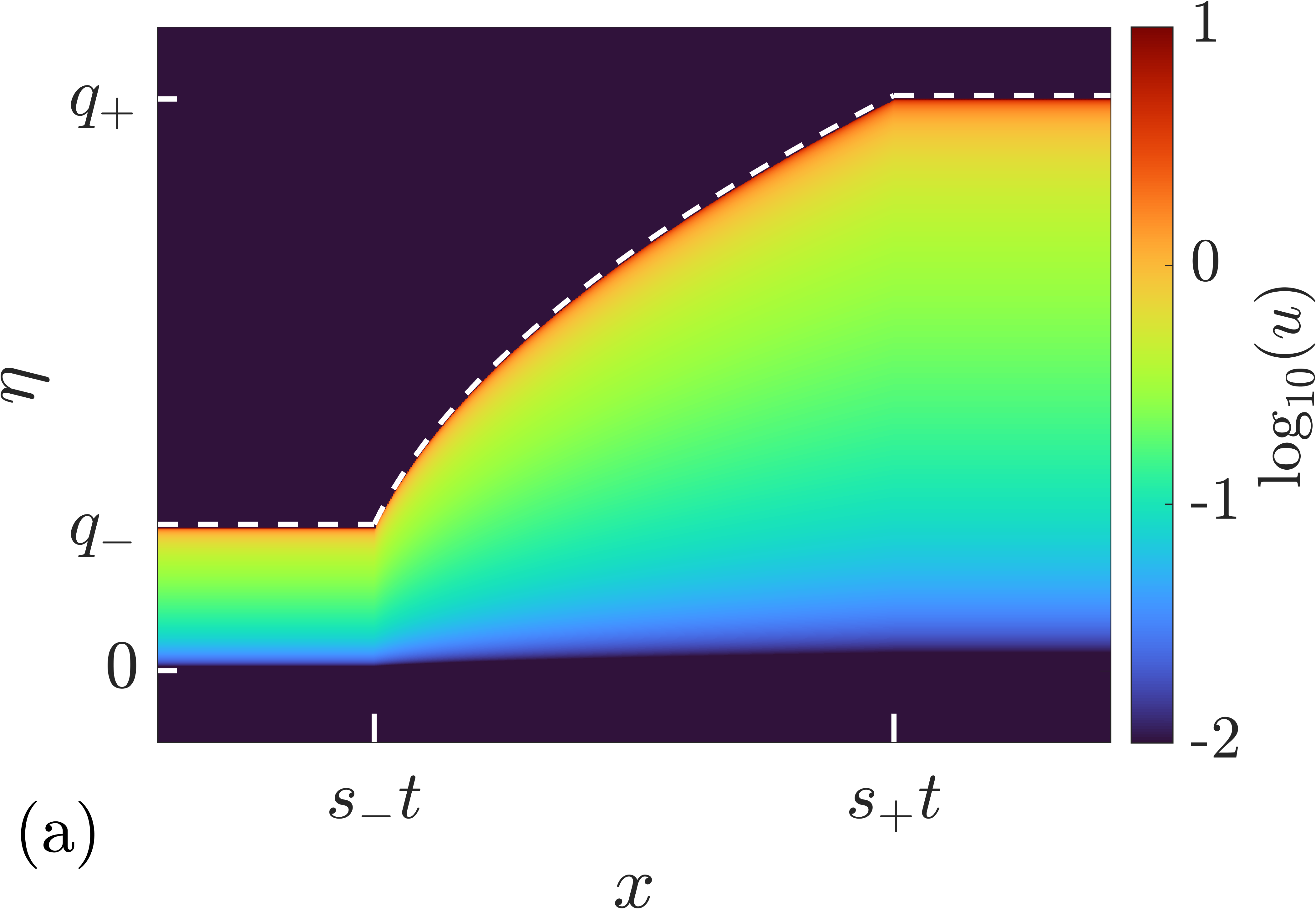}\hspace{1cm} 
  \includegraphics[width=7cm]{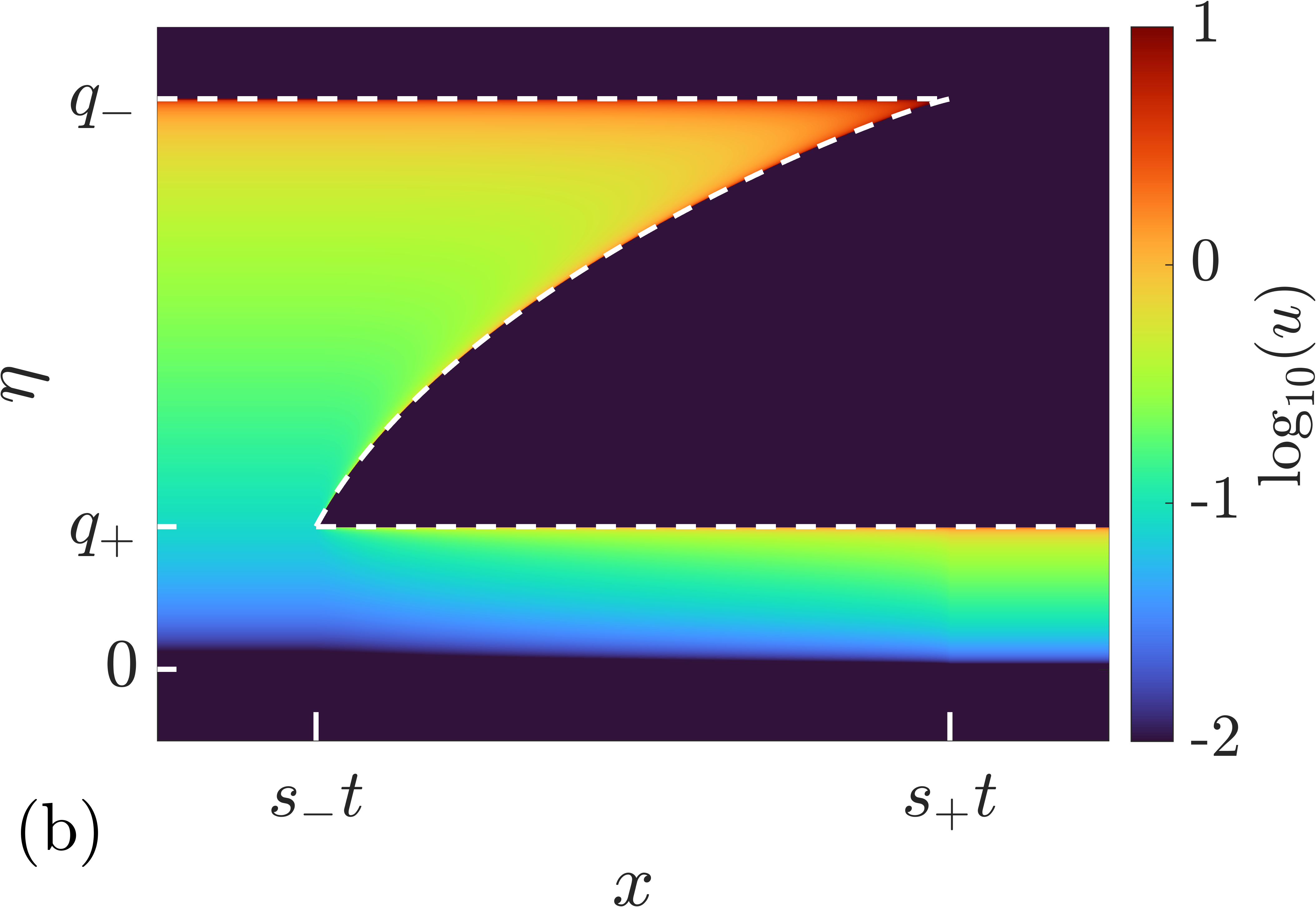}
  \caption{ Solutions to the soliton condensate Riemann problem~\eqref{eq:ustep0}. a) Rarefaction wave (genus $0$) solution~\eqref{eq:rw}, \eqref{eq:sol0} for  $q_- < q_+$. Dashed line: $\la_1(x,t)$,  colors: DOS $u^{(0)}(\eta; \la_1(x,t))$.  b) DSW (genus $1$) solution~\eqref{eq:sol2_dsw}, \eqref{eq:dsw} for $q_->q_+$.
    Dashed line: $\la_1(x,t) \leq \la_2(x,t) \leq \la_3(x,t)$, colors: DOS $u^{(1)}(\eta; q_+, \la_2(x,t), q_-)$.}
  \label{fig:rw}
\end{figure}

The solution of the Riemann problem~\eqref{kineq}, \eqref{eq:ustep0} is given globally (for $t>0$) by  the genus~0 DOS $u^{(0)}(\eta; \la_1)$~\eqref{uv} modulated by the centered rarefaction wave solution
of the Hopf equation~\eqref{Hopf} subject to the step initial condition~\eqref{whitham-riemann0}:
\begin{equation}
  \label{eq:rw}
  \la_1(x,t) =
  \begin{cases}
    q_-,                 & x<s_-t,           \\
    \sqrt{\frac{x}{6t}}, & s_-t < x < s_+ t, \\
    q_+,                 & s_+t<x,
  \end{cases}
\end{equation}
where
\begin{equation}
  \label{eq:4}
  s_- = 6q_-^2t,\quad s_+ = 6q_+^2t.
\end{equation}
Note that the solution~\eqref{eq:rw} is admissible
since $s_-<s_+$. Behavior of $\la_1$ in the solution~\eqref{eq:rw} is shown in Fig.~\ref{fig:rw}a. The evolution of the soliton condensate's DOS
associated with  the spectral rarefaction wave solution~\eqref{eq:rw} is given by
\begin{equation}
  \label{eq:sol0}
  u(\eta; x,t) = \frac{\eta}{\pi\sqrt{\la_1^2 (x,t)-\eta^2}}.
\end{equation}
A contour plot of the DOS~\eqref{eq:sol0} is presented in Fig.~\ref{fig:rw}(a).

\subsubsection{Dispersive shock wave ($q_->q_+$)}

The solution~\eqref{eq:rw}, \eqref{eq:sol0} derived previously is not
admissible for $q_-<q_+$ since $s_->s_+$ in that case. In other words, the compressive discontinuous initial data~\eqref{whitham-riemann0} imply  immediate wavebreaking and necessitate the introduction of the higher genus  DOS
connecting $u^{(0)}(\eta;q_-)$ and $u^{(0)}(\eta; q_+)$. The requisite DOS  is given by equation~\eqref{eq:N1}, which we reproduce here for convenience,
\begin{equation}\label{eq:sol2_dsw}
  u(\eta; x,t) = u^{(1)}(\eta;\lambda_1,\lambda_2,\lambda_3) =
  \frac{i\eta(\eta^2-w^2)}{\pi R(\eta)},
\end{equation}
Here $w(\la_1, \la_2, \la_3)$  is given by~\eqref{m-ans} and  $\lambda_j=\lambda_j(x,t)$, $j=1,2,3$, are slowly modulated according to the Whitham equations~\eqref{wh_g1}, \eqref{eq:mod3}.

The solution of~\eqref{eq:mod3} is self-similar,
$\lambda_j(x/t)$, such that
$u^{(1)}(\eta;\lambda_1,\lambda_2,\lambda_3)$ matches with  $u^{(0)}(\eta;q_-)$ at the left
boundary $x=s_-t$, and with
$u^{(0)}(\eta;q_+)$ at the right boundary $x=s_+t$, with $s_-<s_+$.

The requisite solution is the $2$-wave of the Whitham system~\eqref{wh_g1} (only $\lambda_2$ is non-constant)
\begin{equation}
  \label{eq:dsw}
  \lambda_1 = q_+,\quad V_2(\lambda_1,\lambda_2,\lambda_3) = x/t,\quad
  \lambda_3=q_-,\quad\text{for}\quad s_-t<x<s_+ t,
\end{equation}
where
\begin{equation}
  s_- = V_2(q_+,q_+,q_-) = 12q_+^2-6q_-^2,\quad
  s_+ = V_2(q_+,q_-,q_-) = 2q_+^2+4q_-^2.
\end{equation}
This is the famous GP  solution describing the DSW modulations in the KdV step resolution problem \cite{gurevich_nonstationary_1974}.
Indeed, we have $s_-<s_+$ and, interpreting the GP solution~\eqref{eq:dsw} in terms of soliton condensates the limiting behaviors at the DSW edges is given by
\begin{equation}\label{}
  \begin{split}
    &x \to s_-t,\quad \lambda_2 \to \lambda_1=q_+,\quad
    u^{(1)}(\eta;q_+,\lambda_2,q_-) \to u^{(0)}(\eta;q_-),\\
    &x \to s_+t,\quad \lambda_2 \to \lambda_3=q_-,\quad
    u^{(1)}(\eta;q_-,\lambda_2,q_+) \to u^{(0)}(\eta;q_+).
  \end{split}
\end{equation}

\subsection{ $N_-+N_+=1$}

Before considering the soliton condensate Riemann
problem~\eqref{kineq}, \eqref{eq:ustep} for the case $N_-+N_+=1$ we list the admissible
solutions to the kinetic equation connecting a genus~$0$ distribution $u^{(0)}(\eta;q)$ to a
genus~$1$ distribution $u^{(1)}(\eta;\lambda_1,\lambda_2,\lambda_3)$.
One can easily verify for the next four solutions that
\begin{equation}
  \begin{split}
    &x \to s_-t,\quad
    u^{(1)}(\eta;\lambda_1,\lambda_2,\lambda_3) \to u^{(N_-)}(\eta; \boldsymbol{\lambda}_-),\\
    &x \to s_+t,\quad u^{(1)}(\eta;\lambda_1,\lambda_2,\lambda_3) \to u^{(N_+)}(\eta; \boldsymbol{\lambda}_+),
  \end{split}
\end{equation}
with $s_-<s_+$.

We use the following convention to label the fundamental Riemann problem solutions: we call $j^\pm$-wave,  where
$j$ is the index of the only varying Riemann invariant $\lambda_j$ in the solution, while the remaining invariants are constant; $+$ indicates that $N_+=1$ i.e. the genus $1$ soliton
condensate is initially at $x>0$, and $-$ indicates that $N_-=1$
i.e. the genus $1$ soliton condensate is initially at $x<0$.

\bigskip
\noindent (i) \ {\it $3^+$-wave}

\begin{figure}[h]
  \centering
  \includegraphics[width=7cm]{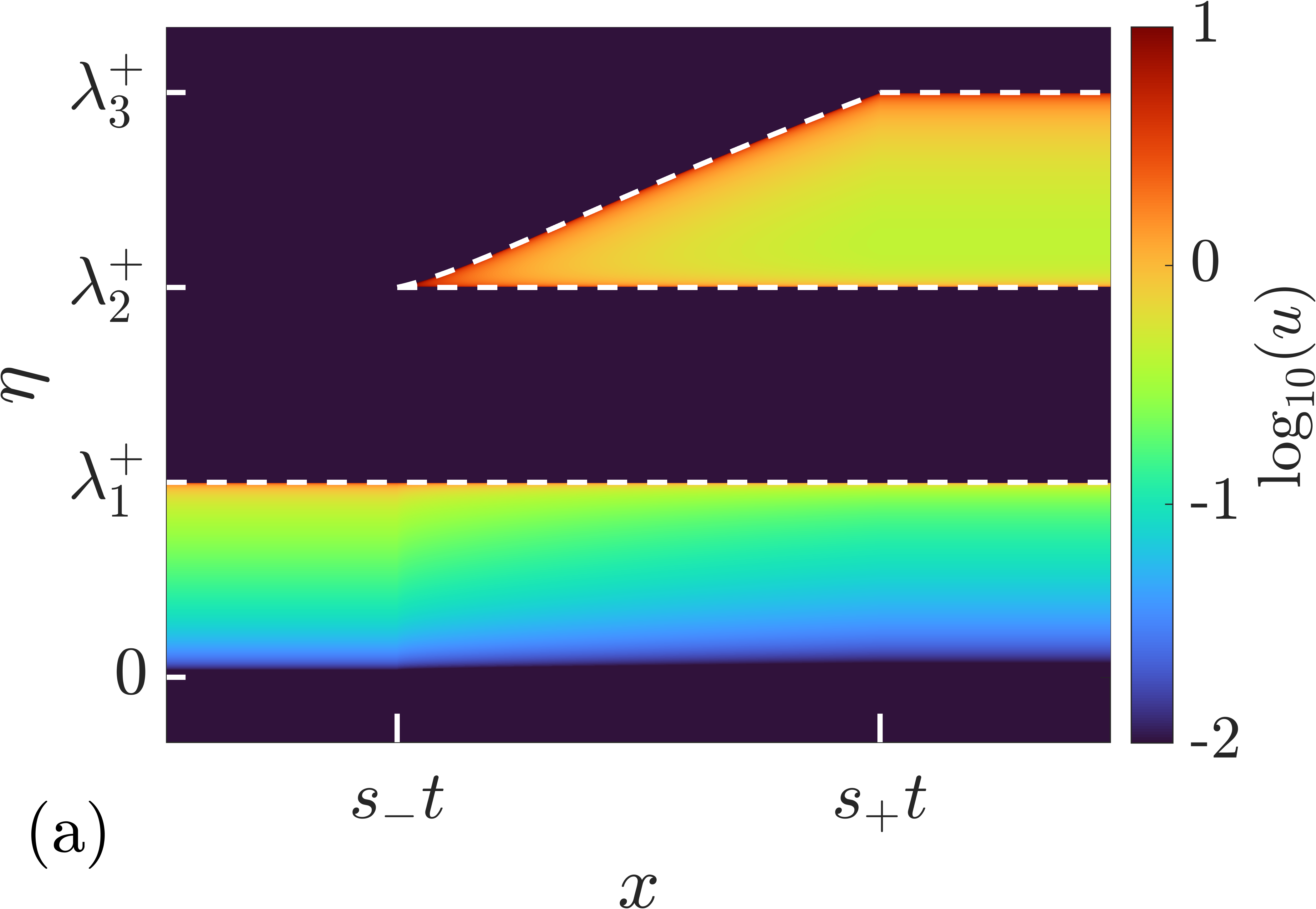}\hspace{1cm}
  \includegraphics[width=7cm]{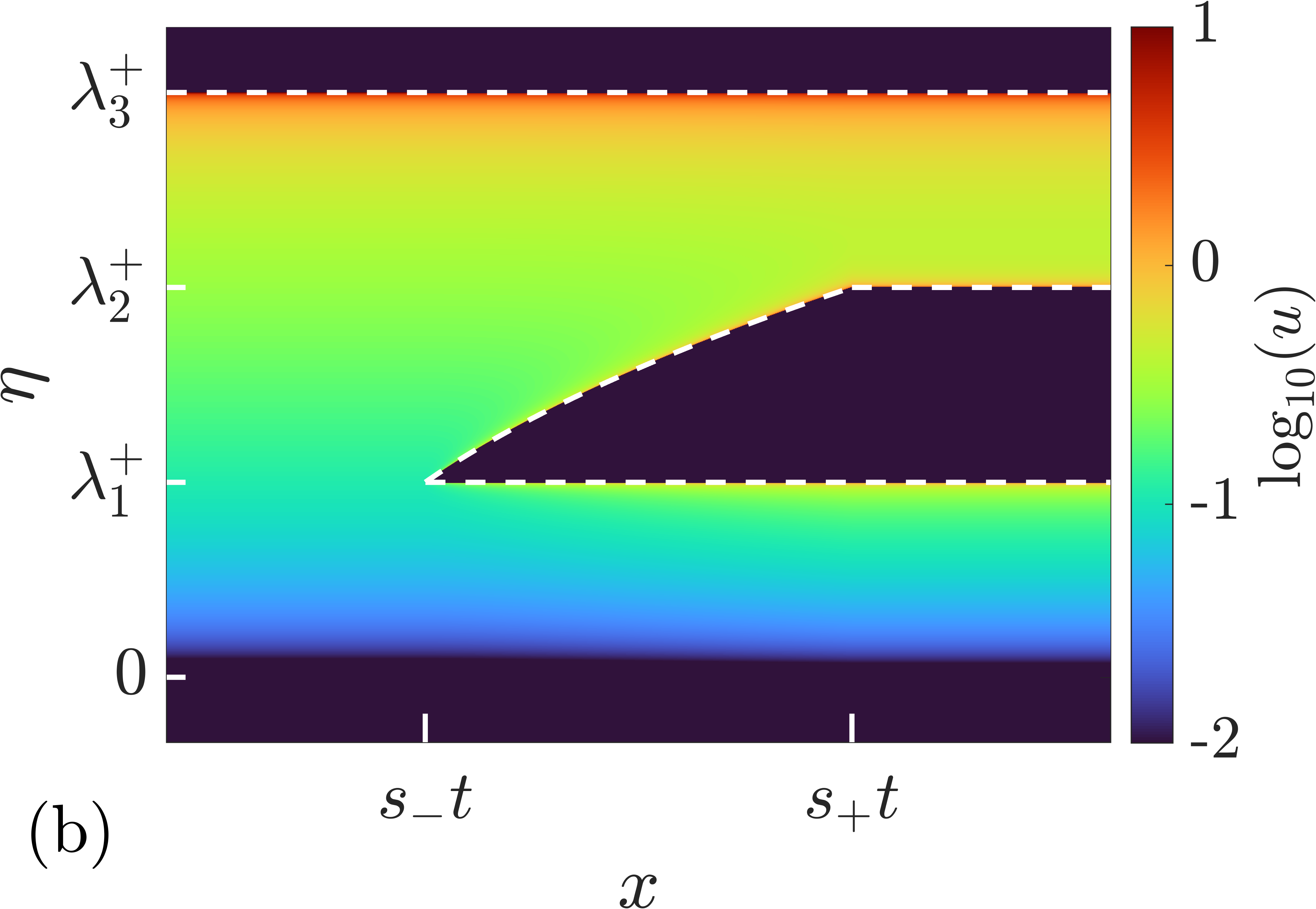}\\[1cm]
  \includegraphics[width=7cm]{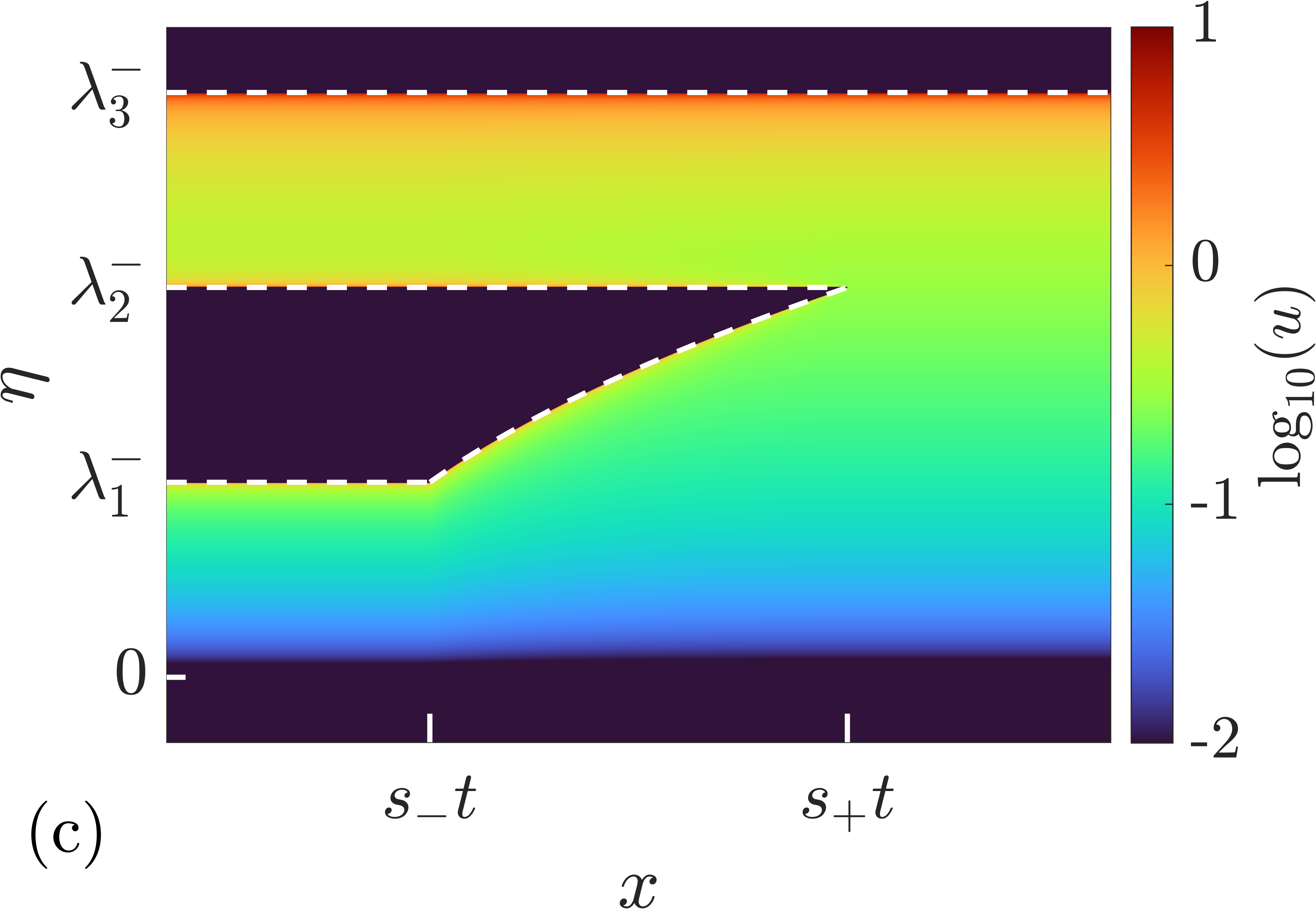}\hspace{1cm}
  \includegraphics[width=7cm]{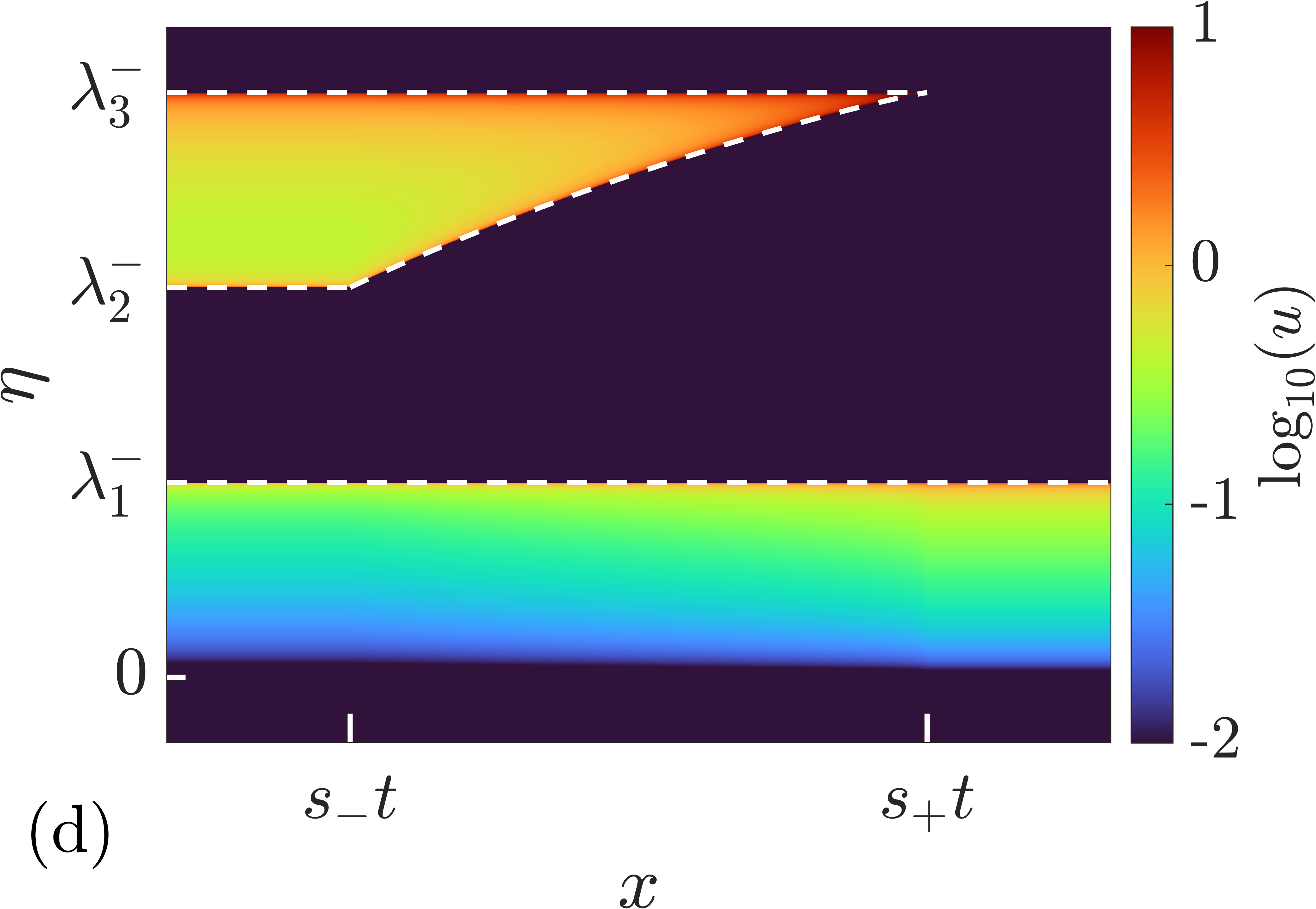}
  \caption{Basic modulation configurations in the Riemann problem~\eqref{kineq}, \eqref{eq:ustep}  for soliton condensates with $N_-+N_+=1$. a) $3^+$-wave solution~\eqref{eq:3-wave}. b)
  $2^+$-wave solution~\eqref{eq:2+-wave}. c) $1^-$-wave
  solution~\eqref{eq:1-wave}. d) $2^-$-wave solution~\eqref{eq:2--wave}. In all cases the dashed lines show the variation of the spectral edges $\la_1\leq  \la_2  \leq \la_3$, and the colors visualize the DOS  $u^{(1)}(\eta; {\bs \la})$.}
  \label{fig:mod_sim}
\end{figure}

\smallskip
Consider the initial condition for  the soliton condensate DOS:
\begin{equation}
  \label{eq:init3+}
  u(\eta, x,t=0) =
  \begin{cases}
    u^{(0)}(\eta;q_-),                                 & x<0, \\
    u^{(1)}(\eta;\lambda_1^+,\lambda_2^+,\lambda_3^+), & x>0,
  \end{cases} \quad\text{with}\quad
  \lambda_1^+=q_-.
\end{equation}
The resolution of the step~\eqref{eq:init3+} is described by
\begin{equation}\label{eq:sol3}
  u(\eta, x,t) =
  \begin{cases}
    u^{(0)}(\eta;q_-),                             & x<s_-t,       \\
    u^{(1)}(\eta; q_-,\lambda_2^+,\lambda_3(x/t)), & s_-t <x<s_+t, \\
    u^{(1)}(\eta; q_-,\lambda_2^+,\lambda_3^+),    & x>s_+t,\end{cases}
\end{equation}
where $\la_3(x/t)$ is given by the $3^+$-wave solution of the modulation equations~\eqref{wh_g1}:
\begin{equation}
  \begin{split}
    \label{eq:3-wave}
    &\lambda_1 =
    q_- ,\quad
    \lambda_2=\lambda_2^+,\quad
    V_3(\lambda_1,\lambda_2,\lambda_3) = x/t,\quad\text{for}\quad
    s_-t<x<s_+ t,\\
    &s_- = V_3(q_-,\lambda_2^+,\lambda_2^+) =
    2(q_-)^2+4(\lambda_2^+)^2,\quad
    s_+ = V_3(q_-,\lambda_2^+,\lambda_3^+).
  \end{split}
\end{equation}
The behavior of the Riemann invariants $\la_j$ in the
$3^+$-wave is shown in
Fig.~\ref{fig:mod_sim}a. The associated soliton condensate  KdV
solution $\varphi(x,t)$ along with the behavior of the mean $\langle
  \varphi \rangle$ are shown in Figs.~\ref{fig:kdv_3+} and~\ref{fig:kdv_3+_2}.

\bigskip
\noindent (ii) \ {\it $2^+$-wave}

\smallskip
Consider the initial condition:
\begin{equation}
  \label{eq:init2+}
  u(\eta, x,t=0) =
  \begin{cases}
    u^{(0)}(\eta;q_-),                                 & x<0, \\
    u^{(1)}(\eta;\lambda_1^+,\lambda_2^+,\lambda_3^+), & x>0,
  \end{cases} \quad\text{with}\quad
  \lambda_3^+=q_-.
\end{equation}
The resolution of the step~\eqref{eq:init2+} is described by
\begin{equation}\label{eq:sol2+}
  u(\eta, x,t) =
  \begin{cases}
    u^{(0)}(\eta;q_-),                            & x<s_-t,       \\
    u^{(1)}(\eta;\lambda_1^+,\lambda_2(x/t),q_-), & s_-t <x<s_+t, \\
    u^{(1)}(\eta;\lambda_1^+,\lambda_2^+,q_-), & x>s_+t,\end{cases}
\end{equation}
where $\la_2(x/t)$ is given by the $2^+$-wave solution of the modulation equations~\eqref{wh_g1}:
\begin{equation}
  \label{eq:2+-wave}
  \begin{split}
    & \lambda_1 = \lambda_1^+,\quad V_2(\lambda_1,\lambda_2,\lambda_3) = x/t,\quad
    \lambda_3=q_-,\quad\text{for}\quad s_-t<x<s_+ t,\\
    & s_- = V_2(\lambda_1^+,\lambda_1^+,\lambda_2^+) =
    12(\lambda_1^+)^2-6(q_-)^2,\quad
    s_+ = V_2(\lambda_1^+,\lambda_2^+,q_-).
  \end{split}
\end{equation}
The behavior of the Riemann invariants $\la_j$ in the
$2^+$-wave is shown in
Fig.~\ref{fig:mod_sim}b.

\medskip
\noindent (iii) \ {\it $1^-$-wave}

\smallskip
Consider the initial condition:
\begin{equation}
  \label{eq:init1-}
  u(\eta, x,t=0) =
  \begin{cases}
    u^{(1)}(\eta;\lambda_1^-,\lambda_2^-,\lambda_3^-), & x>0, \\
    u^{(0)}(\eta;q_+),                                 & x<0,
  \end{cases} \quad\text{with}\quad
  \lambda_3^-=q_+.
\end{equation}
The resolution of the step~\eqref{eq:init1-} is described by
\begin{equation}\label{eq:sol1-}
  u(\eta, x,t) =
  \begin{cases}
    u^{(1)}(\eta;\lambda_1^-,\lambda_2^-,q_+), & x<s_-t,       \\
    u^{(1)}(\eta;\lambda_1(x/t),\lambda_2^-,q_+), & s_-t <x<s_+t, \\
    u^{(0)}(\eta;q_+),                            & x>s_+t,\end{cases}
\end{equation}
where $\la_1(x/t)$ is given by the $1^-$-wave solution of the modulation equations~\eqref{wh_g1}:
\begin{equation}
  \label{eq:1-wave}
  \begin{split}
    &V_1(\lambda_1,\lambda_2,\lambda_3) = x/t,\quad \lambda_2 =
    \lambda_2^- ,\quad
    \lambda_3=q_+,\quad\text{for}\quad s_-t<x<s_+ t,\\
    & s_- = V_1(\lambda_1^-,\lambda_2^-,q_+),\quad
    s_+ = V_1(\lambda_2^-,\lambda_2^-,q_+) = 12(\lambda_2^-)^2-6(q_+)^2.
  \end{split}
\end{equation}
The behavior of the Riemann invariants $\la_j$ in the
$1^-$-wave is shown in
Fig.~\ref{fig:mod_sim}c.

\medskip
\noindent (iv) \ {\it $2^-$-wave}

Consider the initial condition:
\begin{equation}
  \label{eq:init2-}
  u(\eta, x,t=0) =
  \begin{cases}
    u^{(1)}(\eta;\lambda_1^-,\lambda_2^-,\lambda_3^-), & x<0, \\
    u^{(0)}(\eta;q_+),                                 & x>0,
  \end{cases} \quad\text{with}\quad
  \lambda_1^-=q_+.
\end{equation}
The resolution of the step~\eqref{eq:init2-} is described by
\begin{equation}\label{eq:sol2-}
  u(\eta, x,t) =
  \begin{cases}
    u^{(1)}(\eta;q_+,\lambda_2^-,\lambda_3^-),    & x<s_-t,       \\
    u^{(1)}(\eta;q_+,\lambda_2(x/t),\lambda_3^-), & s_-t <x<s_+t, \\
    u^{(0)}(\eta;q_+),                            & x>s_+t,\end{cases}
\end{equation}
where $\la_2(x/t)$ is given by the $2^-$-wave solution of the modulation equations~\eqref{wh_g1}:
\begin{equation}
  \label{eq:2--wave}
  \begin{split}
    &\lambda_1 = q_+,\quad
    V_2(\lambda_1,\lambda_2,\lambda_3) = x/t,\quad
    \lambda_3=\lambda_3^-,\quad\text{for}\quad s_-t<x<s_+ t,\\
    &s_- = V_2(q_+,\lambda_2^-,\lambda_3^-),\quad
    s_+ = V_2(q_+,\lambda_3^-,\lambda_3^-)=
    2(q_+)^2+4(\lambda_3^-)^2.
  \end{split}
\end{equation}
The behavior of the Riemann invariants $\la_j$ in the
$2^-$-wave is shown in
Fig.~\ref{fig:mod_sim}d. The associated soliton condensate  KdV
solution $\varphi(x,t)$ along with the behavior of the mean $\langle
  \varphi \rangle$ are shown in Figs.~\ref{fig:kdv_2-} and~\ref{fig:kdv_2-_2}.

\section{Riemann problem: numerical results}
\label{sec:numeric}

We consider Riemann problems with
$N_-+N_+ \leq 1$. Because of the inherent limitations of the numerical
implementation of soliton gas detailed in Appendix~\ref{sec:numerical}, we restrict the
comparison to the cases $q_-=0$ or $q_+=0$.

\subsection{Rarefaction wave}

\begin{figure}[h]
  \centering
  \includegraphics[width=7cm]{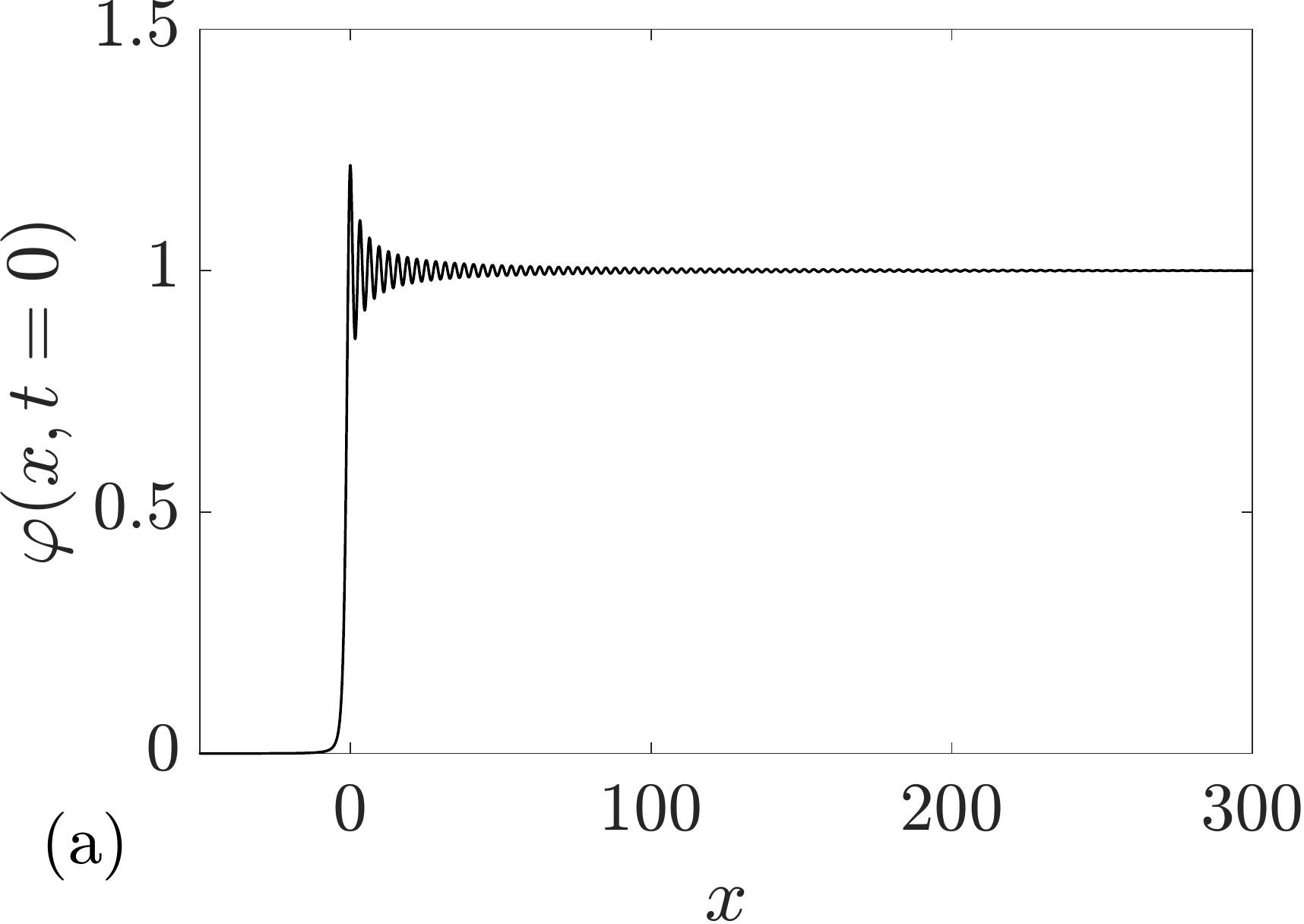}\hspace{1cm}
  \includegraphics[width=7cm]{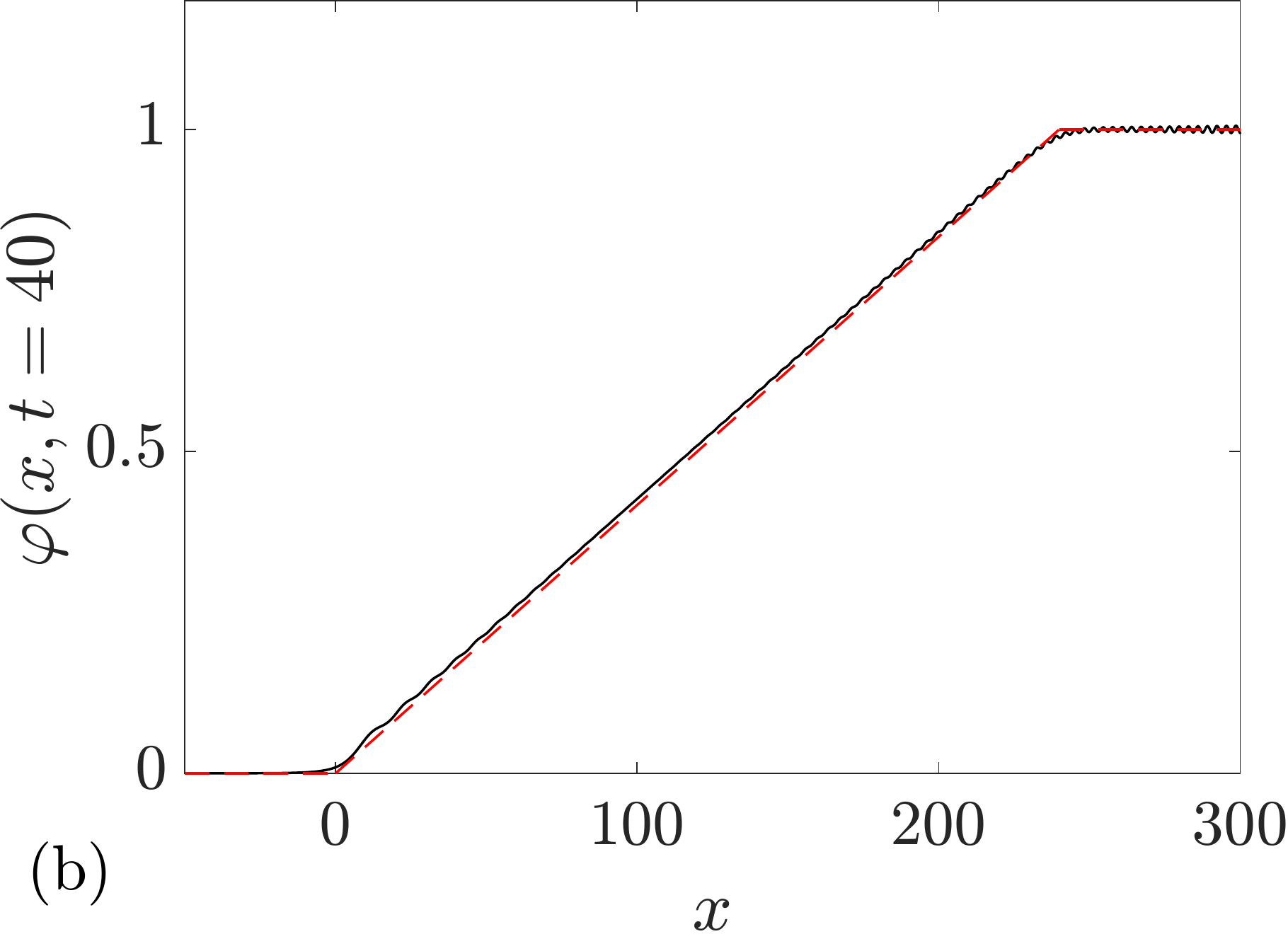}
  \caption{Riemann problem with initial
    condition~\eqref{eq:init_rw} for DOS $u(\eta; x,t)$. The plots
    depict the variation of a condensate's realization $\varphi(x,t)$
    at $t=0$ (a) and $t=40$ (b, solid line). The red dashed
    line in depicts the variation of the rarefaction wave $\varphi = \lambda_1(x/t)^2$~\eqref{eq:rw}.}
  \label{fig:riem0_RW}
\end{figure}

In this first example, we choose
\begin{equation}
  \label{eq:init_rw}
  \{N; {\bs \la}\}( x,t=0) =
  \begin{cases}
    \{0; q_-=0 \}, & x<0, \\
    \{0; q_+=1 \},  & x>0.
  \end{cases}
\end{equation}
A numerical realization of the soliton condensate evolution corresponding to the steplike initial condition~\eqref{eq:init_rw} is displayed in
Fig.~\ref{fig:riem0_RW}. The same figure displays the realization at
$t=40$. The realization corresponds to a $n$-soliton solution with
parameters distributed according to the initial DOS of~\eqref{eq:ustep0}, \eqref{eq:init_rw};
details are given in Appendix~\ref{sec:numerical}. As predicted in
Sec.~\ref{sec:equilibrium}, the realization of the condensate
corresponds to the vacuum $\varphi=0$ at the left of $x=0$, and a
constant $\varphi=1$ at the right of $x=0$. As highlighted in
Appendix~\ref{sec:Riemann_numerical}, the $n$-soliton solution
displays an overshoot at $x=0$, regardless of the number of solitons
$n$, which is reminiscent of Gibbs’ phenomenon in the theory of
Fourier series. This phenomenon has been originally observed in the numerical approximation of the soliton condensate of the focusing NLS equation by a $n$-soliton solution in~\cite{gelash2021}; see for instance the similarities between Figs.~\ref{fig:riem0_RW}a, \ref{fig:riem0_DSW}b and Fig.~2a of~\cite{gelash2021}.
Indeed, in both cases, the IST spectrum of the step distribution
contains a non-solitonic radiative component (cf.~\cite{ablowitz_nonlinear_2011}), which is not taken into account by the $n$-soliton solution; the mismatch between the
exact step and the $n$-soliton solution manifests by the occurrence of the spurious
oscillations observed near $x=0$. 

\

The solution of the Riemann problem with the initial
condition~\eqref{eq:init_rw} is given by $u^{(0)}(\eta; \la_1(x,t))$ where
$\lambda_1(x,t)$ is the rarefaction wave (genus 0) solution~\eqref{eq:rw}. We
have shown in Sec.~\ref{sec:equilibrium} that the genus 0 soliton
condensate is almost surely described by the constant solution
$\varphi = (\lambda_1)^2$. In the context of the evolution of the step \eqref{eq:init_rw} 
$\lambda_1$ varies according to \eqref{eq:rw} so $\lambda(x,t)$  should be treated as a slowly varying (locally constant) condensate solution. In Fig.~\ref{fig:riem0_RW} we compare the numerical realization of the evolution of genus $0$ condensate  with the analytical solution~\eqref{eq:rw}.

\subsection{Dispersive shock wave}

\begin{figure}[h]
  \centering
  \includegraphics[width=7cm]{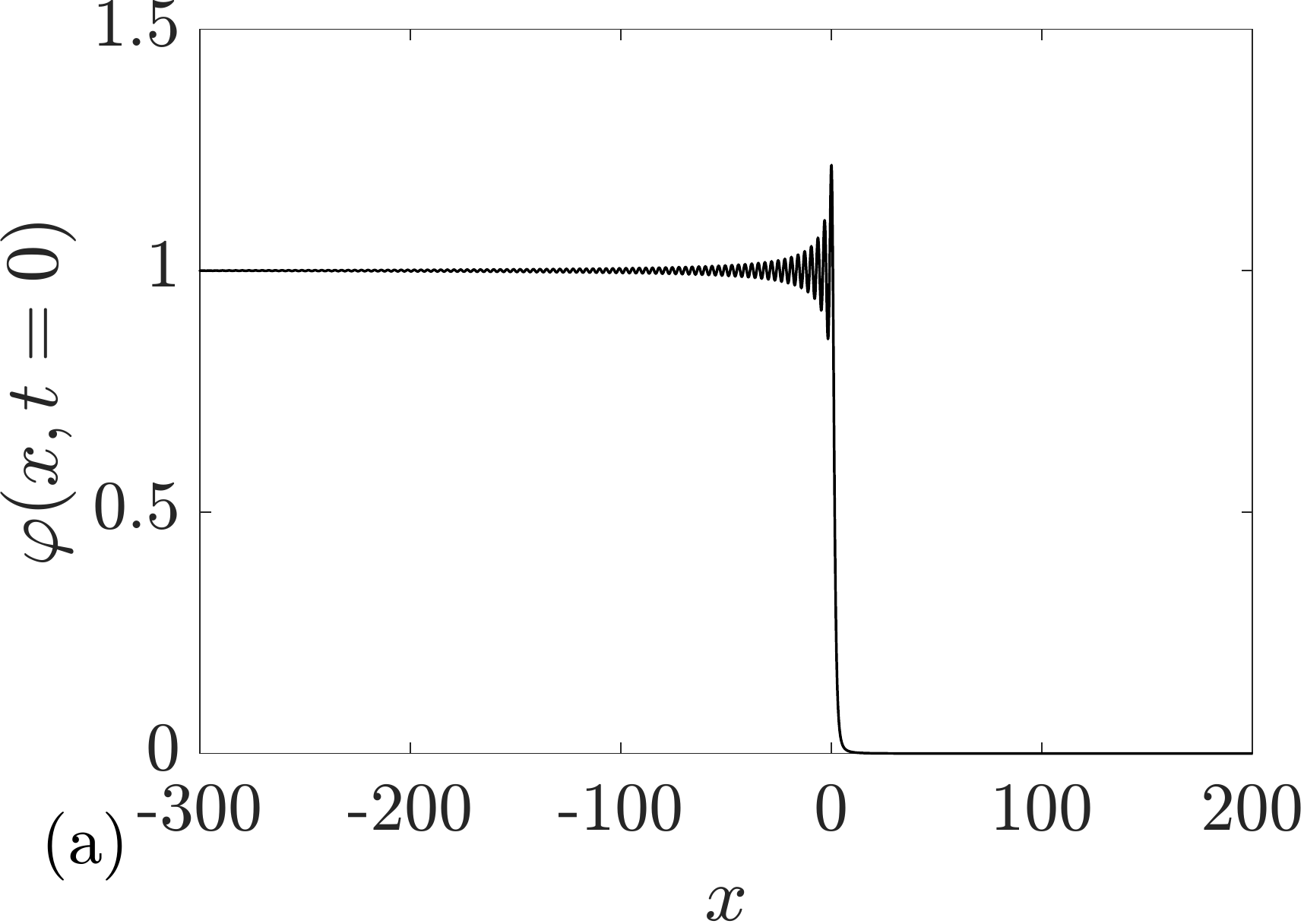}\hspace{1cm}
  \includegraphics[width=7cm]{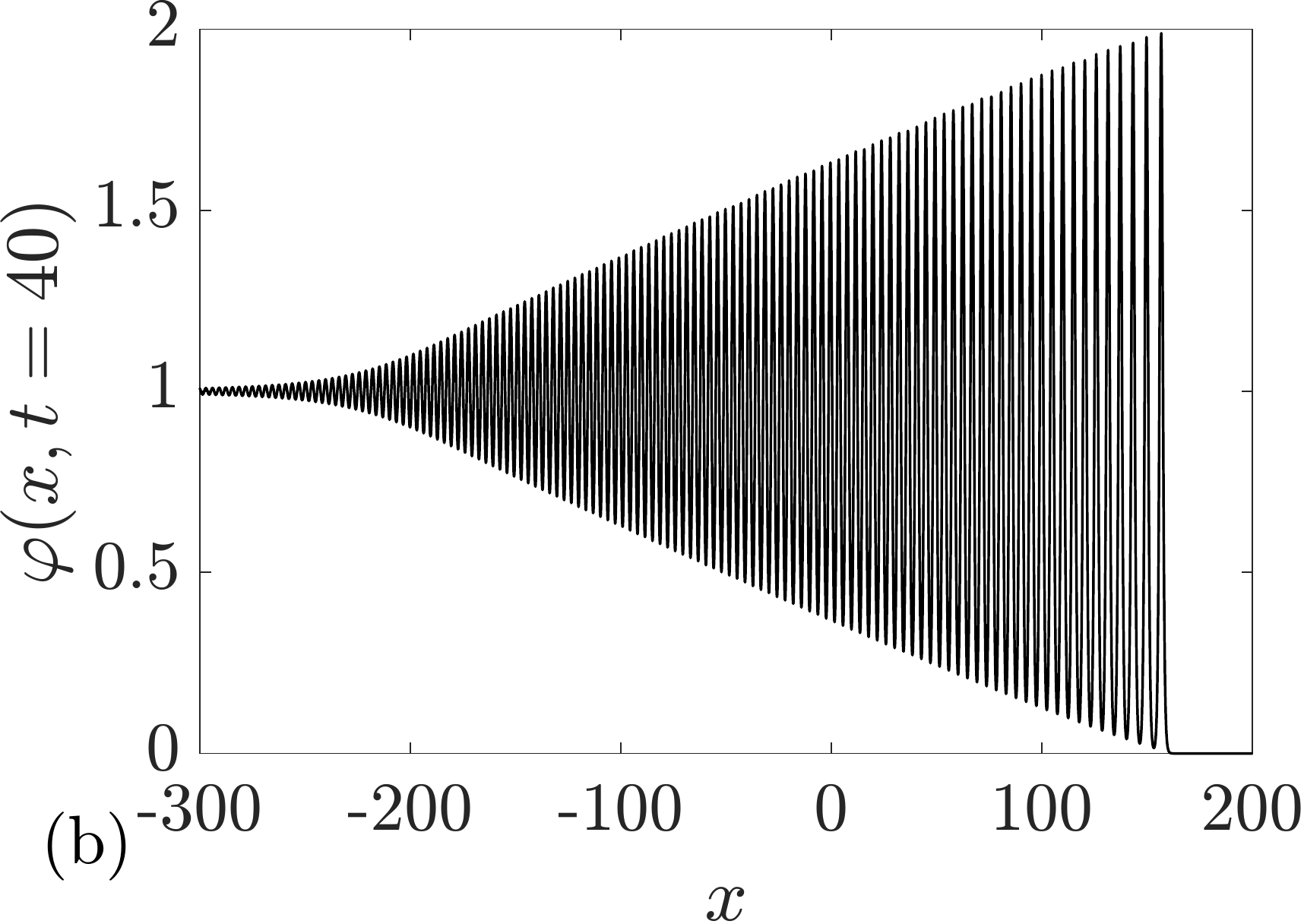}
  \caption{Riemann problem with initial
    condition~\eqref{eq:init_dsw} for DOS $u(\eta; x,t)$. The plots
    depict the variation of a condensate's realization $\varphi(x,t)$
    at $t=0$ (a) and $t=40$ (b).}
  \label{fig:riem0_DSW}
\end{figure}

We now consider
\begin{equation}
  \label{eq:init_dsw}
  \{N; {\bs \la}\}( x,t=0) =
  \begin{cases}
    \{0; q_-=1 \}, & x<0, \\
    \{0; q_+=0 \},  & x>0.
  \end{cases}
\end{equation}
A numerical realization of the genus $0$ soliton  condensate corresponding to the step-initial condition~\eqref{eq:init_dsw} is presented in
Fig.~\ref{fig:riem0_DSW} (a):  it corresponds to the vacuum $\varphi=0$ for $x>0$, and a
constant $\varphi=1$ for  $x<0$. The realization at $t=40$ is shown in Fig.~\ref{fig:riem0_DSW} (b) and it corresponds to a classical DSW solution for the KdV equation.

The solution of the condensate Riemann problem with the initial
condition~\eqref{eq:ustep0},~\eqref{eq:init_dsw} is given by the genus 1 DOS~\eqref{eq:sol2_dsw} modulated by the $2$-wave
solution~\eqref{eq:dsw} of the Whitham equations.  In order to
make a quantitative comparison of this analytical solution with the numerical
evolution of the soliton gas displayed in Fig.~\ref{fig:riem0_DSW}, we
compute numerically the mean $\langle \varphi \rangle$ and the
variance
$\sqrt{\langle \varphi^2 \rangle - \langle \varphi \rangle^2}$, the
latter being an amplitude type characteristic of the cnoidal wave.  We
have conjectured in Sec.~\ref{sec:equilibrium} that any realization of
the uniform genus $1$ condensate corresponds to a cnoidal wave modulo
the initial phase $\theta^0\in[0;2\pi)$. In that case, the ensemble
average of the soliton condensate reduces to an
average over the phase $\theta^0$, or equivalently, over the period of
the cnoidal wave, which can be performed on a single realization.  We
assume here that the result generalizes to non-uniform condensates
so that  the realization computed numerically and displayed
in Fig.~\ref{fig:riem0_DSW}(b) can be consistently compared with a slowly modulated
cnoidal wave solution.  The averages $\langle \varphi(x,t) \rangle$ and
$\langle \varphi(x,t)^2 \rangle$ can be determined via a local phase
average of one realization of the condensate. The local period averages
are obtained via
\begin{equation}
  \label{eq:aver_num}
  \langle \varphi(x,t) \rangle = \frac{1}{L(x,t)} \int_x^{x+L(x,t)}
  \varphi(y,t) dy,\quad
  \langle \varphi(x,t)^2 \rangle = \frac{1}{L(x,t)} \int_x^{x+L(x,t)}
  \varphi(y,t)^2 dy,
\end{equation}
where $L(x,t)$ is the local wavelength extracted numerically.

The comparison between the analytically determined
averages~\eqref{meanphi},\eqref{meansquarephi},\eqref{eq:dsw} and the
averages~\eqref{eq:aver_num} obtained numerically is presented in Fig.~\ref{fig:kdv_3+_2}
and shows a very good agreement.
\begin{figure}[h]
  \centering
  \includegraphics[width=7cm]{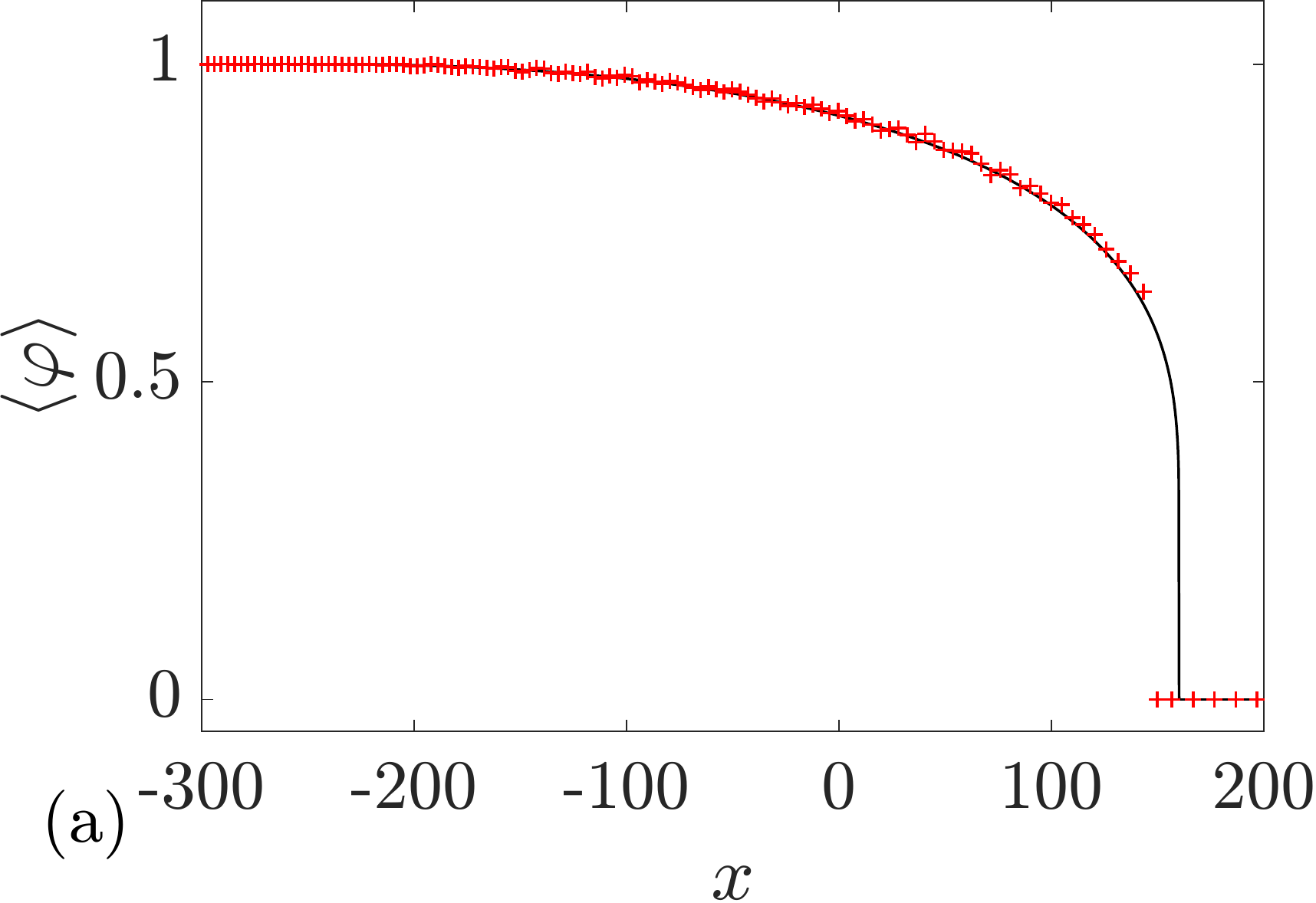}\hspace{1cm}
  \includegraphics[width=7cm]{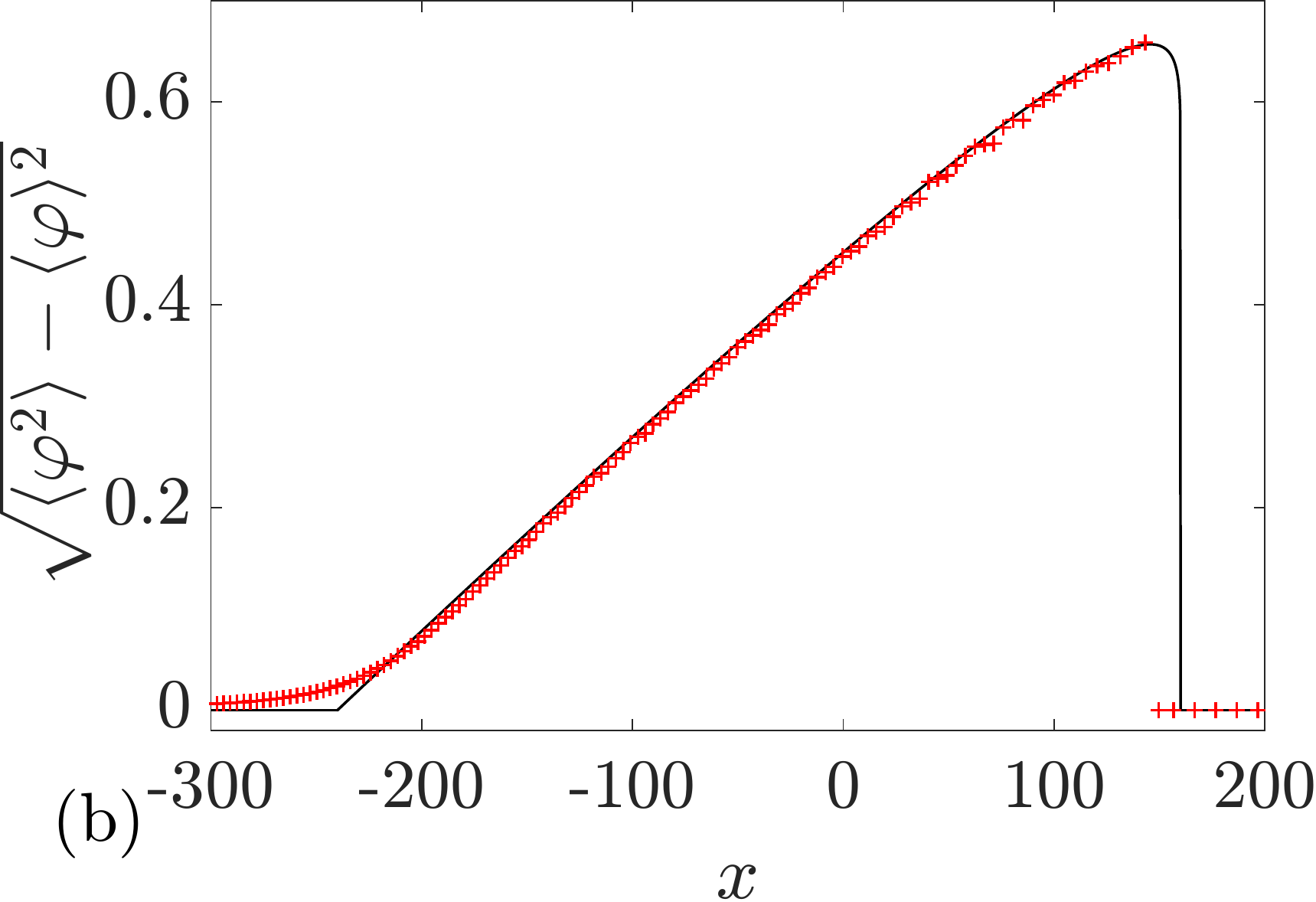}
  \caption{Mean $\langle \varphi \rangle$ (a) and variance
    $\sqrt{\langle \varphi^2 \rangle - \langle \varphi \rangle^2}$
    (b) of the solution of the Riemann problem's
    solution with the initial condition~\eqref{eq:init_dsw}.  The markers
    correspond to averages extracted from the numerical solution
    using~\eqref{eq:aver_num}, and the solid black lines to the
    corresponding analytical
    averages~\eqref{meanphi},\eqref{meansquarephi},\eqref{eq:dsw}.}
  \label{fig:kdv_2}
\end{figure}

\subsection{Generalized rarefaction wave}

\begin{figure}[h]
  \centering
  \includegraphics[width=7cm]{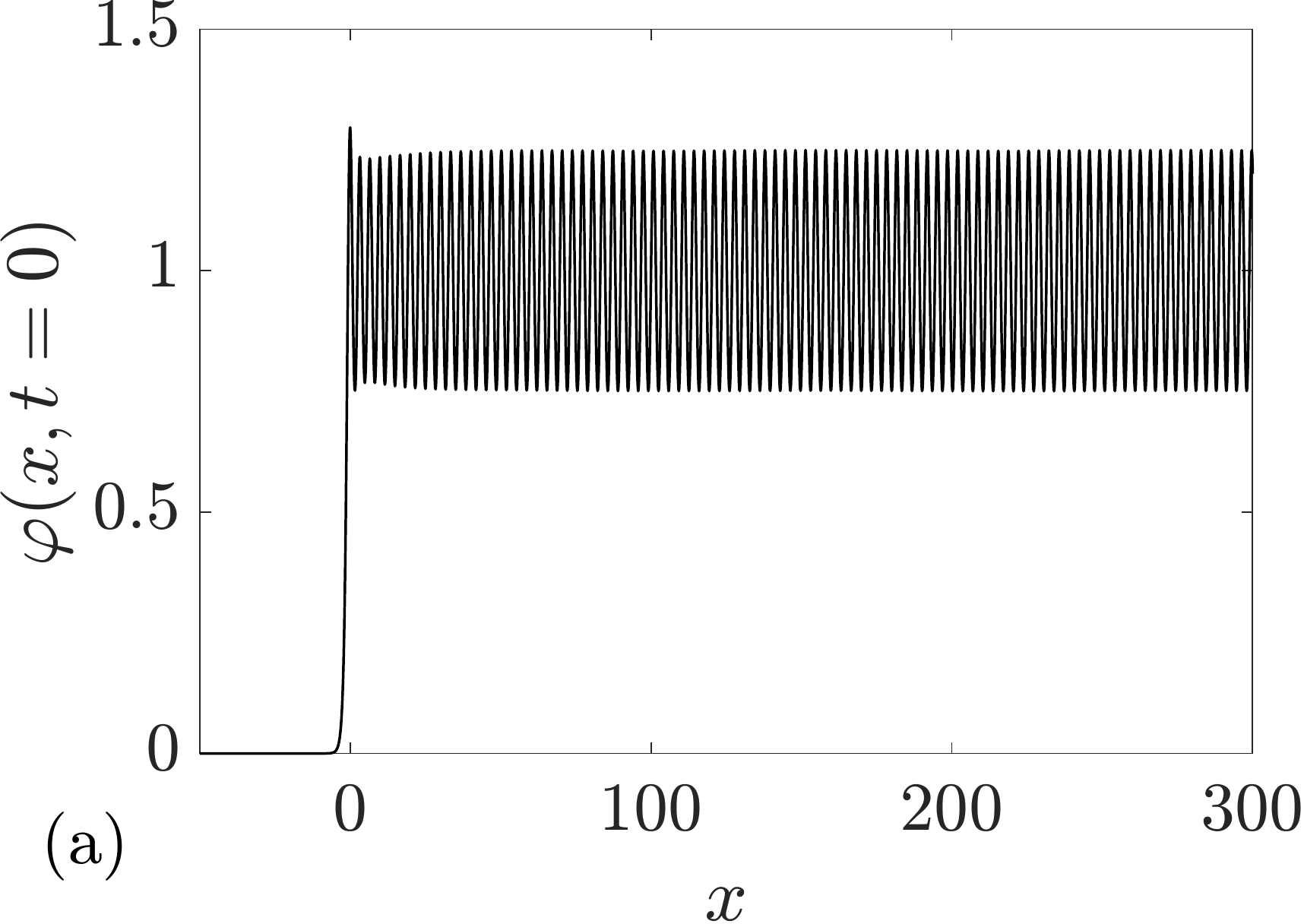}\hspace{1cm}
  \includegraphics[width=7cm]{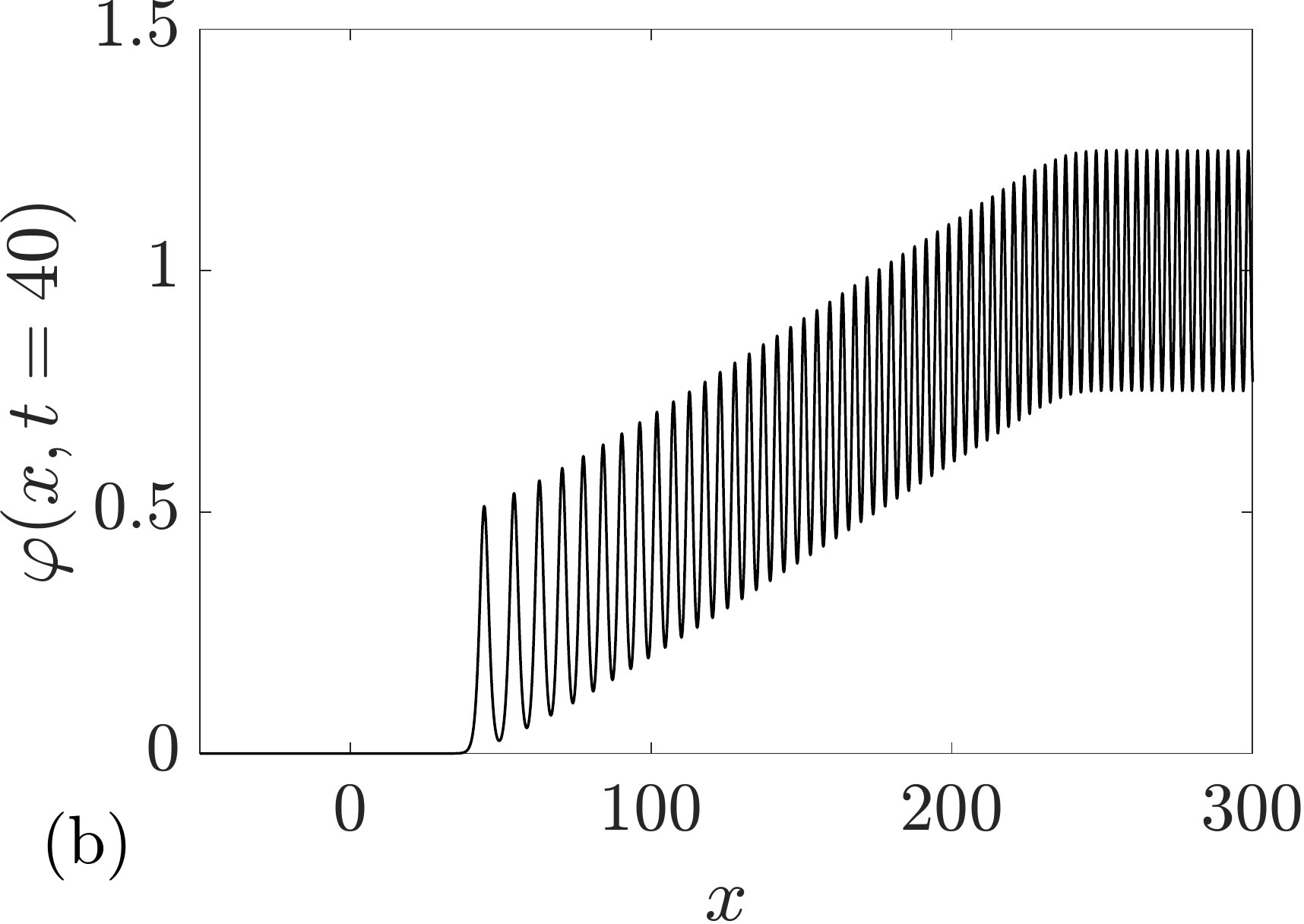}
  \caption{Riemann problem for soliton condensate with initial
    condition~\eqref{eq:init1} for DOS $u(\eta; x,t)$. The plots
    depict the variation of a condensate's realization $\varphi(x,t)$
    at $t=0$ (a) and $t=40$ (b).}
  \label{fig:kdv_3+}
\end{figure}

$N_-+N_+=0$ in the two previous examples. In the next examples, we
choose $N_-+N_+=1$. Let's start with $N_+=1$:
\begin{equation}
  \label{eq:init1}
  \{N; {\bs \la}\}( x,t=0) =
  \begin{cases}
    \{0; q_-=0 \},                                       & x<0, \\
    \{1; (\lambda_1^+=0, \lambda_2^+=1/2,\lambda_2^+=1)\}, & x>0.
  \end{cases}
\end{equation}
A numerical realization of the step-initial condition is displayed in
Fig.~\ref{fig:kdv_3+}. The same figure displays the realization at
$t=40$. The realization of the condensate corresponds to the ``vacuum''
$\varphi=0$ for $x<0$, and a cnoidal wave for $x>0$. Note that the KdV equation does not admit heteroclinic
traveling wave solutions, rendering difficult the numerical
implementation of these ``generalized'' Riemann problems studied for instance in
\cite{sprenger_2020, gavryliuk_2020}. Remarkably here, the solution depicted in
Fig.~\ref{fig:kdv_3+} is an exact, $n$-soliton solution of the KdV
equation. As highlighted previously (see also
Appendix~\ref{sec:Riemann_numerical}), the $n$-soliton solution
exhibits an overshoot at $x=0$, regardless of the number of solitons
$n$.

\begin{figure}[h]
  \centering
  \includegraphics[width=7cm]{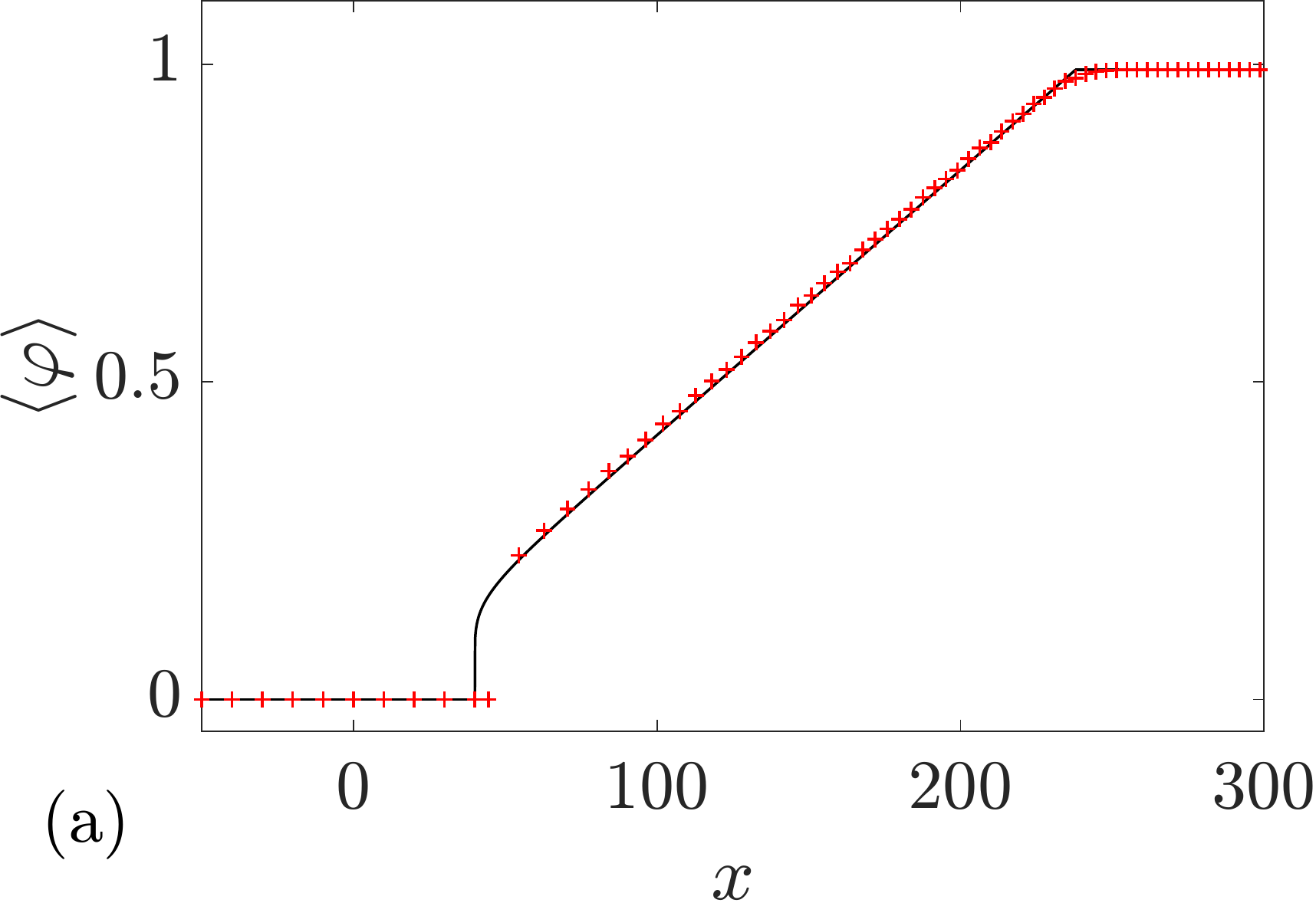}\hspace{1cm}
  \includegraphics[width=7cm]{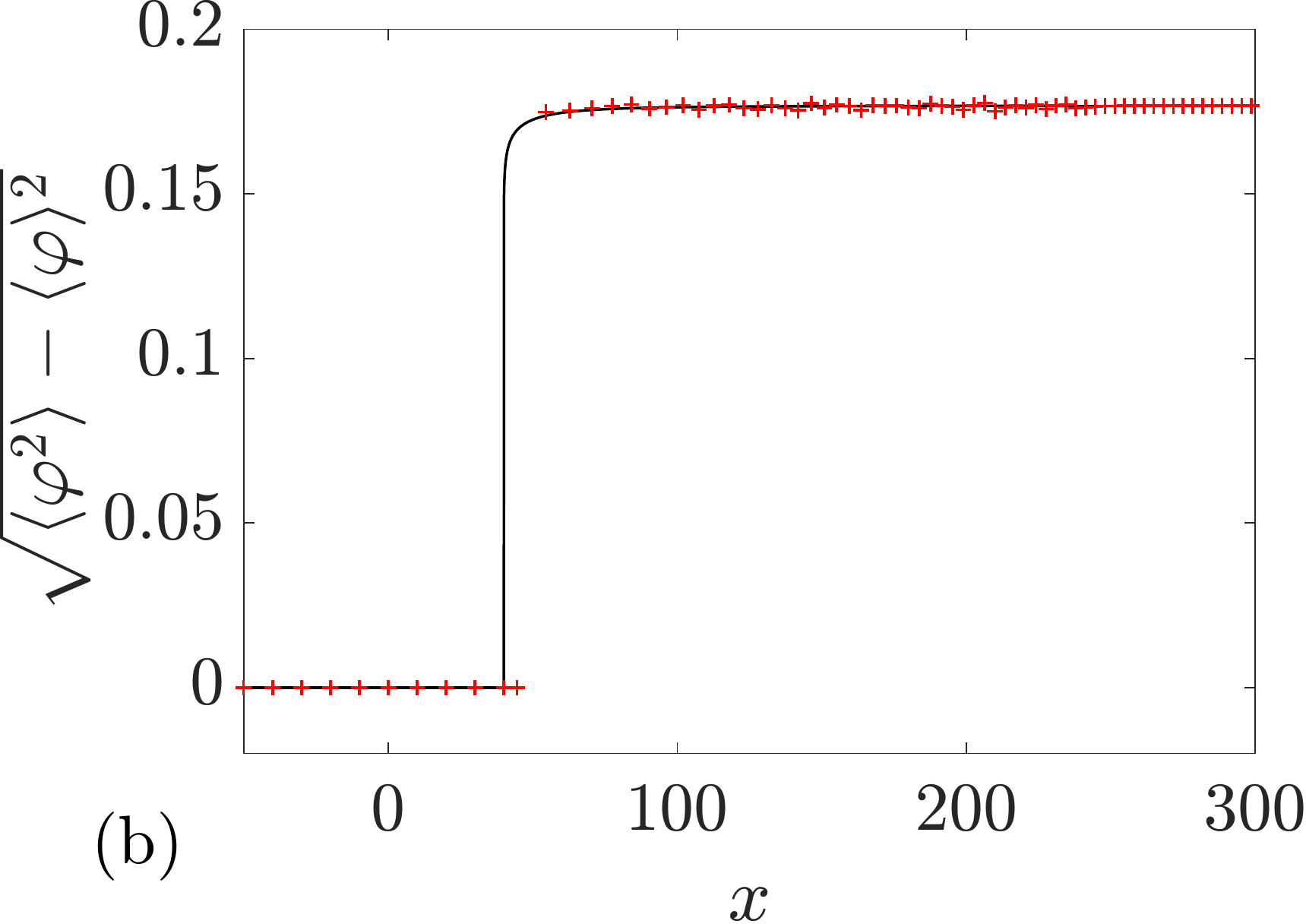}
  \caption{Mean $\langle \varphi \rangle$ (a) and variance
    $\sqrt{\langle \varphi^2 \rangle - \langle \varphi \rangle^2}$
    (b) of the solution of the Riemann problem's
    solution with the initial condition~\eqref{eq:init1}.  The markers
    correspond to the averages extracted from the numerical solution
    using~\eqref{eq:aver_num}, and the solid black lines to the
    corresponding analytical
    averages~\eqref{meanphi},\eqref{meansquarephi},\eqref{eq:3-wave}.}
  \label{fig:kdv_3+_2}
\end{figure}

The solution of the Riemann problem for the kinetic equation with the initial
condition~\eqref{eq:init3+}, \eqref{eq:init1} is given by the $3^+$-wave ~\eqref{eq:sol3},\eqref{eq:3-wave}.  The comparison between
the analytical
averages~\eqref{meanphi},\eqref{meansquarephi},\eqref{eq:3-wave} and
the averages obtained numerically is shown in
Fig.~\ref{fig:kdv_3+_2} and shows a very good agreement.  The
modulation depicted in Figs.~\ref{fig:kdv_3+}b and~\ref{fig:kdv_3+_2}a resembles the
modulation of a cnoidal wave of an almost constant amplitude but with a
varying mean. The variation of the mean $\langle \varphi \rangle$ is
similar to the variation of the field in a classical rarefaction wave, so we call the
corresponding structure shown in Fig.~\ref{fig:kdv_3+}b a {\it generalized rarefaction wave}. 
The variance of the wavefield $\varphi$ in the generalized rarefaction wave is shown in Fig.~\ref{fig:kdv_3+_2}b.

\subsection{Generalized dispersive shock wave}

\begin{figure}[h]
  \centering
  \includegraphics[width=7cm]{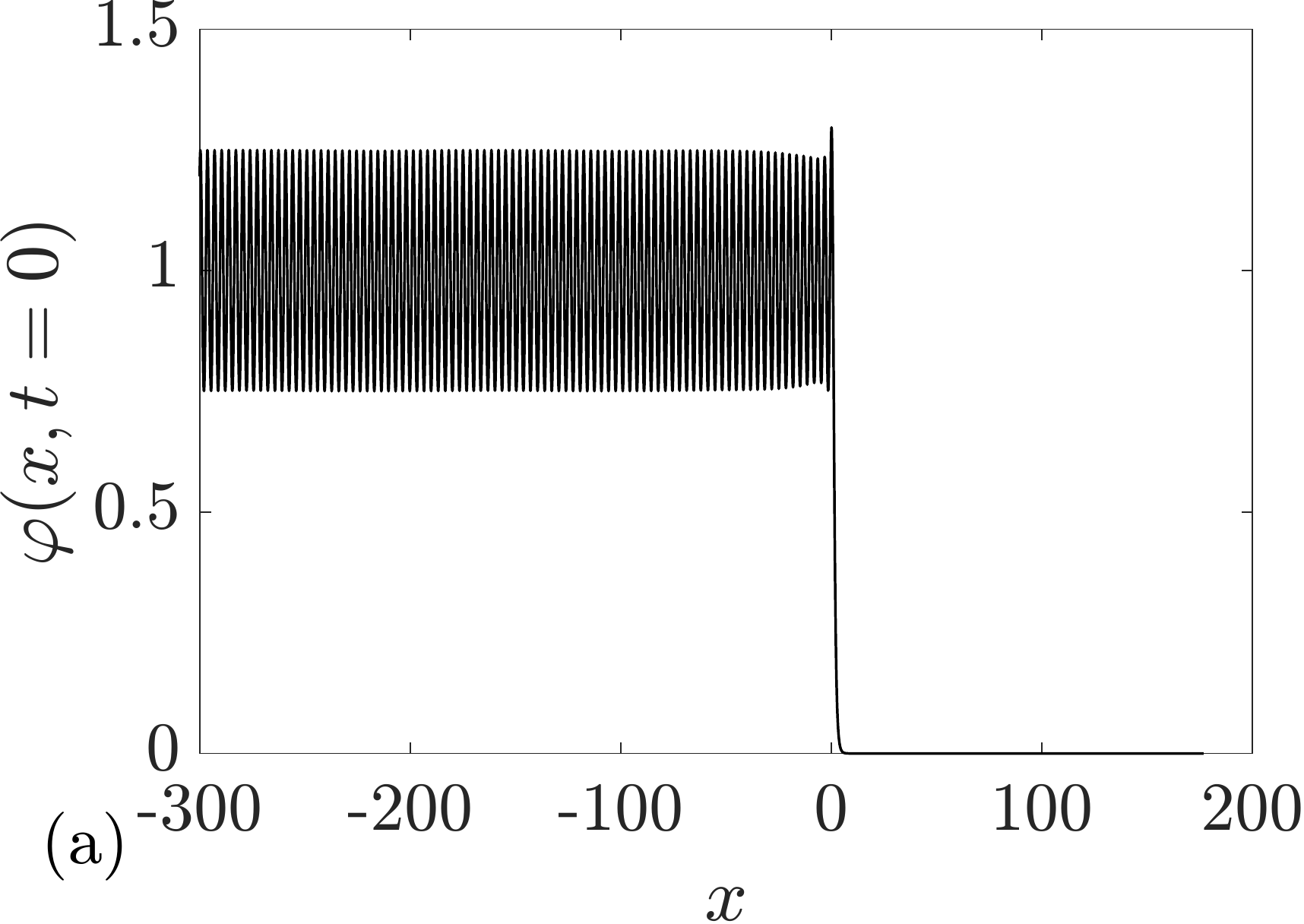}\hspace{1cm}
  \includegraphics[width=7cm]{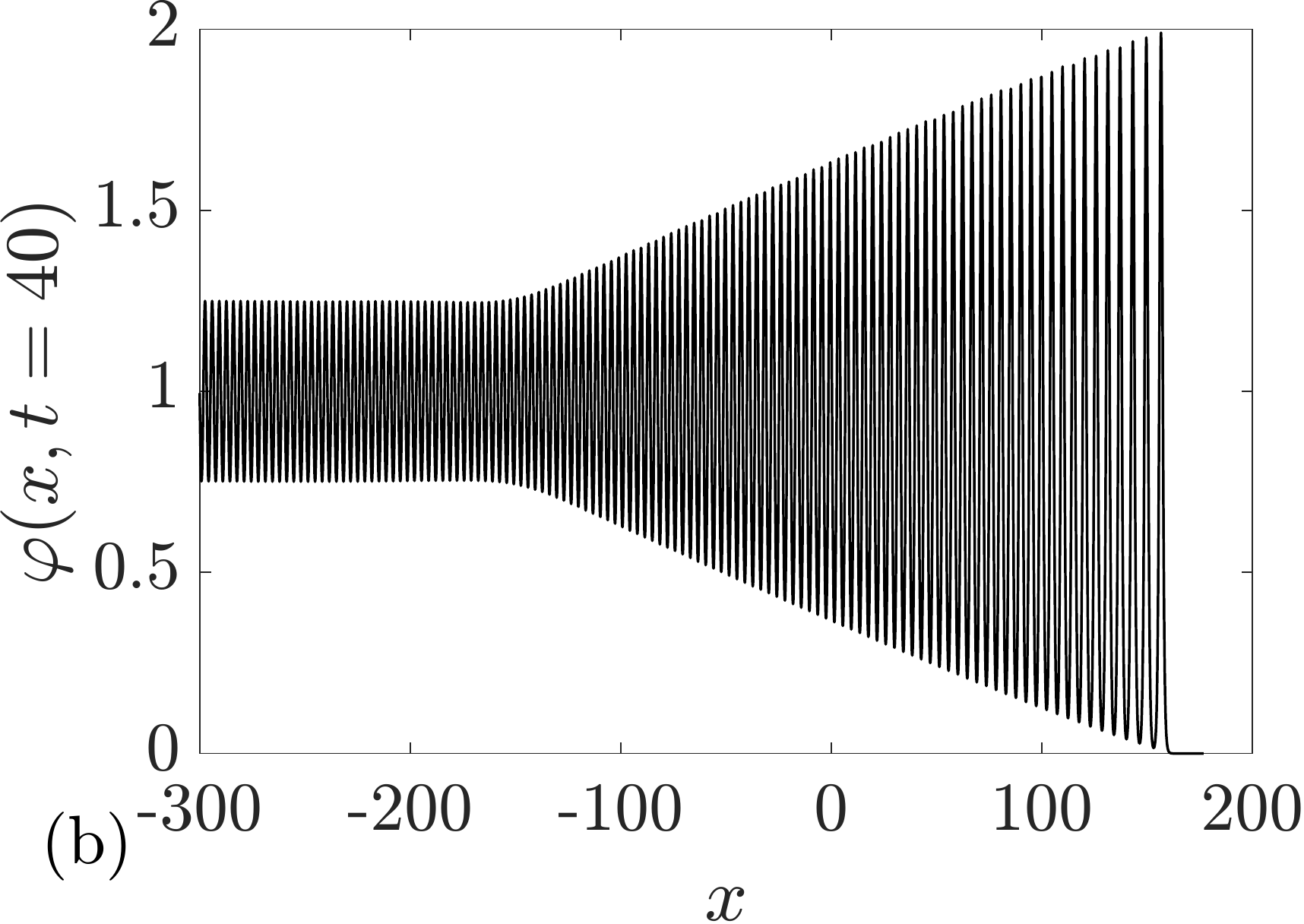}
  \caption{Riemann problem with initial
    condition~\eqref{eq:init2} for DOS $u(\eta; x,t)$. The plots
    depict the variation of a condensate's realization $\varphi(x,t)$
    at $t=0$ (a) and $t=40$ (b).}
  \label{fig:kdv_2-}
\end{figure}

We now consider the ``complementary'' initial condition
\begin{equation}
  \label{eq:init2}
  \{N; {\bs \la}\}( x,t=0) =
  \begin{cases}
    \{1; (\lambda_1^-=0, \lambda_2^-=1/2,\lambda_2^-=1)\}, & x<0, \\
    \{0; q_+=0 \},                                       & x>0.
  \end{cases}
\end{equation}
An example of the numerical realization of the soliton gas step-initial condition and its
evolution at $t=40$ are displayed in
Fig.~\ref{fig:kdv_2-}.

\

\begin{figure}[h]
  \centering
  \includegraphics[width=7cm]{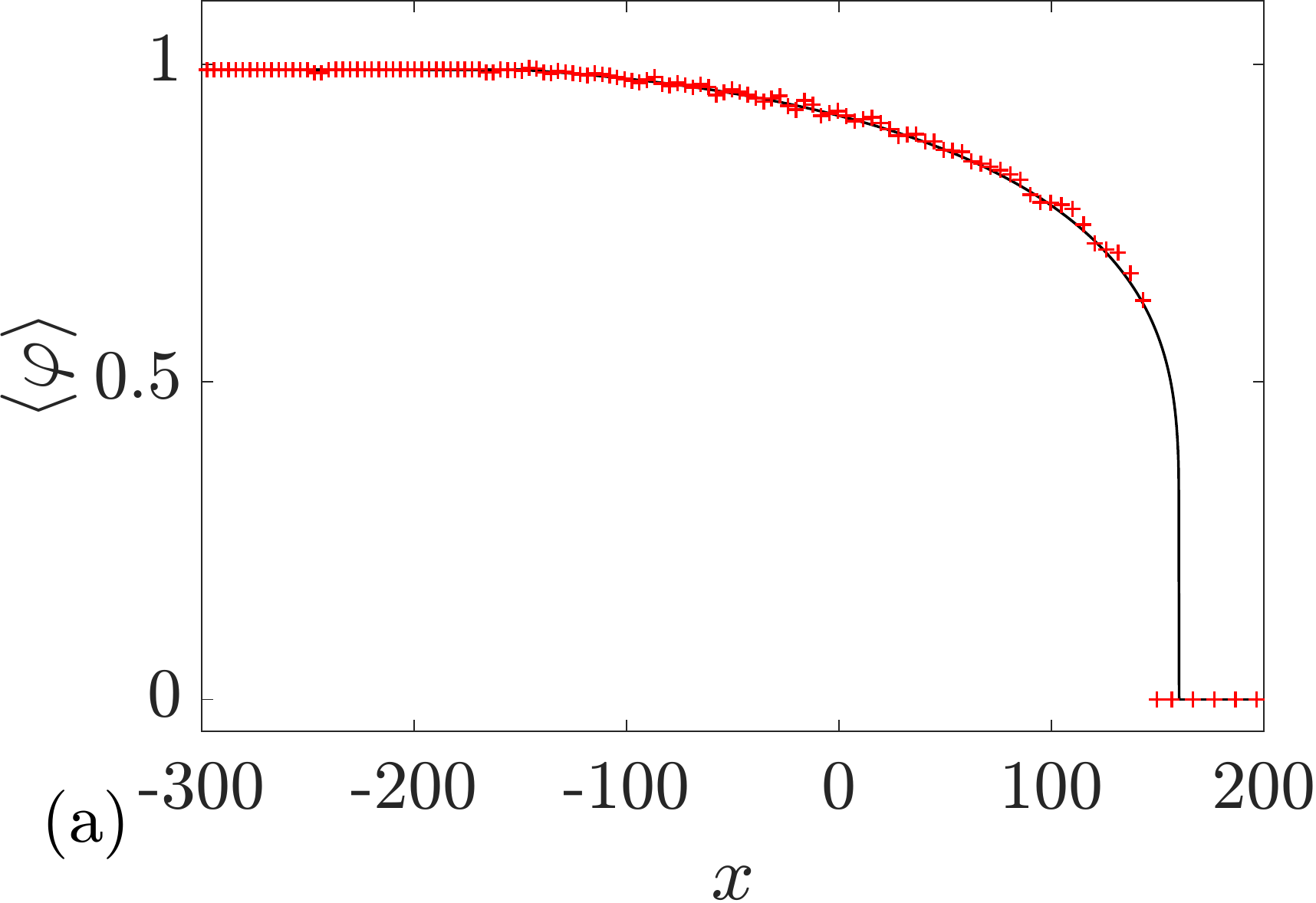}\hspace{1cm}
  \includegraphics[width=7cm]{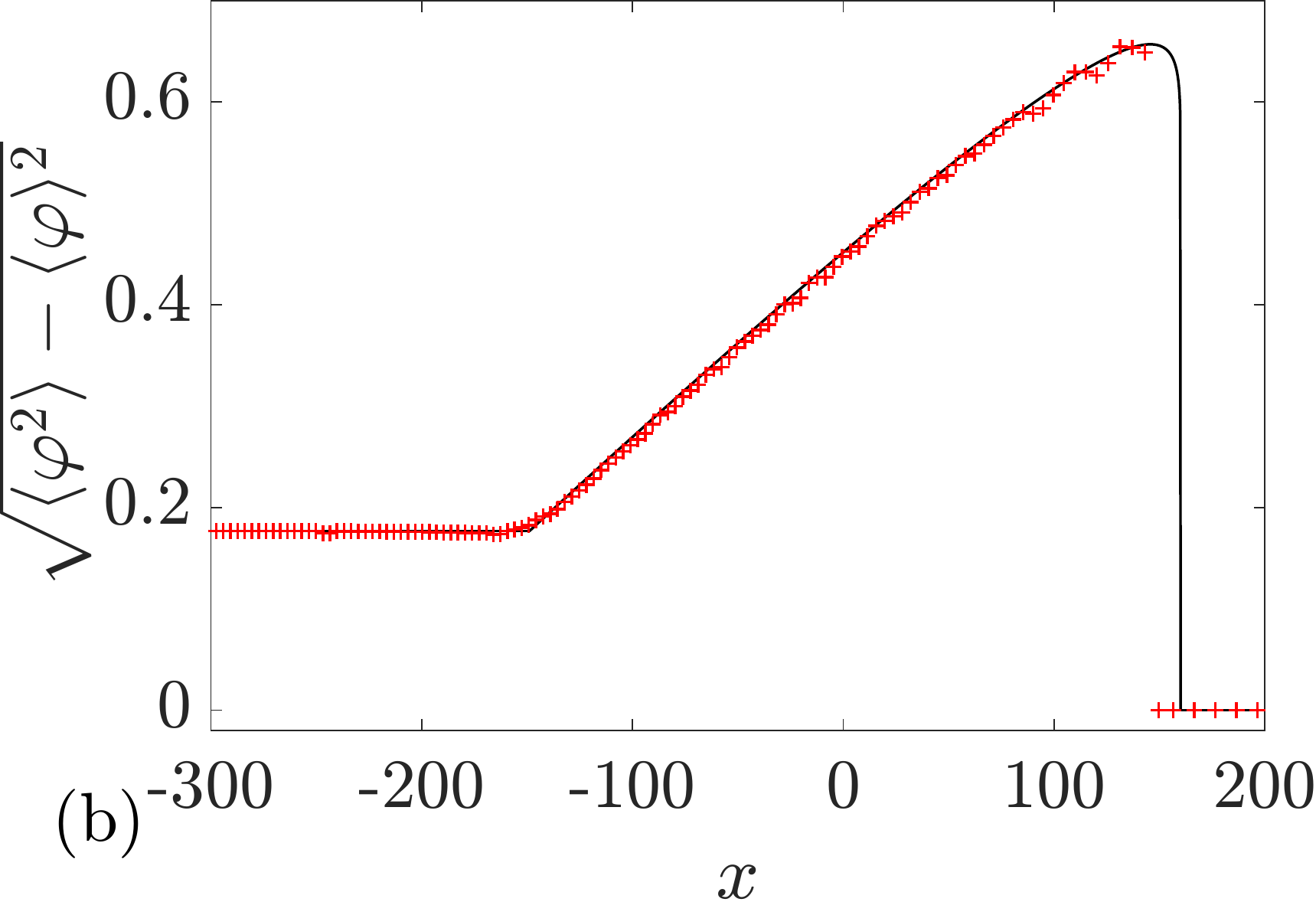}
  \caption{Mean $\langle \varphi \rangle$ (a) and variance
    $\sqrt{\langle \varphi^2 \rangle - \langle \varphi \rangle^2}$
    (b) of the solution of the Riemann problem's
    solution with the initial condition~\eqref{eq:init2}.  The markers
    correspond to averages extracted from the numerical solution
    using~\eqref{eq:aver_num}, and the solid black lines to the
    corresponding analytical
    averages~\eqref{meanphi},\eqref{meansquarephi},\eqref{eq:3-wave}.}
  \label{fig:kdv_2-_2}
\end{figure}

The solution of the Riemann problem with the initial
condition~\eqref{eq:init2} is given by the $2^-$-wave ~\eqref{eq:sol2-}, \eqref{eq:2--wave}.  The comparison between the analytically derived
averages~\eqref{meanphi},\eqref{meansquarephi},\eqref{eq:2--wave} and
the averages obtained numerically is displayed in
Fig.~\ref{fig:kdv_2-_2}, and shows a very good agreement.  The
modulation observed in Figs.~\ref{fig:kdv_2-}, ~\ref{fig:kdv_2-_2} resembles the
modulation of partial dispersive shock wave: the modulated cnoidal
wave reaches the soliton limit $m=1$ for $x \to s_+ t$ but terminates at  $m \neq 0$ for $x \to s_-t$. The solution then continues as a non-modulated cnoidal wave for $x < s_- t$. This structure differs from the
celebrated dispersive shock wave solution of the KdV equation involving the entire range $0 \leq m \leq 1$
\cite{el_dispersive_2016}. We call the described structure connecting a constant state (a genus 0 condensate) at $x \to + \infty$ with a periodic solution (a genus 1 condensate) at $x \to - \infty$ a {\it generalized DSW}.  We note that the soliton condensate structure shown in Fig.~\ref{fig:kdv_2-}b exhibits strong similarity  to the ``deterministic KdV soliton gas'' solution constructed in~\cite{girotti_rigorous_2021}.

\section{Diluted soliton condensates }
\label{sec:dilute}

\subsection{Equilibrium properties}
\label{sec:dilute_eq}
We now introduce the notion of a ``diluted'' soliton condensate by considering DOS $u(\eta)=C u^{(N)} (\eta)$, where
$u^{(N)}(\eta)$ is the condensate DOS of genus $N$, and $0<C<1$ is the
``dilution constant''.

\begin{figure}[h]
  \centering
  \includegraphics[width= 7cm]{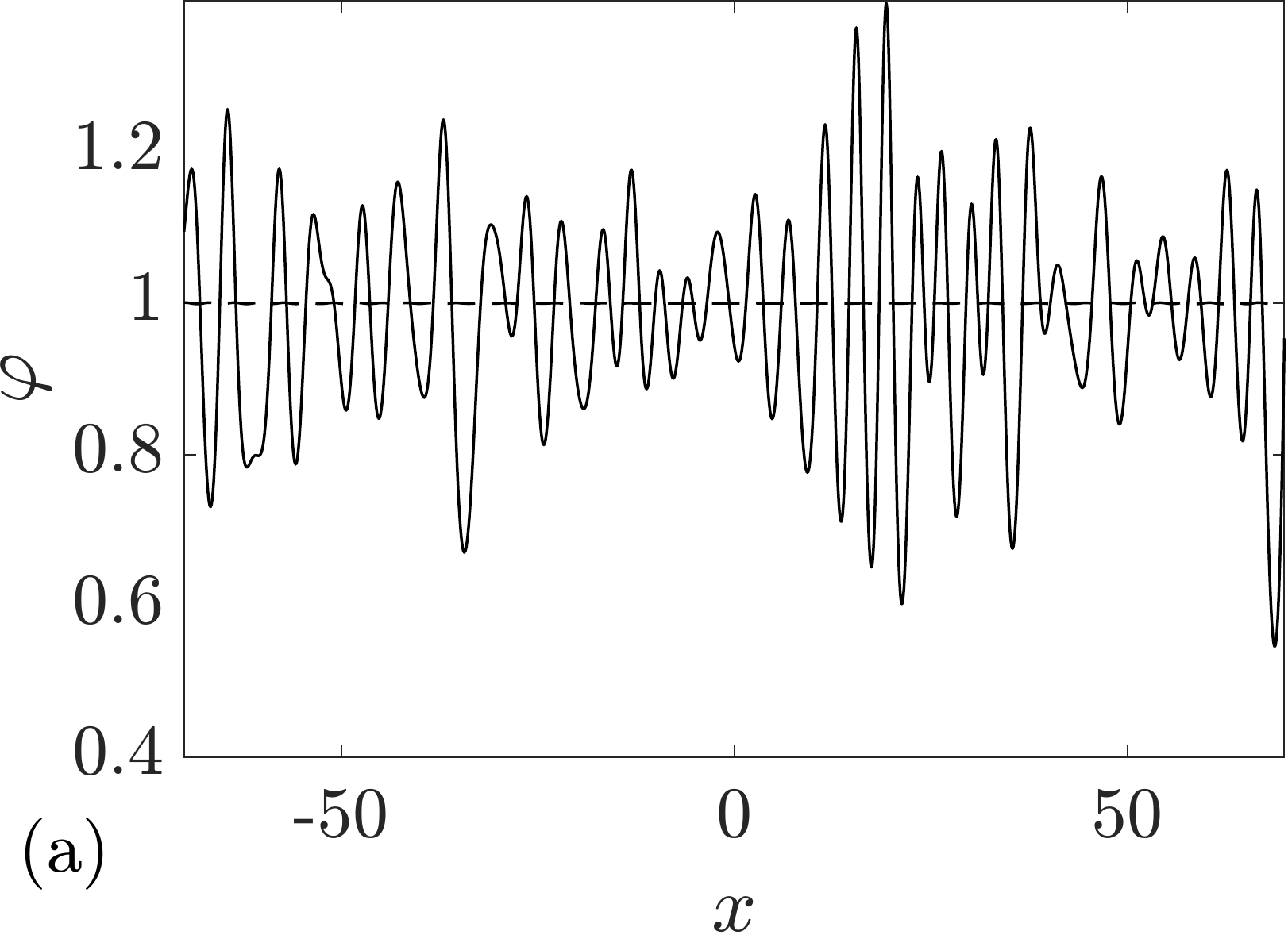}\hspace{1cm} \includegraphics[width= 7cm]{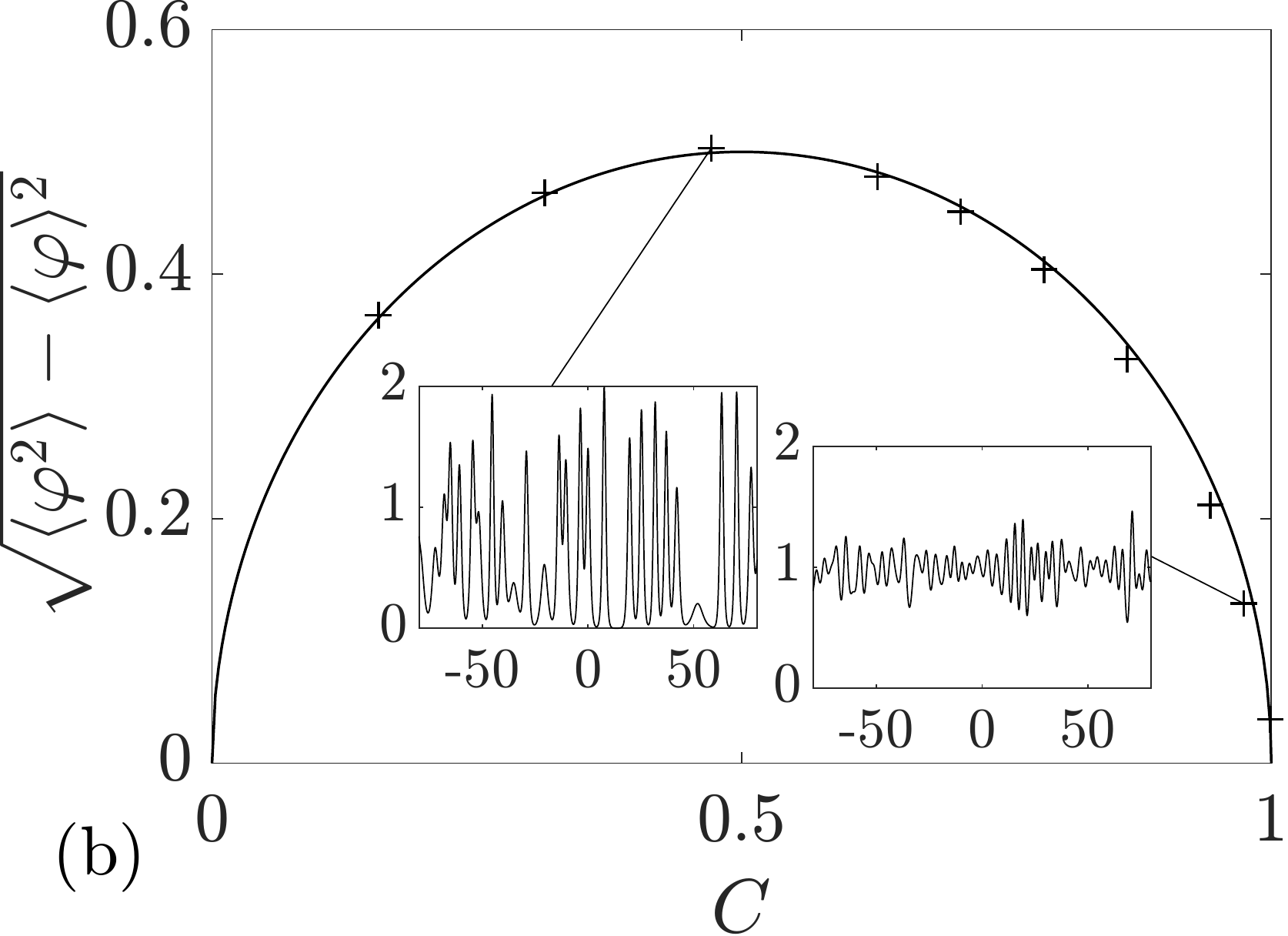}
  \caption{a) Realizations soliton gas with the DOS
    ~\eqref{diluted_conds} and $\lambda_1=1$: $C=1$ in dashed line
    (genus 0 condensate) vs $C= 0.97$ in solid line (diluted genus
    zero condensate); in both cases the gas is realized numerically
    with $N=100$ solitons. b) Variance for diluted condensates
    $C<1$. Solid line: formula~\eqref{eq:variance}; markers:
    numerically extracted values of the variance; insets: typical
    realizations of the KdV wave field $\varphi(x,t)$ in diluted
    condensates.}
  \label{fig:compare_dilute}
\end{figure}

E.g. the diluted soliton condensate of genus 0 is characterized by DOS
\begin{equation}\label{diluted_conds}
  u(\eta)= C \frac{\eta}{\pi \sqrt{\la_1^2-\eta^2}}, \quad 0 < C < 1.
\end{equation}
We recover  the genus $0$ condensate  DOS~\eqref{uv} by setting $C=1$. As $C$ decreases, the ``averaged spacing'' between the solitons
\begin{equation}
  \kappa^{-1} = \left(\int u(\eta) d\eta \right)^{-1} \propto C^{-1}
\end{equation}
increases and the condensate gets ``diluted''. Comparison between the most probable
realization of the condensate ($C=1$) and a typical realization of a slightly dilute
condensate ($C=0.97$) is given in
Fig.~\ref{fig:compare_dilute}. Remarkably, one can see that a slight increase of the average spacing between the solitons within the condensate results in the emergence of  significant random oscillations of the KdV wave field.

\begin{figure}[h]
  \centering
  \includegraphics[width= 7cm]{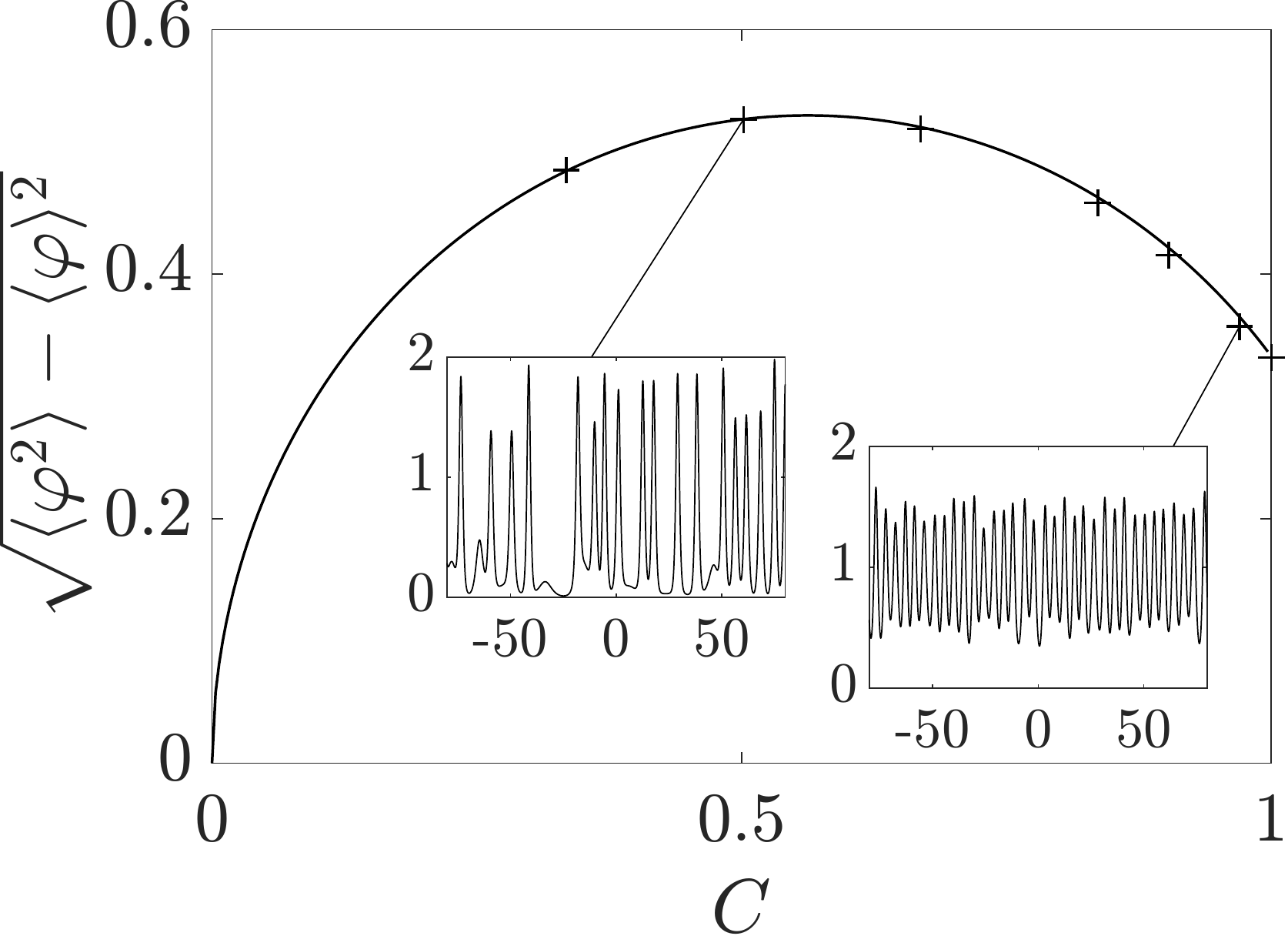}
  \caption{Variance for diluted genus 1 condensates with DOS~\eqref{diluted_conds2} and $(\lambda_1,\lambda_2,\lambda_3)=(0.5,0.85,1)$;
    insets: typical realizations of the KdV wave field $\varphi(x,t)$ in
    diluted condensates.}
  \label{fig:compare_dilute2}
\end{figure}

As follows from~\eqref{eq:moments} we have $\langle \varphi \rangle =  \langle \varphi^2
  \rangle = C$ for the diluted genus 0 condensate so that the variance is given by:
\begin{equation}
  \label{eq:variance}
  \Delta = \sqrt{\langle
    \varphi^2 \rangle - \langle
    \varphi \rangle^2} = \sqrt{C(1 - C)}.
\end{equation}
The comparison between~\eqref{eq:variance} and the variance obtained numerically
by averaging over different diluted condensates
is presented in Figure~\ref{fig:compare_dilute}. Assuming  ergodicity
of a generic uniform  soliton gas, the ensemble average
$\langle \dots \rangle$ in Fig.~\ref{fig:compare_dilute}a (and
Fig.~\ref{fig:compare_dilute2}) is computed here numerically with a
spatial average of one, spatially broad, gas realization.

More generally, the diluted soliton condensate of genus $N$ is characterized by DOS
\begin{equation}\label{diluted_conds2}
  u(\eta)= C u^{(N)}(\eta;\lambda_1,\dots,\lambda_{2N+1}) , \quad 0 < C < 1.
\end{equation}
We have in the general case
\begin{equation}
  \langle \varphi \rangle = C \langle \varphi_{\rm c}^{(N)} \rangle,\quad
  \langle \varphi^2 \rangle = C \langle \big(\varphi_{\rm c}^{(N)} \big)^2 \rangle,
\end{equation}
where $\langle \varphi_{\rm c}^{(N)} \rangle$, $\langle \big(\varphi_{\rm c}^{(N)} \big)^2\rangle$ are
the ensemble averages obtained for the genuine condensate ($C=1$), functions of
$\lambda_1,\dots,\lambda_{2N+1}$ only; for instance
$\langle \varphi_{\rm c}^{(1)} \rangle$, $\langle \big(\varphi_{\rm c}^{(1)} \big)^2 \rangle$ are given
by~\eqref{meanphi},\eqref{meansquarephi}. Since
$\langle \big(\varphi_{\rm c}^{(N)} \big)^2\rangle \neq \langle \varphi_{\rm c}^{(N)} \rangle^2$ for
$N \geq 1$ and distinct $\lambda_i$'s, the variance of diluted genus
condensates
\begin{equation}
  \Delta = \sqrt{C \langle \big(\varphi_{\rm c}^{(N)} \big)^2 \rangle - C^2 \langle \varphi_{\rm c}^{(N)} \rangle^2}
\end{equation}
never vanishes if $N \geq 1$, as can be seen in the example $N=1$ shown in
Fig.~\ref{fig:compare_dilute2}. Thus, in contrast with the genus 0 case, the transition from the genus 1 condensate
($C=1$) to diluted genus 1 condensate ($C<1$) does not see a drastic change in the
oscillations' amplitude. In particular, the oscillations seem to
remain ``almost'' coherent -- i.e. an average period can be identified -- for
the dilution factors $C$ close to 1 as depicted in the inset of
Fig.~\ref{fig:compare_dilute2}.

\begin{figure}[h]
  \centering
  \includegraphics[width= 14cm]{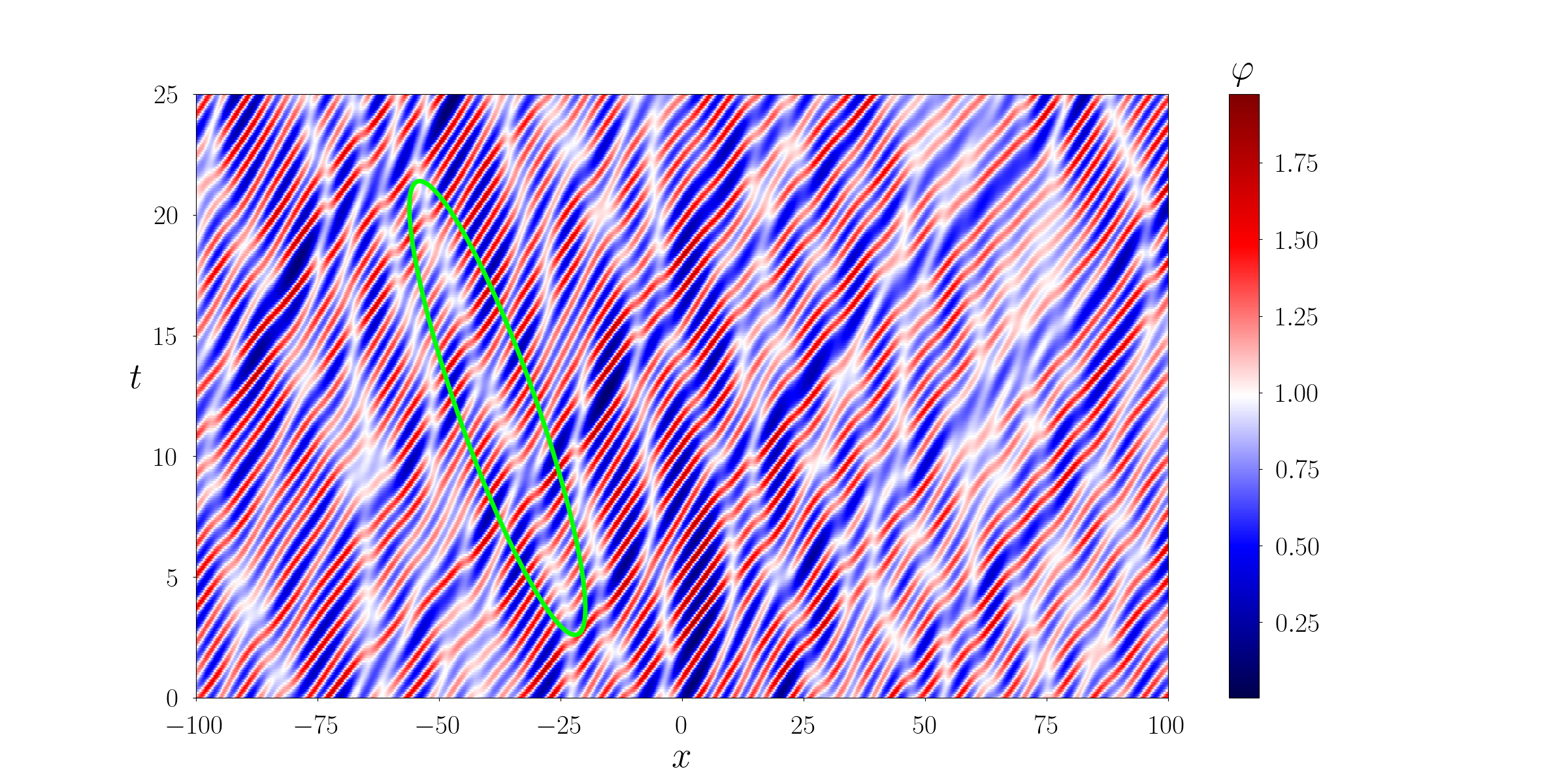}
  \caption{Soliton trajectories in a diluted genus 0 soliton condensate with $C=0.9$. Highlighted is a small-amplitude tracer soliton moving backwards.}
  \label{fig:backflow}
\end{figure}

Diluted condensates present a convenient  framework to verify the prediction formulated in Remark~\ref{r:backflow} regarding  the ``backflow'' effect (i.e. the existence of tracer KdV solitons moving in negative direction) in sufficiently dense soliton gases. A numerical simulation of the diluted genus $0$ condensate with $C=0.9$ where one can clearly see the soliton trajectory with a negative slope is presented in Fig.~\ref{fig:backflow}.

\subsection{Riemann problem}
\label{sec:riemann_diluted}

We can now consider the soliton condensate Riemann problem for
diluted condensates for which the initial DOS~\eqref{eq:ustep} is replaced by
\begin{equation}
  \label{eq:ustep2}
  u(\eta, x,t=0) =
  \begin{cases}
    C_- \,u^{(N_-)}(\eta;\lambda_1^-, \dots, \lambda_{2N_-+1}^-), & x<0, \\
    C_+ \,u^{(N_+)}(\eta;\lambda_1^+, \dots, \lambda_{2N_++1}^+), & x>0,
  \end{cases}
\end{equation}
where $0<C_\pm<1$.

\begin{figure}[h]
  \centering
  \includegraphics[width=7cm]{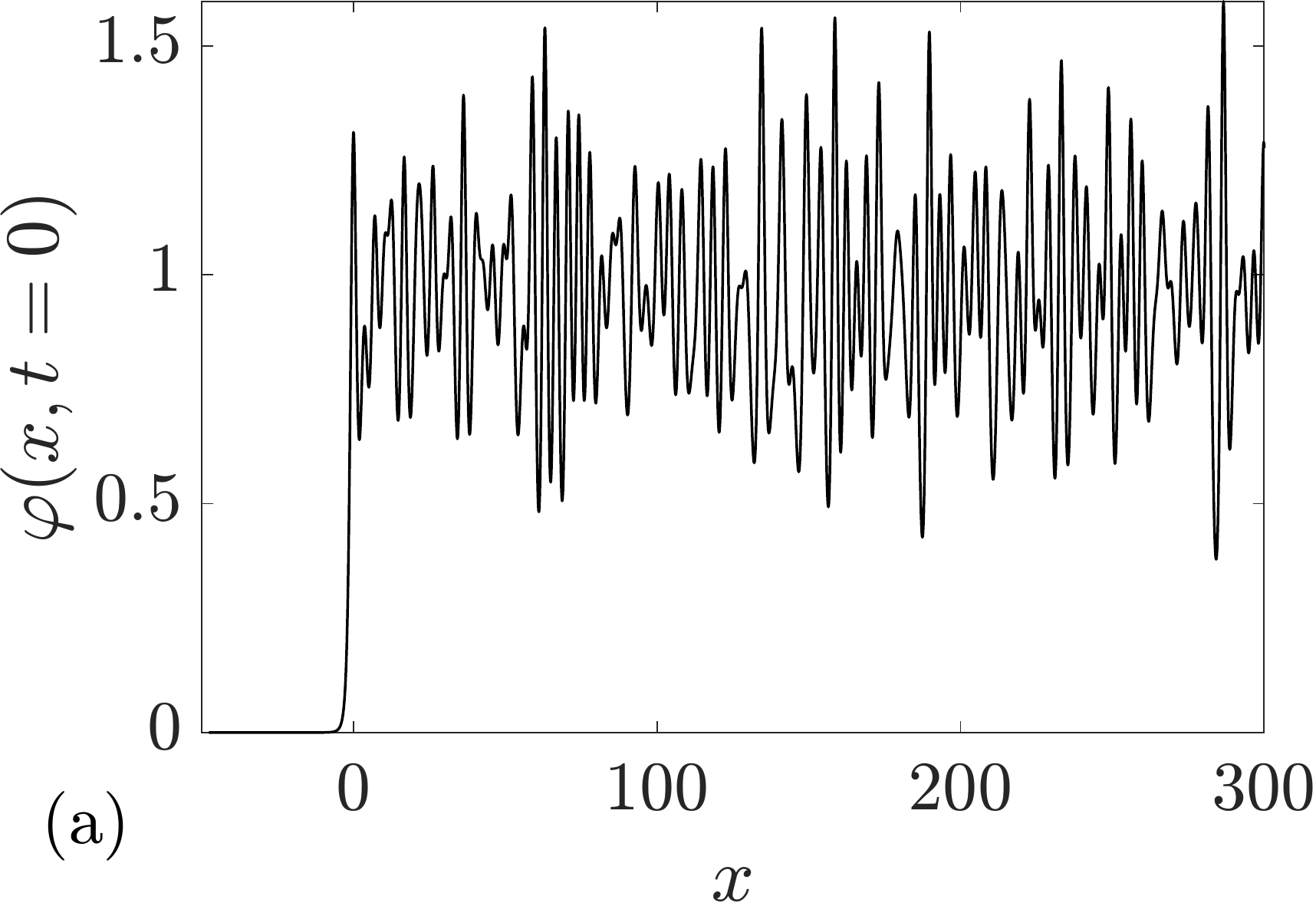}\hspace{1cm}
  \includegraphics[width=7cm]{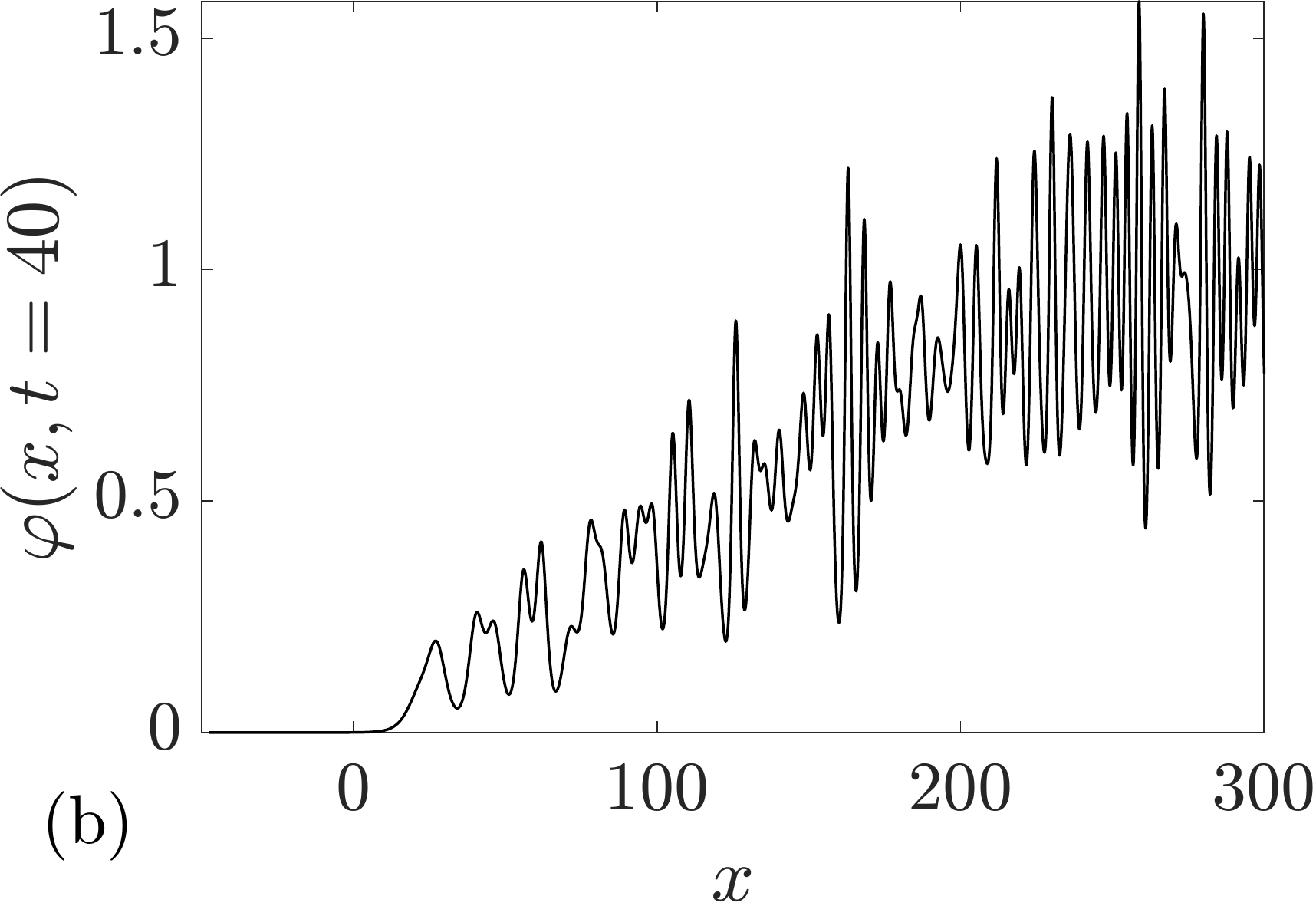}\\[1cm]
  \includegraphics[width=7cm]{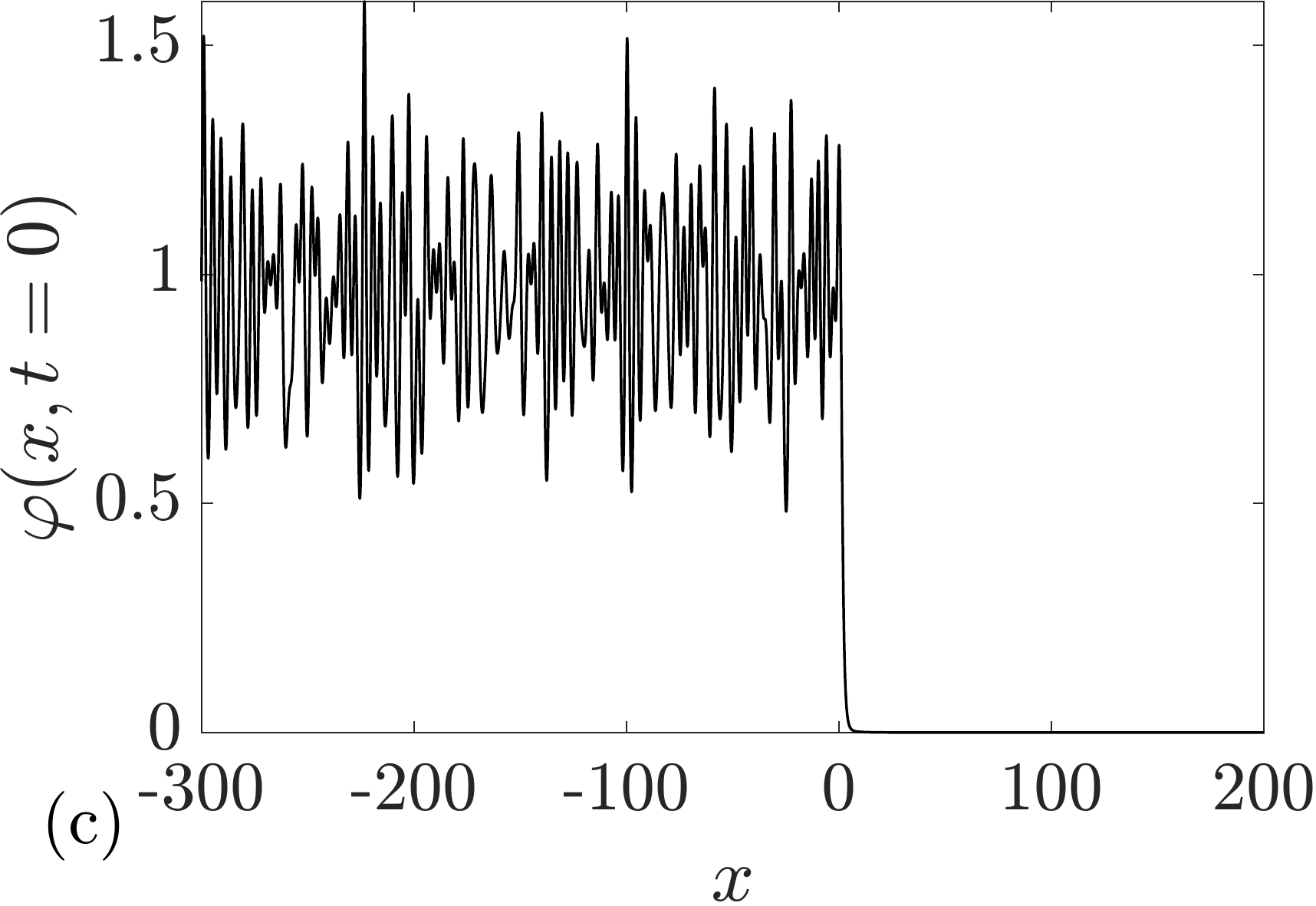}\hspace{1cm}
  \includegraphics[width=7cm]{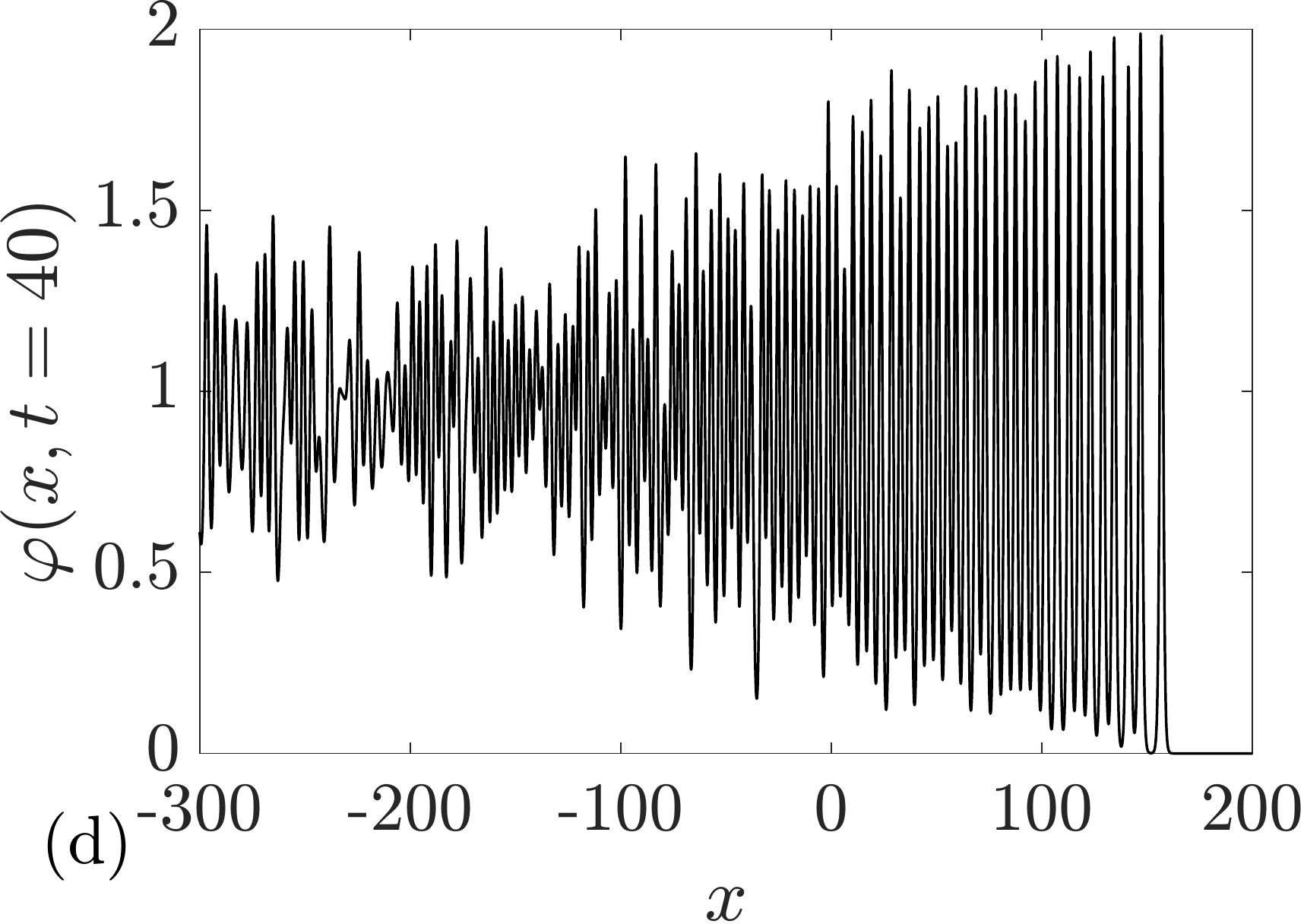}\\[1cm]
  \includegraphics[width=7cm]{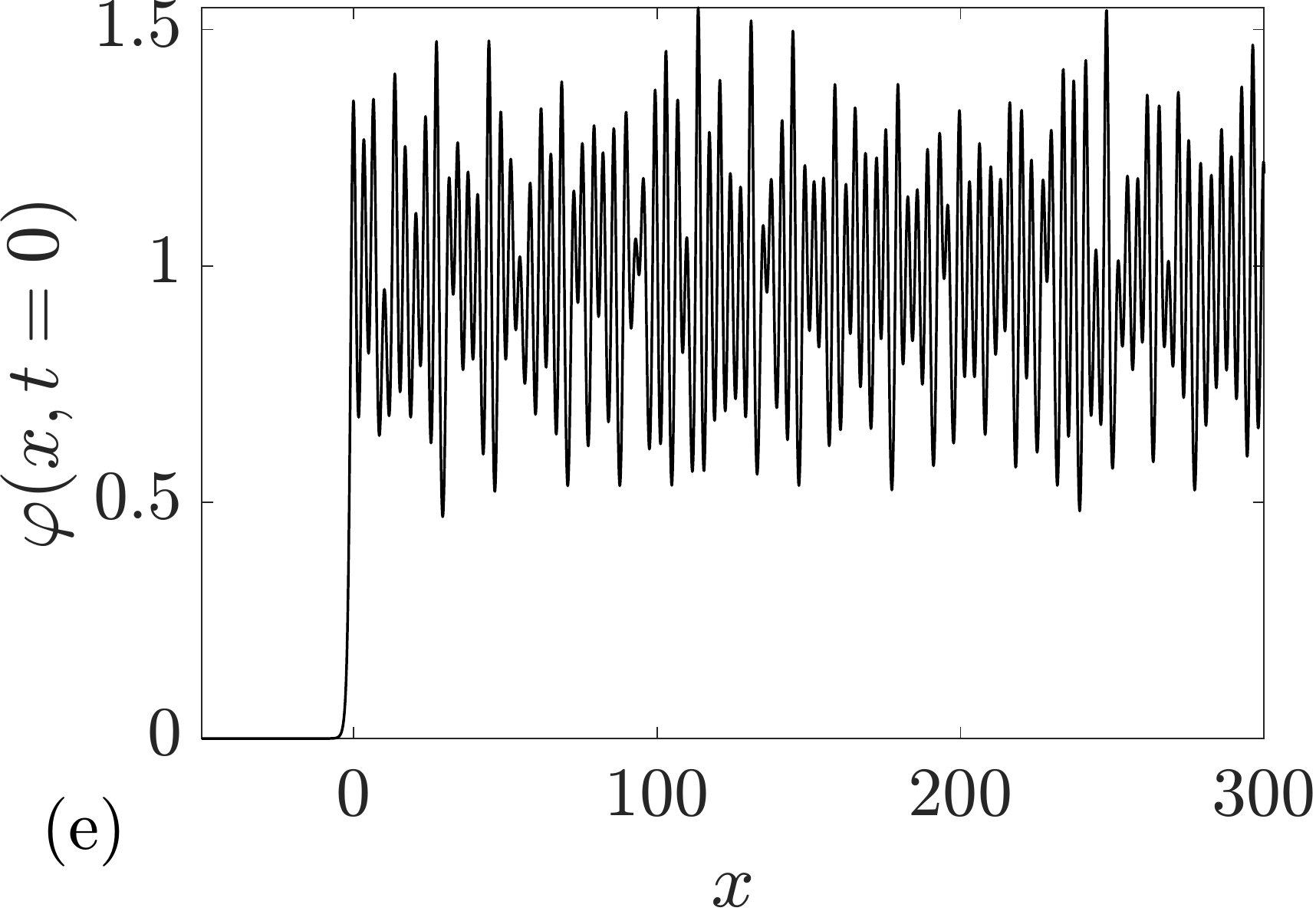}\hspace{1cm}
  \includegraphics[width=7cm]{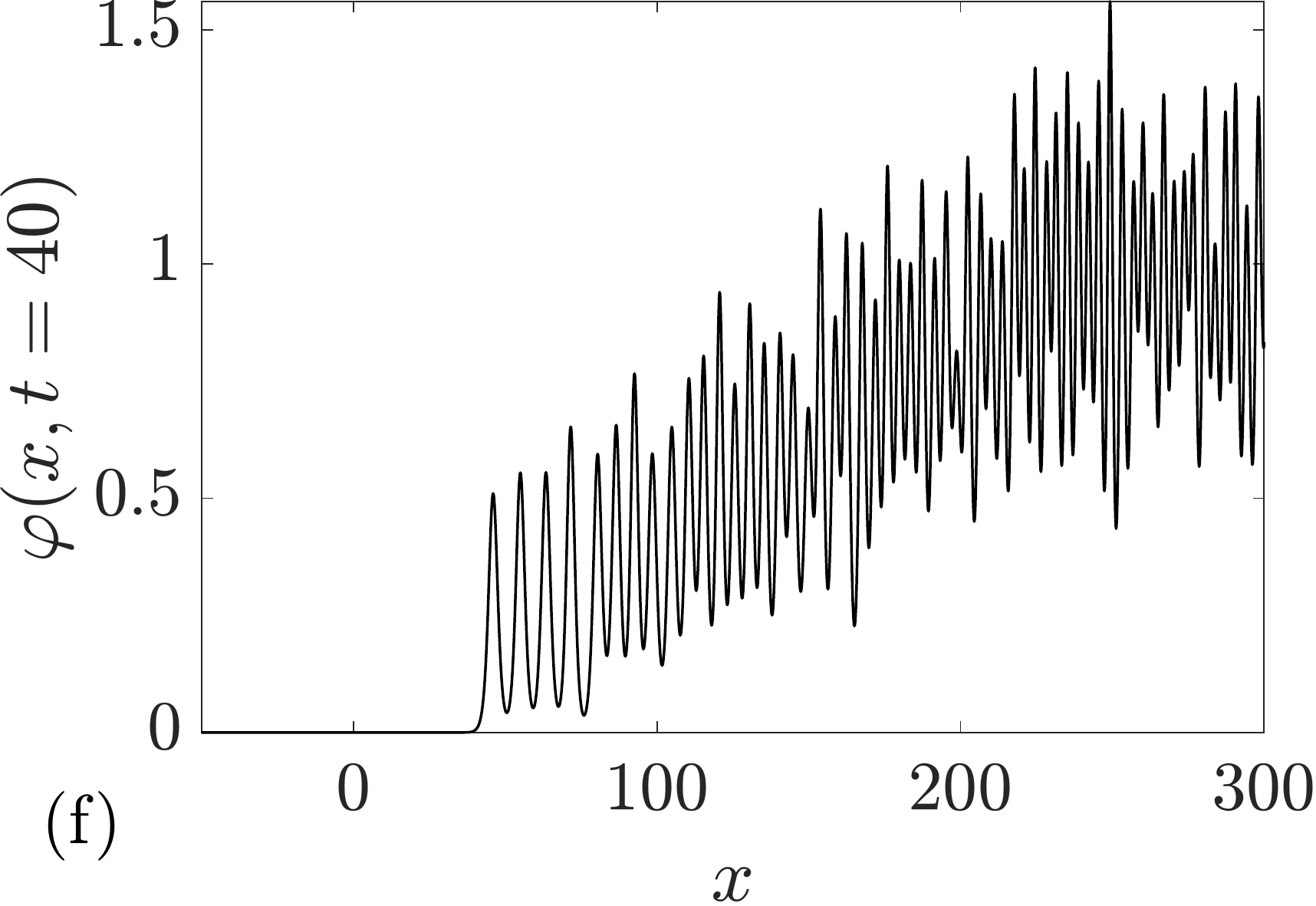}\\[1cm]
  \includegraphics[width=7cm]{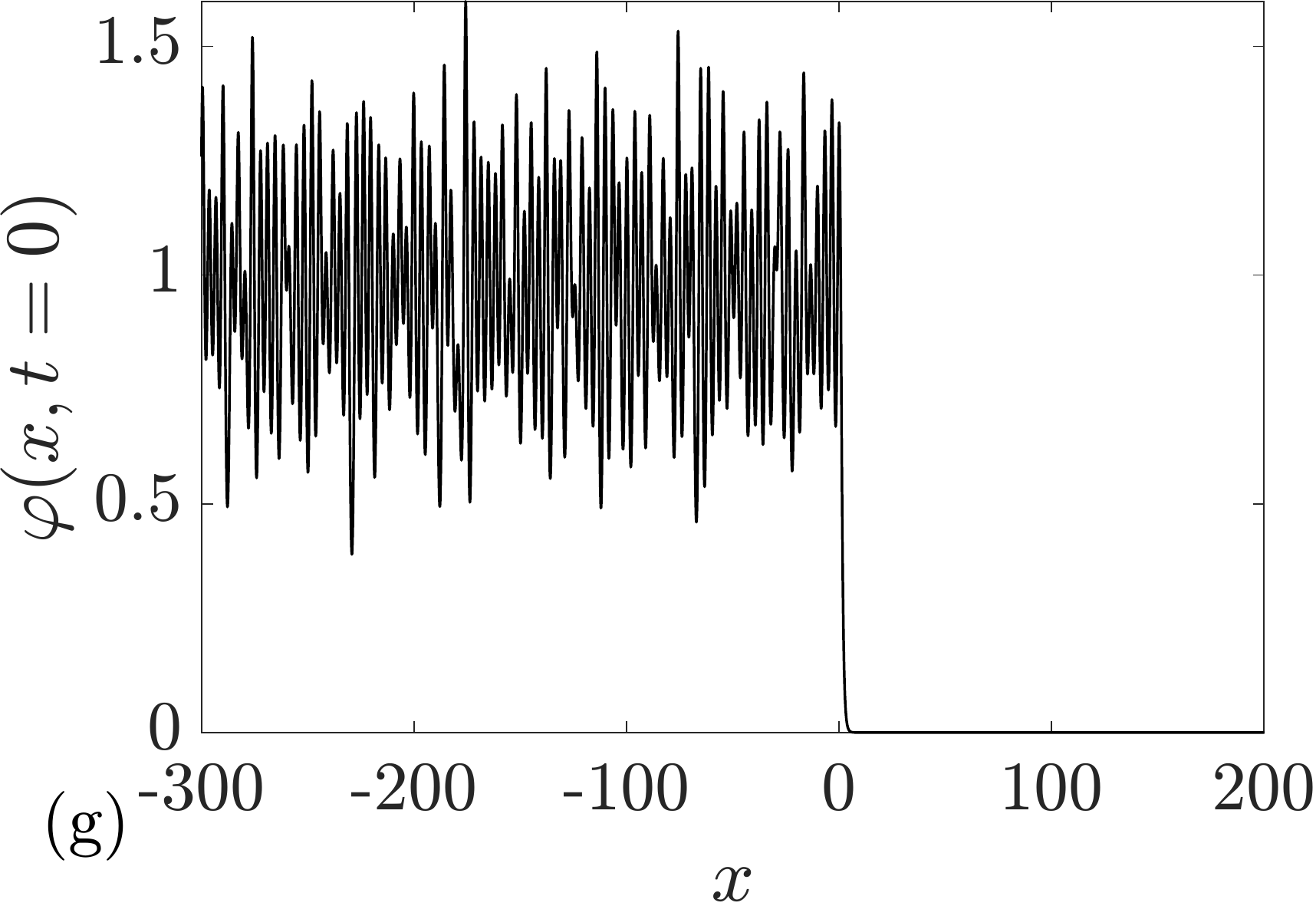}\hspace{1cm}
  \includegraphics[width=7cm]{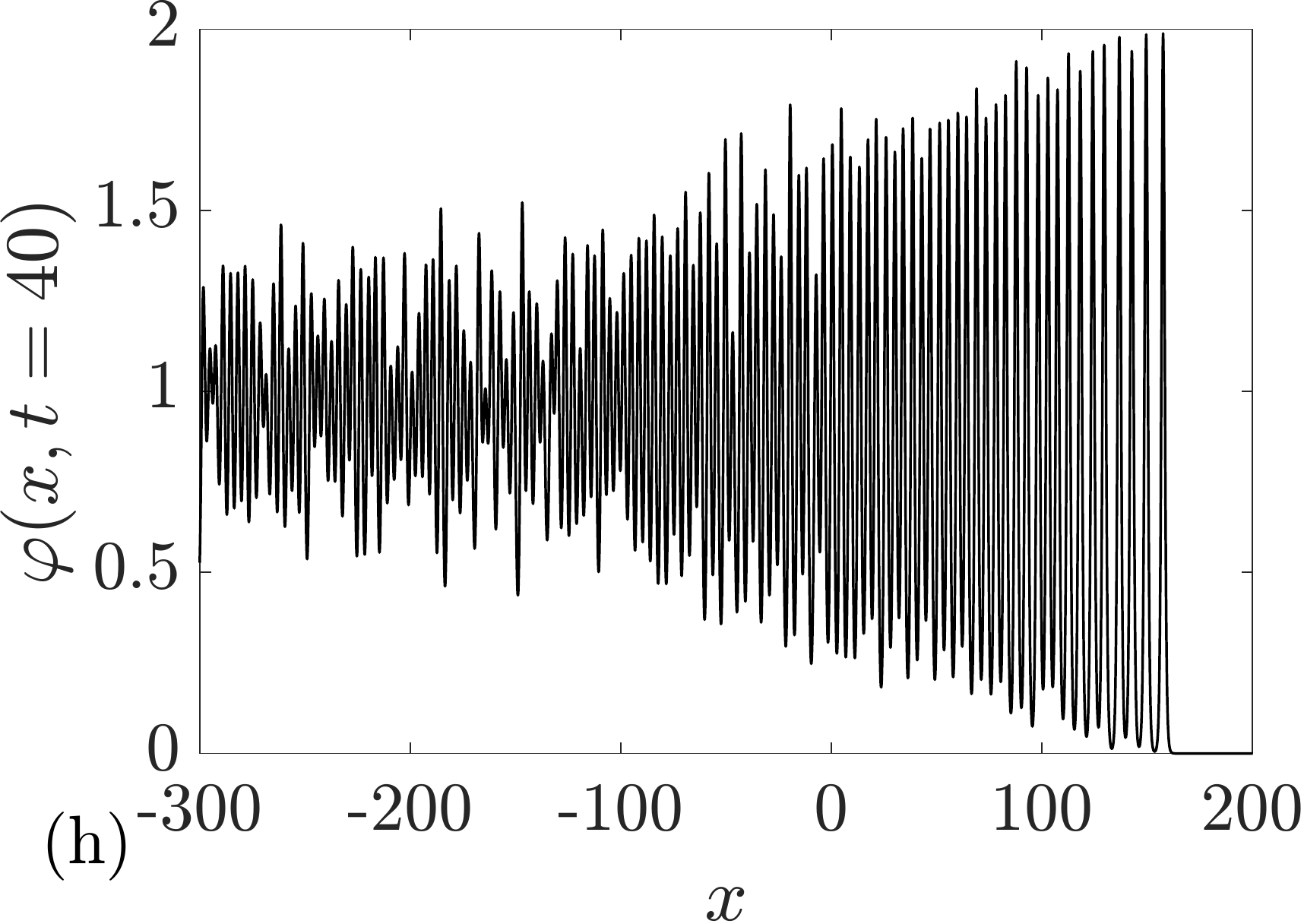}
  \caption{Riemann problem for diluted soliton condensates with initial condition~\eqref{eq:ustep2}
    with $C_\pm =0.95$. (a)-(d) $N_-+N_+=0$ and $\lambda_1=1$; (e)-(h)
    $N_-+N_+=1$ and $(\lambda_1,\lambda_2,\lambda_3)=(0,1/2,1)$. The diluted condensates are realized with exact $n$-soliton solutions ($n=200$) configured spectrally according to the respective scaled condensate DOS's. The evolution results in the generation of incoherent rarefaction and dispersive shock waves.}
  \label{fig:riem_dilute}
\end{figure}

To be specific, we investigate numerically the evolution of the diluted condensate
initial conditions~\eqref{eq:ustep2} with $N_-+N_+\leq 1$ and $\lambda_i$ chosen from
the examples presented in Sec.~\ref{sec:numeric}.
Numerical realizations of the
step-initial condition and their evolution in time are presented in
Fig.~\ref{fig:riem_dilute}.  One can see that generally, realizations of the diluted
soliton condensate do not exhibit a macroscopically coherent structure as observed in Sec.~\ref{sec:numeric}.
However,  in the case $N_-+N_+=1$, the evolution of the diluted condensate
realizations, despite the visible incoherence, still qualitatively resembles the evolution of the ``genuine'' condensates depicted in
Figs.~\ref{fig:kdv_3+},~\ref{fig:kdv_2-}. One can see that the recognizable patterns
of  the generalized rarefaction wave (see Fig.~\ref{fig:riem_dilute}f) and the
generalized DSW (see Fig.~\ref{fig:riem_dilute}h)
persist even if $C < 1$. Indeed, as shown in Sec.~\ref{sec:dilute_eq},
the oscillations in a realization of the diluted genus 1 condensate
appear almost coherent for a small dilution factor. The persistence of
coherence can also be observed in the case $N_-+N_+=0$ when
$\lambda_1^- > \lambda_1^+$ (Fig.~\ref{fig:riem_dilute}d): a
DSW develops if $C=1$, and coherent, finite amplitude
oscillations still develop for $C\neq 1$ at the right edge of the
structure where the amplitudes of oscillations are large.
In connection with the above, it is important to note that, although the initial condition~\eqref{eq:ustep2} is given by the discontinuous diluted condensate DOS,  $u(\eta; x,0)=Cu^{(N)}(\eta)$, the kinetic equation evolution  does not imply that the  DOS will remain to be of the same form for $t>0$. In other words, unlike genuine condensates, the diluted condensates do not retain the spectral ``diluted condensate'' property during the evolution.

\clearpage

\section{Conclusions and Outlook}

We have considered a special kind of soliton gases for the KdV equation, termed soliton condensates, which are defined by the property of vanishing the spectral scaling function $\sigma(\eta)$ in the soliton gas nonlinear dispersion relations \eqref{ndr_kdv1}, \eqref{ndr_kdv2}.   As a result, the density of states $u(\eta)$  in a soliton condensate is uniquely determined by its spectral support $\Gamma^+ \in \mathbb{R}^+$. By considering $\Gamma^+$ to be a union of $N+1$  disjoint intervals, $[0, \la_1]\cup [\la_2, \la_3] \cup \dots \cup [\la_{2N}, \la_{2N+1}]$,  and allowing the endpoints $\{\la_j\}_{j=1}^{2N+1}$ vary slowly in space-time we prove that the kinetic equation for soliton gas reduces in the condensate limit to  the genus $N$ KdV-Whitham modulation for  $\la_j(x,t)$.  The KdV-Whitham equations were originally derived via the wave averaging procedure in \cite{whitham_non-linear_1965}, \cite{flaschka_multiphase_1980}  and via the semicalssical limit of the KdV equation in \cite{lax_small_1983}. These equations have been extensively used for the description of dispersive shock waves \cite{el_dispersive_2016}, particularly in the context of dispersive Riemann problem originally introduced by Gurevich and Pitaevskii \cite{gurevich_nonstationary_1974}. 

Along with the characterization of the large scale, modulation dynamics of soliton condensates, our work suggests that they represent ``coherent'' or ``deterministic'' soliton gases whose typical realizations are  given by finite-gap potentials. We prove this conjecture for genus zero condensates and present a strong numerical evidence  for $N=1,2$. 

By invoking the results from the modulation theory of dispersive shock waves we have constructed analytical solutions to several Riemann problems for the soliton gas kinetic equation
subject to discontinuous condensate initial data. These solutions
describe the evolution of generalized rarefaction and dispersive shock 
waves in soliton condensates.  We performed numerical simulations of the Riemann problem for the KdV soliton condensates  by constructing an exact $n$-soliton solutions with $n$ large and the spectral parameters distributed according to the condensate densty of states.   A comparison of the numerical simulations with analytical predictions from the solutions of the kinetic equation showed excellent agreement.

 Finally, we considered the basic properties of
 ``diluted'' soliton condensates having a scaled condensate density of states and exhibiting rich
incoherent behaviors. 

There are several avenues for  future work suggested by our results. One pertinent problem would be to consider near-condensate soliton gas dynamics by assuming the spectral scaling function $\sigma$ to be ``small'' ($\sigma(\eta) = \epsilon \tilde \sigma (\eta)$, $\epsilon \ll 1$, $\tilde \sigma = \mathcal{O}(1)$). Another area of major interest is the extension of the developed KdV soliton condensate theory to other integrable equations, particularly the focusing NLS equation, where a number of important theoretical and experimental results on the soliton gas dynamics have been obtained recently, see \cite{gelash_prl2019, el_spectral_2020, roberti_pre2021,gelash2021}. And last but not least, one of the most intriguing open questions is the possibility of phase transitions in soliton gases, i.e. the formation of a soliton condensate from non-condensate initial data. 
The generalized hydrodynamics approach to the thermodynamics of soliton gases \cite{bonnemain_generalised_2022} provides a promising framework to explore this  possibility.  At the same time,   this direction of research could require some departure from integrability and the development of the soliton gas theory for perturbed integrable equations.

\section*{Acknowledgments}

The authors would like to thank the Isaac Newton Institute for Mathematical Sciences for support and hospitality during the programme ``Dispersive hydrodynamics: mathematics, simulation and experiments, with applications in nonlinear waves''  when the work on this paper was undertaken. This work was supported by EPSRC  Grant Number EP/R014604/1.
GE's and GR's work was also supported by EPSRC  Grant Number EP/W032759/1 and  AT's work was supported by NSF Grant DMS 2009647.  TC and GR thank Simons Foundation for partial support. All authors thank T.~Bonnemain, S.~Randoux and P.~Suret for numerous useful discussions.
\appendix


\section{Numerical implementation of soliton gas}
\label{sec:numerical}

\subsection{Riemann problem}
\label{sec:Riemann_numerical}

The realizations of the soliton gas are approximated numerically by the $n$-soliton
solution
\begin{equation}
  \label{eq:soliton-sol}
  \varphi \equiv
  \varphi_n(x,t;\eta_1,\dots,\eta_n,x^0_1,\dots,x^0_n),\quad n \in \mathbb{N},
\end{equation}
where the $\eta_i$'s and $x^0_i$'s correspond respectively to the
spectral parameters and the ``spatial phases'' of the solitons;
$\eta_i<\eta_{i+1}$ by convention. The numerical implementation
of~\eqref{eq:soliton-sol} is described in Sec.~\ref{sec:app_algo}
below. The numerical solutions presented in this work are all
generated with $n=200$ solitons, unless otherwise stated.

Since $n$ is finite, the $n$-soliton solution reduces to a sum of
separated solitons in the limit $|t| \to \infty$. By construction, we
have in the limit $t\to \pm\infty$
\begin{equation}
  \varphi_n(x,t) \sim
  \sum_{i=1}^n 2\eta_i^2 {\rm sech}^2 \left[\eta_i(x - 4 \eta_i^2 t - x_i^\pm) \right],
\end{equation}
where $x_i^\pm$ are the spatial phases of the $i$-th soliton at $t \to \pm \infty$. We then take 
the spatial phase in \eqref{eq:soliton-sol} to be  $x_i^0 = (x_i^-+x_i^+)/2$.

Consider a uniform soliton gas with
the density of states $u(\eta)$. Let the
 spectral parameters $\eta_i$ be distributed on $\Gamma^+$ with density
\begin{equation}
  \label{eq:rho}
  \phi(\eta) = \frac{u(\eta)}{\kappa},\quad \kappa = \int_{\Gamma_+} u(\eta) d \eta,
\end{equation}
where the normalization by the spatial density of solitons $\kappa$ ensures that $\phi(\eta)$ is normalized
to $1$.
It was shown in \cite{mazur}  that the spatial density $\kappa$ is obtained if the phases $x^0_i$  are uniformly
distributed on the interval (denoted ``$S$-set'' in~\cite{mazur}):
\begin{equation}
  \label{eq:kappas}
  I_s = \left[-\frac{n}{2\kappa_s},+\frac{n}{2\kappa_s} \right],\quad \kappa_s = \int_{\Gamma_+ } \frac{\eta}{\sigma(\eta)} d \eta,
\end{equation}
where $\sigma(\eta)$ is the spectral scaling function in the NDRs ~\eqref{ndr_kdv1}, \eqref{ndr_kdv2}; $y(\eta)$ in \cite{mazur} is given here by $y(\eta) = u(\eta) \sigma(\eta)/\eta$. The derivation of~\eqref{eq:kappas} has been revisited recently in the context of generalized hydrodynamics \cite{bonnemain_generalised_2022}: it was shown that $\kappa_s$ corresponds to the density of spatial phases $x_i^0$, or equivalently  $x_i^\pm$ which are well defined asymptotically ($t \to \pm \infty$) where the solitons are ``non-interacting'' and their position are given by $x_i(t) \sim 4\eta_i^2 t + x_i^\pm$.  
In the rarefied gas limit the interaction term in the NDR \eqref{ndr_kdv1} is small and therefore $\sigma(\eta)u(\eta) \approx \eta$ so that we obtain
$\kappa_s = \kappa$ as expected. In the general case though the density $\kappa_s$ of
non-interacting phases  is different from the ``physical'' density $\kappa$, as demonstrated with the soliton condensate examples
below. In the thermodynamic limit $n \to \infty$, the soliton
solution~\eqref{eq:soliton-sol} represents a realization of the uniform soliton gas.

Since the number $n$ of solitons  is finite, the $n$-soliton solution
has a finite spatial extent. By distributing the phases $x^0_i$ uniformly
on the interval $I_s$, the
$n$-soliton solution $\varphi_n(x,t=0)$ approximates a realization of
the uniform soliton gas for $x \in [-\ell/2,\ell/2]$ where $\ell=n/\kappa$;
$\varphi_n(x,t=0) \sim 0$ outside of this interval. This naturally
generates the box-like initial condition for the kinetic equation
\begin{equation}
  u(\eta;x,t=0) \sim
  \begin{cases}
    0, & x<-\ell/2,\\
    u(\eta), & -\ell/2<x<\ell/2,\\
    0, & \ell/2<x.
  \end{cases}
\end{equation}
Note that $u(\eta;x,t=0) = 0$ can be seen as the genus $0$ condensate
where the end point of the central s-band $\lambda_1 \to 0$. This limits the type of initial
condition that can be implemented for the Riemann problem and we
choose in practice $(N_-=0,q_-=0)$ or $(N_+=0,q_+=0)$. For
convenience, we shift the $x$-axis by $\pm \ell/2$ to obtain one of the
discontinuities at the position $x=0$.

The evolution in time of the soliton gas realization is obtained by
varying the parameter~$t$. Contrarily to a direct resolution of the
KdV equation, via finite difference or spectral method, the
time-evolution presented here is instantaneous and does not accumulate
any numerical errors since the $n$-soliton solution is an exact
solution. For the Riemann problem, the maximal time is bounded by the
finite extent of the $n$-soliton solution: after a sufficently long time, the two
hydrodynamic states originating from the discontinuities  at $x=-\ell/2$ and $x=\ell/2$ start
interacting. Longer times can be reached by choosing a larger number
of solitons $n$.

\

\begin{figure}[h]
  \centering
  \includegraphics[width=7cm]{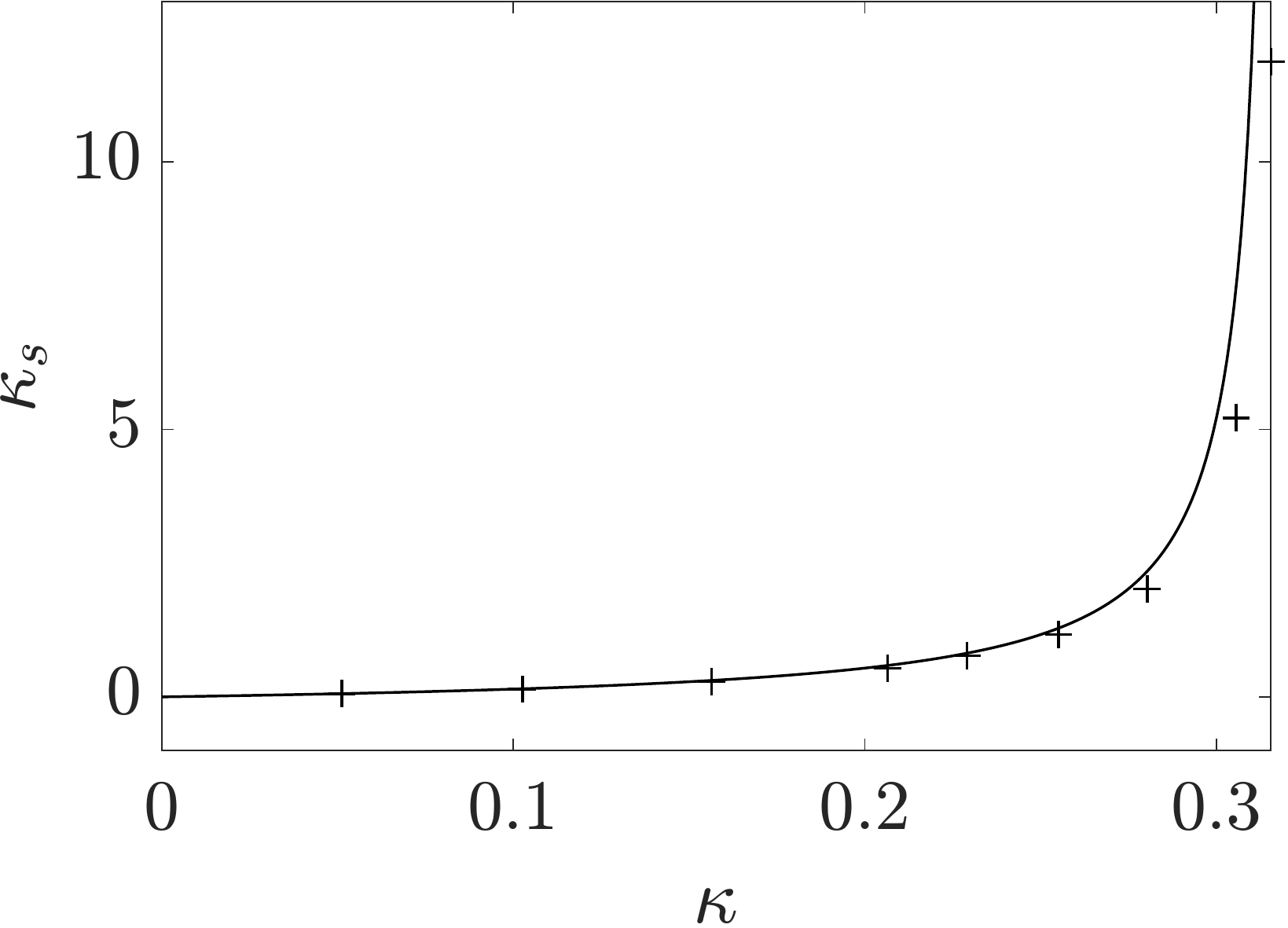}
  \caption{The solid line represents the variation of $\kappa_s$ with
    respect to $\kappa$ for a diluted genus $0$ condensate, cf.~\eqref{eq:ks}. The markers
    are obtained using the $100$-soliton solution: $\kappa=\ell/n$ where
    $\ell$ corresponds to the spatial extension of the $n$-soliton solution.}
  \label{fig:variance_comp}
\end{figure}

We consider now the density of states of interest for this work:
\begin{equation}
  u(\eta) = C u^{(N)}(\eta; \la_1, \dots \la_{2N+1}),\quad C\leq 1,
\end{equation}
where $u^{(N)}$ is density of states of the condensate defined in Sec.~\ref{sec:gen_sol_cond}. \eqref{eq:kappas} rewrites
\begin{equation}
  \label{eq:ks}
  \kappa_s = \frac{\kappa(C)}{1-C},\quad \kappa(C) = C \int_{\Gamma_+} u^{(N)}(\eta; \la_1, \dots, \la_{2N+1}) d \eta.
\end{equation}
Fig.~\ref{fig:variance_comp} shows the comparison between the spatial
density of solitons $\kappa$ and the density of phases $\kappa_s$ for
the genus $0$ case where $\kappa(C) = C/\pi$. The phases density $\kappa_s$ diverges
in the condensate limit $C\to 1$, and $x_i^0$'s are all
equal to the same phase $x^0$ ($I_s \to \{x^0\}$). This limit is in agreement with the results obtained in Sec.~\ref{sec:equilibrium} for
genus $0$ and genus $1$ condensates: each realisation of the condensate ($C=1$) is approximated with a {\it coherent} $n$-soliton solution where $x_i^0=x^0={\rm cst},\; \forall i$. 

Examples of numerical realizations of soliton condensates and diluted soliton condensates
are given in Sections~\ref{sec:equilibrium}, \ref{sec:numeric}, \ref{sec:dilute} and \ref{sec:gen_two_cond}. Figs.~\ref{fig:cond0}, \ref{fig:cnoidal3} and~\ref{fig:genus23} shows that numerical
approximations of condensate via the $n$-soliton solution are not
exactly uniform; realizations become more uniform near the center
of the interval $[-\ell,\ell]$ as the number of soliton $n$ increases.

\begin{figure}[h]
  \centering
  \includegraphics[width=7cm]{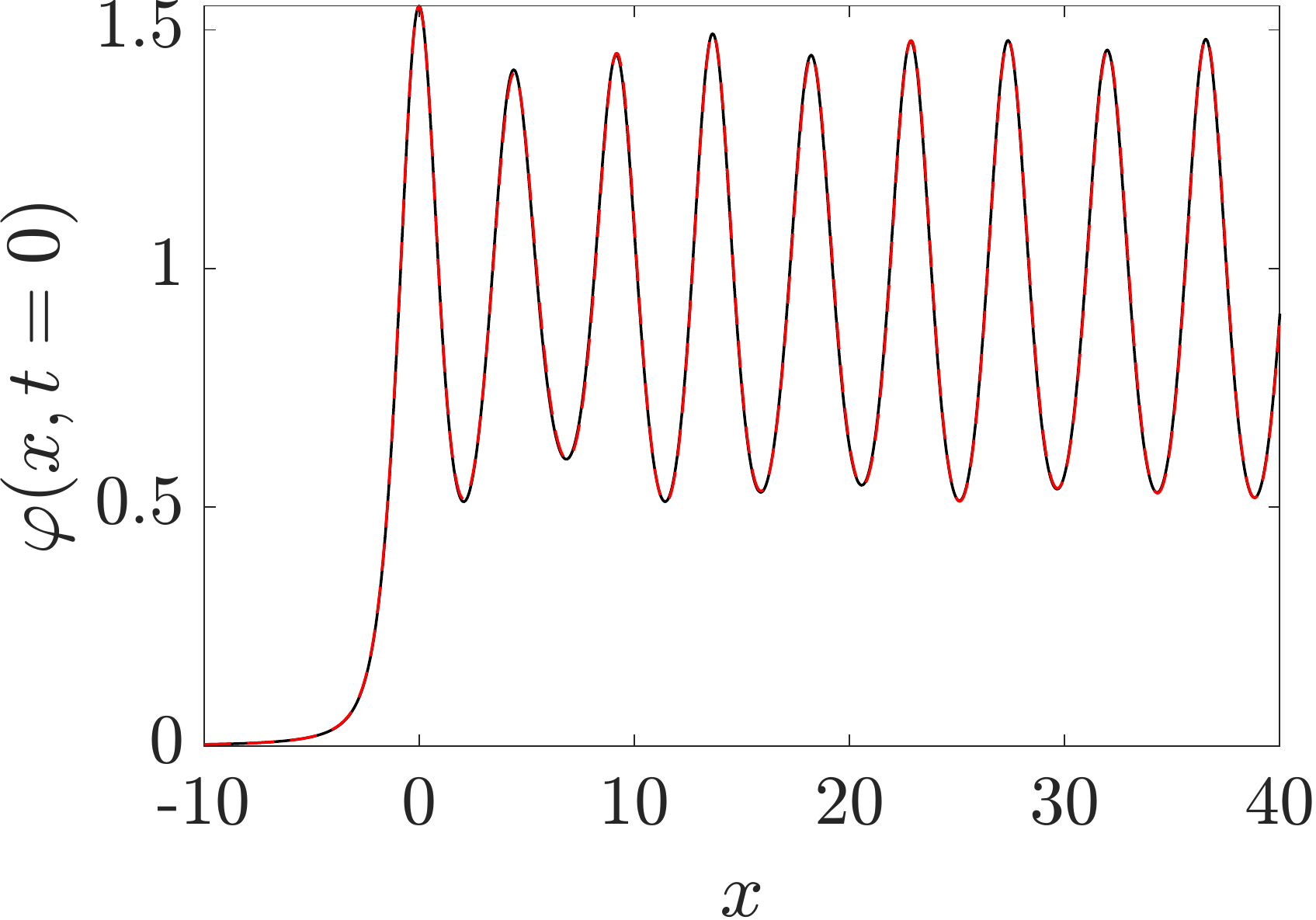}
  \caption{$n$-soliton solution approximating a realization of the
    condensate $(N; \lambda_1,\lambda_2,\lambda_3) = (1; 0,0.5,0.85)$.
    The solid black line represent the solution $n=100$ and the red
    dashed line the solution $n=200$. Both solutions have been shifted
    such that the maximum of the solution is located at $x=0$.}
    \label{fig:overshoot_compare}
\end{figure}

The realization at $t=0$ in Figs.~\ref{fig:riem0_RW}, \ref{fig:riem0_DSW}, \ref{fig:kdv_3+}
and~\ref{fig:kdv_2-} also displays the ``border effects'' observed at the
discontinuities of the Riemann problem initial condition (located at
$x=0$). 
These border effects,
manifesting as overshoots of the realization, persist regardless of
the number of solitons $n$ as shown by the comparison between the
$100$-soliton and $200$-soliton solutions in Fig.~\ref{fig:overshoot_compare}. However, because of their finite size, the
observed border effects seem to have no effect on the
asymptotic dynamics of the condensate as demonstrated by the very good
agreement between the theory and the numerical solution in Sec.~\ref{sec:numeric}.

\subsection{Generation of spectral parameters $\eta_i$}

The spectral parameters of the $n$-soliton solution are distributed
with probability density $\phi(\eta)$, cf.~\eqref{eq:rho}. This can be achieved by
choosing the solutions of the nonlinear equation
\begin{equation}
  \int_0^{\eta_i} \phi(\mu) d\mu =
  \frac{i}{n},\quad i=1 \dots n.
\end{equation}

For genus 0 condensate whose DOS is given by~\eqref{uv}, this equation reduces to
\begin{equation}
  1-\sqrt{1-\frac{\eta_i^2}{\lambda_1^2}}= \frac{i}{n}.
\end{equation}

For genus 1 condensate with DOS~\eqref{eq:N1}, this equation reads
\begin{equation}
  \frac{U(\eta_i)}{U(\lambda_3)} = \frac{i}{n},\quad
  \text{where:}\quad
  U(\eta) = \int_0^\eta
  u^{(1)}(\mu;\lambda_1,\lambda_2,\lambda_3) d\mu.
\end{equation}
We have
\begin{equation}
  U(\eta) = \int_0^\eta \frac{i(\mu^2-p^2)}{2\pi\sqrt{R(\mu)}}
  d(\mu^2) = \int_0^{\eta^2} \frac{i(x-p^2)}{2\pi\sqrt{(x-\lambda_1^2)(x-\lambda_2^2)(x-\lambda_3^2)}}
  dx,
\end{equation}
which yields
\begin{equation}
  U(\eta) = U_0 +
  \begin{cases}
    \displaystyle
    - \frac{1}{\pi} \left[\sqrt{\frac{(\lambda_3^2-\eta^2)(\lambda_1^2-\eta^2)}{\lambda_2^2-\eta^2}} -  \frac{\lambda_1^2-p^2}{\sqrt{\lambda_3^2-\lambda_1^2}}F(\beta,q)-
      \sqrt{\lambda_3^2-\lambda_1^2}E(\beta,q) \right] ,
           & 0<\eta<\lambda_1,          \\
    0,                                                                                     & \lambda_1<\eta<\lambda_2,  \\
    \displaystyle
    \frac{1}{\pi} \left[ \frac{\lambda_1^2-p^2}{\sqrt{\lambda_3^2-\lambda_1^2}}F(\kappa,q) +
    \frac{\lambda_2^2-\lambda_1^2}{\sqrt{\lambda_3^2-\lambda_1^2}}\Pi(q,\kappa,q) \right], & \lambda_2< \eta<\lambda_3.
  \end{cases}
\end{equation}
where:
\begin{align}
   & \beta = \sin^{-1} \left(\sqrt \frac{\lambda_1^2-\eta^2}{\lambda_2^2-\eta^2} \right),\quad \kappa =\sin^{-1} \left(\sqrt \frac{(\lambda_3^2-\lambda_1^2)(\eta^2-\lambda_2^2)}{(\lambda_3^2-\lambda_2^2)(\eta^2-\lambda_1^2)} \right), \\
   & q = \frac{\lambda_3^2-\lambda_2^2}{\lambda_3^2-\lambda_1^2},                                                                                                                                                                         \\
   & U_0 = \frac{1}{\pi} \left[\sqrt{\frac{\lambda_3^2\lambda_1^2}{\lambda_2^2}} - \frac{\lambda_1^2-p^2}{
      \sqrt{\lambda_3^2-\lambda_1^2}}F(\beta_0,q)-\sqrt{\lambda_3^2-\lambda_1^2}E(\beta_0,q) \right],\quad\beta_0
  = \sin^{-1} \left(\frac{\lambda_1}{\lambda_2} \right).
\end{align}

\subsection{Algorithm for the $n$-soliton solution }
\label{sec:app_algo}

The algorithm generating the exact $n$-soliton~\eqref{eq:soliton-sol}, originally developed in \cite{huang1992darboux},
relies on the Darboux transformation.  This scheme is subject to
roundoff errors during summation of exponentially small and large
values for a large number of solitons $n$. We improve it following
\cite{gelash2018strongly}, with the implementation of  high
precision arithmetic routine to overcome the numerical accuracy problems and
generate solutions with a number of solitons $n \gtrsim 10$.

In order to simplify the algorithm, it is suggested to consider
simultaneously the KdV equation~\eqref{kdv} and its equivalent form
\begin{equation}
  \varphi-6\varphi\varphi_x+\varphi_{xxx}=0
  \label{eq:kvd_2}
\end{equation}
obtained from ~\eqref{kdv} by the reflection $\varphi \to - \varphi$.
The Darboux transformation presented here relates the Jost solution
associated with the $(n-1)$-soliton solution of one equation, to the
$n$-soliton solution of the other equation. 

Considering the direct scattering problem for the Lax pair in the matrix  form
\begin{equation}
  \Phi_{x}=\left(\begin{array}{cc}
      \eta    & \mp 1  \\
      \varphi & - \eta
    \end{array}\right) \Phi,
\end{equation}
with $-1$ corresponding to~\eqref{kdv} and $+1$ to~\eqref{eq:kvd_2}, the Jost solutions $J,\tilde{J}\in \mathbb{R}^{2 \times 2}$ are defined recursively by the Darboux transformations $D(\eta)$ and $\tilde{D}(\eta)$ such that:
\begin{equation}
  J_{n}(\eta)=D_{n}(\eta) J_{n-1}(\eta),\quad \textrm{with:} \quad D_{n}(\eta)=I+\frac{2\eta_{n}}{\eta-\eta_{n}} P_{n},
\end{equation}
\begin{equation}
  \tilde{J}_{n}(\eta)=\tilde{D}_{n}(\eta) \tilde{J}_{n-1}(\eta), \quad \textrm{with:} \quad \tilde{D}_{n}(\eta)=I-\frac{2\eta_{n}}{\eta+\eta_{n}} \tilde{P}_{n}.
\end{equation}
$P_n(x,t)$ and $\tilde{P}_n(x,t)$ are independent of $\eta$ and
have the form:
\begin{equation}
  P_{n}=\sigma_2\tilde{P}^{ \textrm{T}}_{n}\sigma_2=\frac{J_{n-1}\left(-\eta_{n}\right)\left(\begin{array}{c}
        -b_n \\
        1
      \end{array}\right)\left(\begin{array}{ll}
        b_n & 1
      \end{array}\right) \tilde{J}_{n-1}^{-1}\left(\eta_{n}\right)}{\left(\begin{array}{ll}
        b_n & 1
      \end{array}\right) \tilde{J}_{n-1}^{-1}\left(\eta_{n}\right) J_{n-1}\left(-\eta_{n}\right)\left(\begin{array}{c}
        -b_n \\
        1
      \end{array}\right)}
\end{equation}
with the real constants $b_n$ depending on the spatial phases
\begin{equation}
  b_n=\left(-1\right)^{n}\exp\left(2\eta_n x^{0}_{n}\right).
\end{equation}
The Jost solutions for the initial seed solution $\varphi_0=0$ are
given by
\begin{equation}
  J_{0}(\eta)=\tilde{J}_{0}(\eta)=\left(\begin{array}{cc}
      \exp \left[\eta x-4 \eta^{3} t\right] & -\exp \left[-\eta x+4 \eta^{3} t\right]         \\
      0                                     & - 2 \eta \exp \left[-\eta x+4 \eta^{3} t\right]
    \end{array}\right),
\end{equation}
and one can show that at each recursion step
\begin{equation}
  \varphi_{n}=\varphi_{n-1}+4\eta_{n}\left(P_{n}\right)_{21},
\end{equation}
where $\varphi_n$ is the $n$-soliton solution of~\eqref{kdv} for $n$
even and solution of~\eqref{eq:kvd_2} for $n$ odd. Recently,  a more efficient and accurate  algorithm has been proposed in \cite{Prins_2021} to generate the $n$-soliton KdV solution employing a $2$-fold Crum transform.

\section{Genus $2$ condensate}
\label{sec:gen_two_cond}

\begin{figure}[h]
  \centering
  \includegraphics[width=7cm]{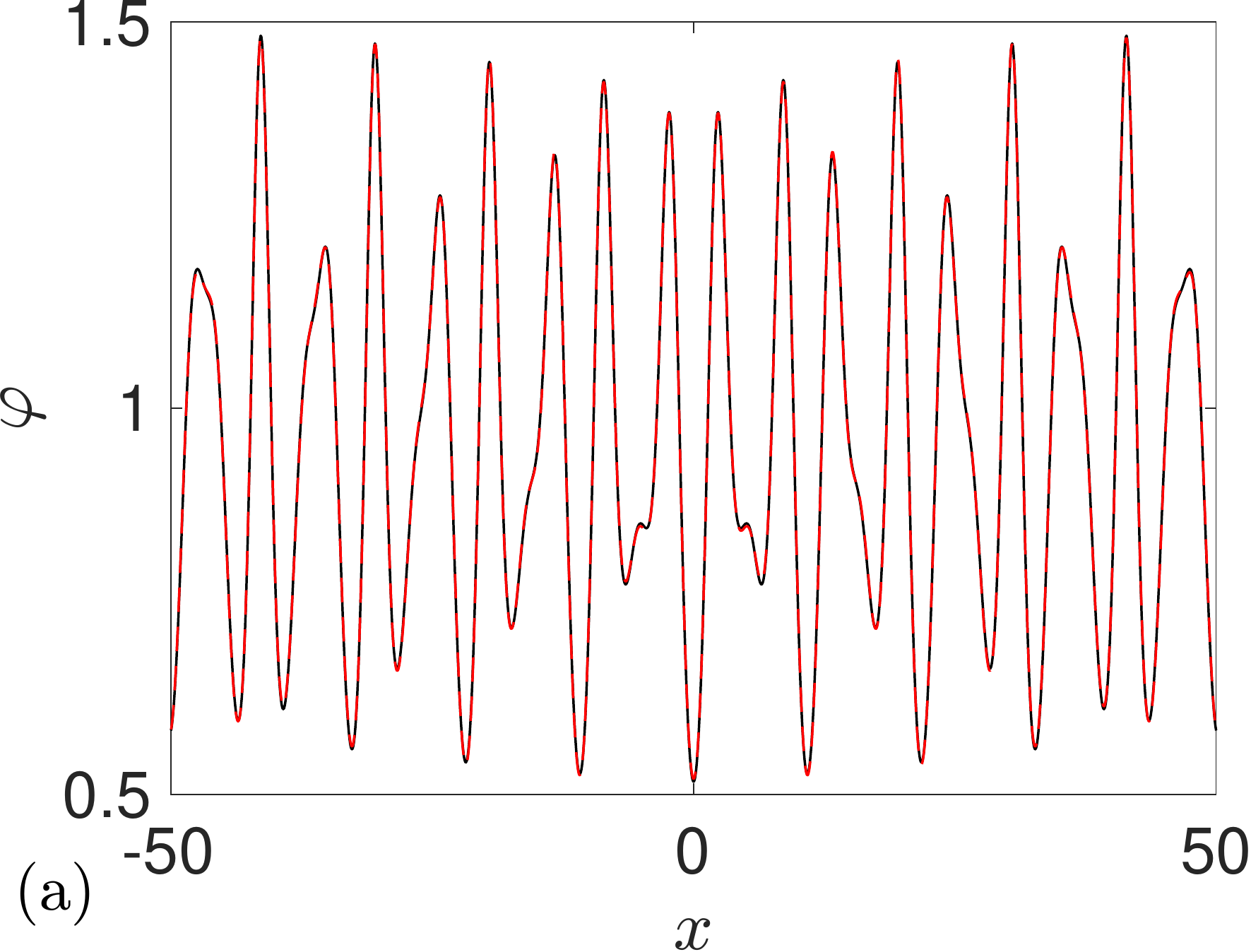}\hspace{1cm}
  \includegraphics[width=7cm]{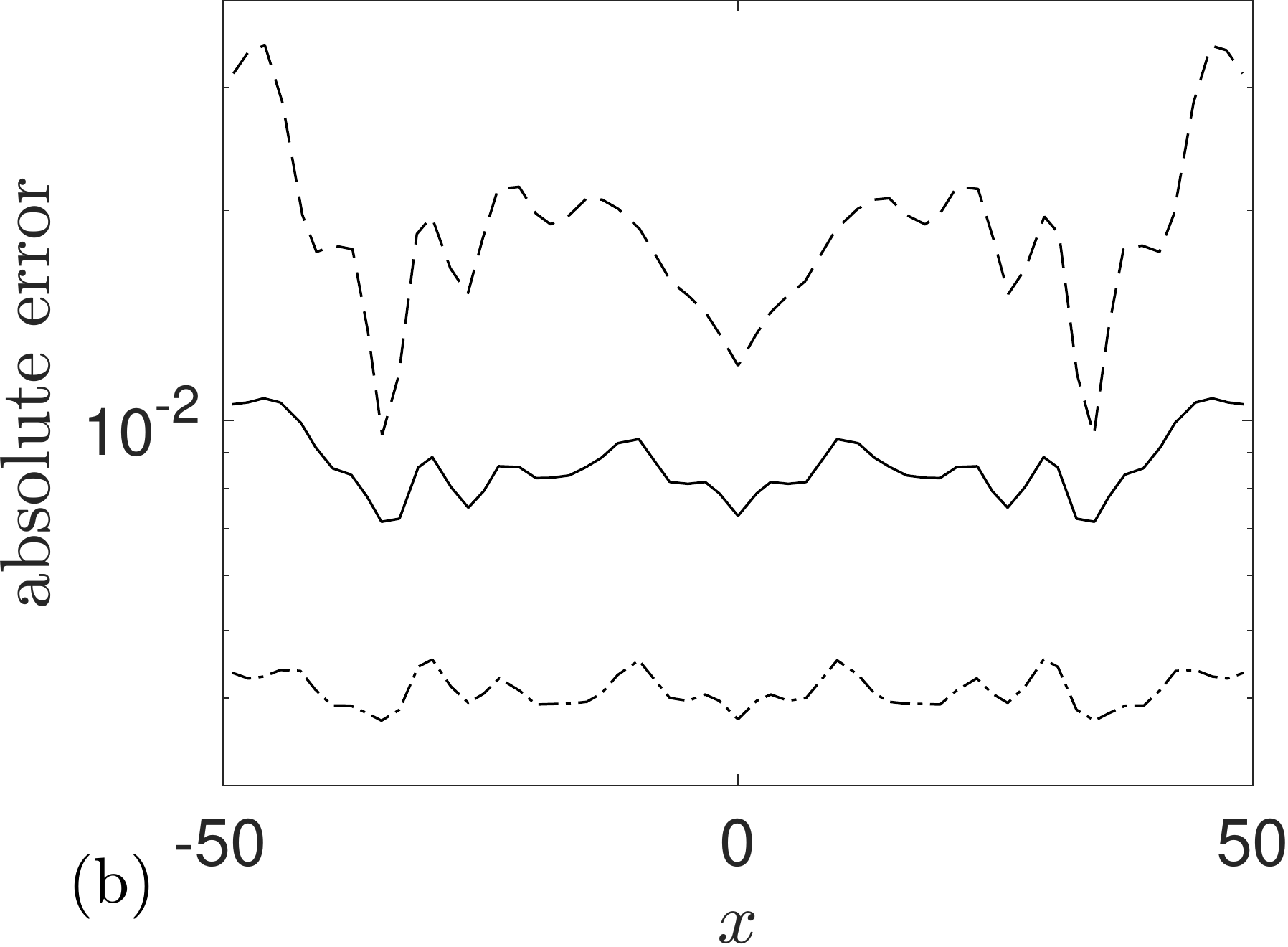}
  \caption{a) Comparison between KdV genus 2 soliton condensate realized via $n=204$-soliton solution  (solid line) and the exact 2-phase KdV solution  (dashed line) for
    $\lambda_1=0.3,\lambda_2=0.5,\lambda_3=0.7,\lambda_4=0.9,\lambda_5=1.$; the two plots are visually indistinguishable from one another; b) Absolute
    error for the condensate generated with $50$ solitons (dashed line),
    $100$ solitons (solid line) and $204$ solitons (dash-dotted line);  for
    readability the absolute error is evaluated at the extrema of the solutions.
  }
  \label{fig:genus23}
\end{figure}

In Fig.~\ref{fig:genus23} 	a realization of the genus 2 soliton condensate is compared with the two-phase KdV solution associated with the same spectral surface and equipped with an appropriately chosen initial phase vector. The two-phase solution has been computed numerically using the so-called trace formula \cite{novikov_theory_1984}:
\begin{equation}
  \varphi(x, t)=\lambda_{1}^2-2 \sum_{j=1}^{2}\left(\mu_{j}^2(x, t)-\frac{\lambda_{2 j}^2+\lambda_{2 j+1}^2}{2}\right),
\end{equation}
where  the auxiliary spectra $\mu_j(x,t)$ satisfy  Dubrovin's ordinary differential equations:
\begin{equation}
  \frac{\partial \mu_{j}^2}{\partial x}=\frac{2 \sigma_{j} R \left(\mu_{j}\right)}{\prod_{j \neq k}^{2}\left(\mu_{j}^2-\mu_{k}^2\right)}, \quad j=1,2
\end{equation}
with $\sigma=\pm 1$ and
\begin{equation}
  R(\mu)=\sqrt{ \left(\mu^2-\lambda_{1}^2\right)\left(\mu^2-\lambda_{2}^2\right) \left(\mu^2-\lambda_{3}^2\right)\left(\mu^2-\lambda_{4}^2\right)\left(\mu^2-\lambda_{5}^2\right)}.
\end{equation}
Each $\mu_j$ oscillates in the corresponding s-gap $[\lambda_{2j-1},\lambda_{2j}]$ so that the sign of $\sigma_j$ changes every time $\mu_j$
changes direction of motion at the gap end point. We observe that, to compute the KdV finite-gap solution corresponding to a given realization soliton condensate, all the initial phases $\mu_j(x_0,t)$ must be placed at the edges of the corresponding s-gaps while the choice of the gap's edge (right/left) is determined by the number of discrete eigenvalues (odd/even) that are located within the s-band $[\lambda_{2j},\lambda_{2j+1}]$ in the numerical construction of the condensate.

\clearpage

\bibliographystyle{plain}

\end{document}